\documentclass{article}
\usepackage[a4paper]{geometry}
\usepackage{amssymb,amsthm}
\usepackage{authblk}
\usepackage{bbm}
\usepackage{bm}
\usepackage{upgreek}
\usepackage{multirow}
\usepackage{natbib}
\usepackage{amsmath}
\usepackage{graphicx}
\usepackage{subcaption}
\usepackage{algorithm}
\usepackage{algorithmic}
\usepackage{url}
\usepackage{tikz}
\usepackage{rotating}
\usetikzlibrary{positioning,chains,fit,shapes,calc,arrows}
\newtheorem{definition}{Definition}[section]
\newtheorem{prop}{Proposition}[section]
\newtheorem{example}{Example}[section]
\newtheorem{thm}{Theorem}[section]
\newtheorem{lemma}{Lemma}[section]
\newcounter{dummy}
\usepackage{enumitem}
\makeatletter
\newcommand\myitem[1][]{\item[#1]\refstepcounter{dummy}\def\@currentlabel{#1}}
\makeatother

\title{ Latent Space Modelling of Hypergraph Data} 
\author[1]{Kathryn Turnbull}
\author[1]{Sim\'{o}n Lunag\'{o}mez}
\author[1]{Christopher Nemeth}
\author[2]{Edoardo Airoldi}
\affil[1]{Department of Mathematics and Statistics, Lancaster University}
\affil[2]{Fox School of Business, Temple University}
\date{}

\definecolor{mycol1}{RGB}{100, 160, 250}
\definecolor{mycol2}{RGB}{250, 140, 50}
\definecolor{hl}{RGB}{230, 5, 218}

\begin{document}

\maketitle

\begin{abstract}
The increasing prevalence of relational data describing interactions among a target population has motivated a wide literature on statistical network analysis. In many applications, interactions may involve more than two members of the population and this data is more appropriately represented by a hypergraph. In this paper, we present a model for hypergraph data which extends the well established latent space approach for graphs and, by drawing a connection to constructs from computational topology, we develop a model whose likelihood is inexpensive to compute. A delayed-acceptance MCMC scheme is proposed to obtain posterior samples and we rely on Bookstein coordinates to remove the identifiability issues associated with the latent representation. We theoretically examine the degree distribution of hypergraphs generated under our framework and, through simulation, we investigate the flexibility of our model and consider estimation of predictive distributions. Finally, we explore the application of our model to two real-world datasets. \\

\end{abstract}


\section{Introduction}
\label{sec:introduction}

\definecolor{myblue}{RGB}{0,0,0}
\definecolor{mygreen}{RGB}{0,0,0}

\begin{figure}[t!]
\begin{subfigure}[t]{.48\textwidth}
  \centering
  \includegraphics[width=.2\textwidth]{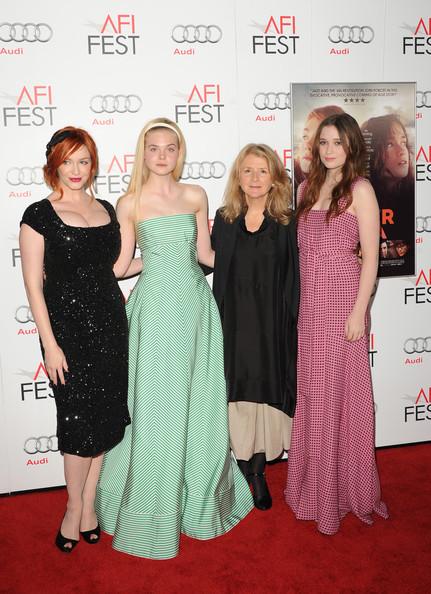}
  \hspace{1cm}
  \includegraphics[width=.2\textwidth]{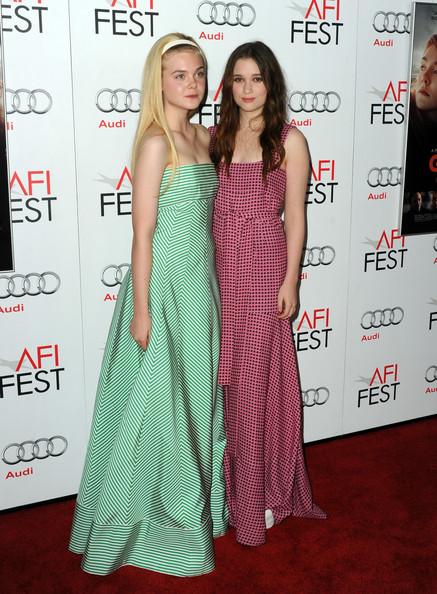}
\caption{Photographs from \cite{zhong2018}} \label{fig:photos}
\end{subfigure}
\begin{subfigure}[t]{.48\textwidth}
  \centering
\begin{tikzpicture}
    \node (u1) at (0,0) {};
    \node (u2) at (0, .8) {};
    \node (u3) at (.8, .8) {};
    \node (u4) at (.8, 0) {};

    \begin{scope}[fill opacity=0.8]
    \filldraw[fill=mycol1!70] ($(u1)+(-.75,0)$) 
        to[out=90,in=180] ($(u2) + (0,.75)$)
        to[out=0,in=0] ($(u3) + (0, .75)$)
        to[out=0,in=0] ($(u4) + (0, -.75)$)
        to[out=180,in=270] ($(u1) + (-.75,0)$) ;
    \filldraw[fill=mycol2!70] ($(u1)+(0,-.5)$) 
        to[out=180,in=180] ($(u2) + (0,.5)$)
        to[out=0,in=0] ($(u1) + (0, -.5)$) ;
    \end{scope}

    \foreach \u in {1,2,3,4} {
        \fill (u\u) circle (0.1);
    }

    \fill (u1) circle (0.1) node [left] {$1$};
    \fill (u2) circle (0.1) node [left] {$2$};
    \fill (u3) circle (0.1) node [right] {$3$};
    \fill (u4) circle (0.1) node [right] {$4$};
\end{tikzpicture}
\caption{Hypergraph from photographs in Figure \ref{fig:photos}.} \label{fig:photo_hyp}
\end{subfigure}
~
  \begin{subfigure}[t]{.32\textwidth}
  \centering
  \includegraphics[width=.7\textwidth, trim={2cm 2cm 1.8cm 2cm}, clip]{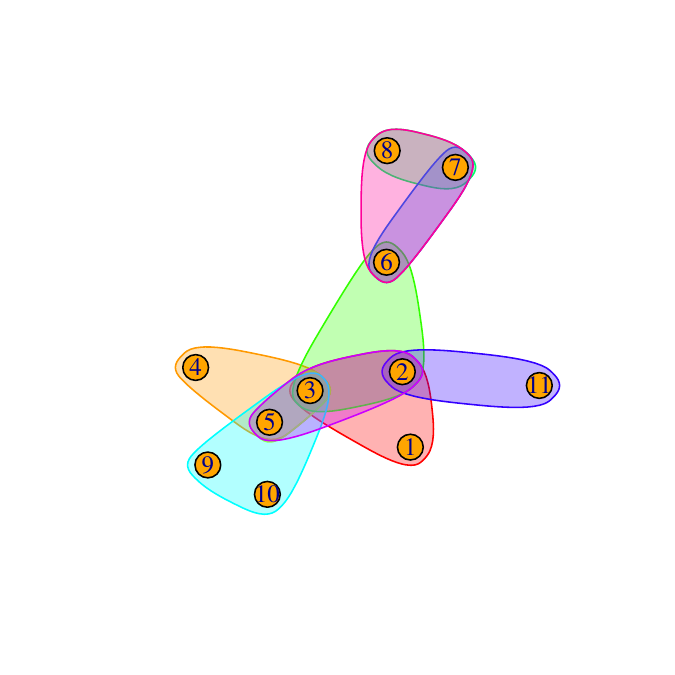}
\caption{Hypergraph subsample obtained via random walk.} \label{fig:coauth_sub_hyp}
\end{subfigure}
~
\begin{subfigure}[t]{.32\textwidth}
  \centering
  \includegraphics[width=.7\textwidth, trim={2cm 2cm 1.8cm 2cm}, clip]{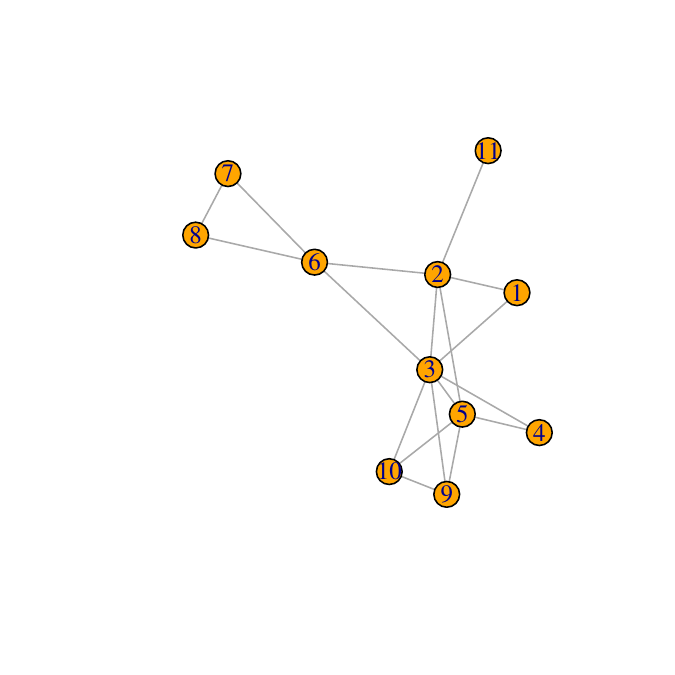}
\caption{Projection of hypergraph from Figure \ref{fig:coauth_sub_hyp} onto a graph.} \label{fig:coauth_sub_graph}
\end{subfigure}
~
  \begin{subfigure}[t]{.32\textwidth}
  \centering
  \includegraphics[width=.7\textwidth, trim={2cm 2cm 1.8cm 2cm}, clip]{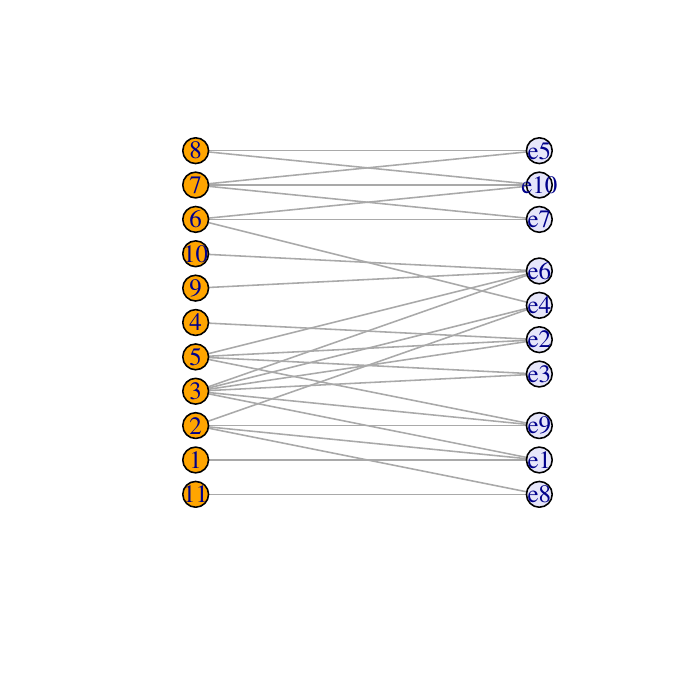}
\caption{Bipartite representation of hypergraph from Figure \ref{fig:coauth_sub_hyp}.} \label{fig:coauth_sub_bipartite}
\end{subfigure}
\caption{Examples of hypergraph datasets. Figure \ref{fig:photo_hyp} represents co-tagging data as a hypergraph and Figure \ref{fig:coauth_sub_hyp} shows a subsample of the coauthorship network of \cite{ji2016}. All but Figure \ref{fig:photos} have been made using  R packages \texttt{HyperG} \citep{hypergpackage} and \texttt{igraph} \citep{csardi2006}. } \label{fig:examples}
\end{figure}

A hypergraph \citep{bretto2013} describes interactions which occur among arbitrary subsets of a population of interest. This data structure is comprised of a node set, which indexes the population, and a hyperedge set, which indicates which members of the population interact. Examples of hypergraphs occur in image co-tagging (see Figure \ref{fig:photos}), where hyperedges indicate users who appear in a same photograph in an online platform, and coauthorship (see Figure \ref{fig:coauth_sub_hyp}), where hyperedges represent the set of authors who collaborated on an academic article. A key feature of a hypergraph representation is that is able to express higher-order interactions and, whilst these data could be analysed according to an extensive graph modelling literature (see \cite{kolaczyk2009}, \cite{goldenberg2010}, \cite{barabasi2016} and \cite{saltertownshend2012}), this results in a loss of structural information which cannot be recovered (see Figures \ref{fig:coauth_sub_hyp} and \ref{fig:coauth_sub_graph}). This represents the main motivation of this article where we develop a novel model for hypergraph data, detail a procedure for inference and present theoretical results.

More formally, a hypergraph $G = (V, E)$ consists of a set of $N$ node labels $V$ and $M$ hyperedges $E$, where $e \in E$ contains no repeated elements and $e \subseteq V$ (see Figure \ref{fig:coauth_sub_hyp}). This data type can also be equivalently represented as a bipartite graph with an $(N \times M)$ adjacency tensor in which each hyperedge is indexed by a node and an edge from a population node to a hyperedge node indicates presence in a hyperedge (see Figure \ref{fig:coauth_sub_bipartite}). Whilst a significant portion of the existing literature relies on the bipartite representation, we focus on the former representation since it allows us to develop a model which allows for an arbitrary number of hyperedges by avoiding conditioning on $M$.

We consider extending the latent distance model of \cite{hoff2002} to the hypergraph setting using the representation in Figure \ref{fig:coauth_sub_hyp}. In this approach, a low-dimensional coordinate is associated with each node and nodes whose latent coordinates are close in terms of Euclidean distance are assumed more likely to share a tie. This modelling choice is appealing since the latent representation offers an intuitive visualisation of the data, encourages transitive relationships due to the metric property of Euclidean distance and allows control over the joint distribution of subgraph counts. Furthermore, once estimated, models of this type allow a practitioner to quantify uncertainty in interaction probabilities and predict future connections. When developing our model, we wish to take advantage of hypergraph analogues of these properties and we focus on the following inferential questions.
\begin{enumerate}[noitemsep]
  \myitem[(Q1)]  \label{item:q1} \emph{``How can we determine an intuitive visualisation of an observed hypergraph?''} 
\myitem[(Q2)]  \label{item:q2} \emph{``How do we expect new nodes to interact with the observed hypergraph relationships?''} 
\end{enumerate}
Latent space network models are well-established in the graph modelling literature and have been extended to an array of network data types \citep{mccormick2015, kim2017, saltertownshend2017, dangelo2019} and address inferential questions ranging from community detection \citep{handcock2007, krivitsky2009} to graph comparisons \citep{asta2014, tang2017}. Whilst we do not consider these questions in this article, they suggest promising avenues for future work and further motivate our interest in this class of models.

There exists a growing literature concerned with hypergraph modelling \citep{battiston2020}, and several well-adopted graph models have been extended to the bipartite setting. Examples include exponential random graphs \citep{wang2013}, blockmodels \citep{doreian2004}, preferential attachment \citep{ergun2002, guillaume2006}, configuration models \citep{fosdick2018} and latent space network models \citep{friel2016, kitsak2017, vasques2020}. We note that, in contrast to our approach, these models assume that the number of hyperedges $M$ is known and fixed. 

Examples of models that are appropriate for the full generality of hypergraphs include the $\beta$ model of \cite{stasi2014}, models based on preferential attachment \citep{wang2010, liu2013} and the clustering-based model of \cite{ng2018}. A rich set of models have also been developed for uniform and simplicial classes of hypergraphs. In a $m$-uniform hypergraph all hyperedges contain exactly $m$ elements, and extensions of stochastic blockmodels (see \cite{ghoshdastidar2014}, \cite{kim2018}, \cite{chien2018}, \cite{pal2019}, \cite{ahn2019}) and graphons \citep{balasubramanian2021} have been studied for this hypergraph type. Simplicial hypergraphs, or rather simplicial complexes (see \cite{edelsbrunner2010} for an introduction), are a significant class of hypergraphs in which the presence of a hyperedge implies the presence of all subsets of that hyperedge. There is a rich literature that deals with modelling simplicial complexes and this includes an analogue of exponential random graphs \citep{zuev2015}, preferential attachment based models \citep{bianconi2017, wu2015}, simplicial configuration models \citep{courtney2016, young2017} and models based on recursive procedures in which hyperedges are included with increasing order \citep{kahle2014, costa2016}. We refer the reader to recent reviews for an in-depth discussion of random simplicial complexes \citep{kahle2014, kahle2016} and highlight here the work on geometric simplicial complexes \citep{bobrowski2018, bobrowski2019} which is most related to our proposed procedure. However, we note that our focus on inference marks a clear distinction between our work and much of the existing literature. Finally, we comment that simplicial complexes appear more broadly in the statistics literature \citep{pronzato2019, lunagomez2009, chazal2017} and that our terminology `simplicial hypergraph' is non-standard in this literature.

The main contributions of this article are as follows. First, using the representation shown in Figure \ref{fig:coauth_sub_hyp}, we develop a computationally tractable latent space model for non-simplicial hypergraph data by relying on tools from computational geometry \citep{edelsbrunner2010}. Second, to remove non-identifiability present in our model, we define the latent representation on the space of Bookstein coordinates which have so far not been explored in this context. {Third, we perform inference on hypergraph data that is not within the reach of competing methods and present a novel application of delayed-acceptance MCMC methodology.} Finally, we theoretically investigate the properties of the degree distribution of our model and, whilst this proves challenging, our discussion provides an outline which can be explored for other modelling choices.

The rest of this paper is organised as follows. Section \ref{sec:netlit} provides the background for the hypergraph model presented in Section \ref{sec:lshyps}. Section \ref{sec:theory} considers the degree distribution of our model and Section \ref{sec:postsmp} describes a procedure for obtaining posterior samples. The simulation studies and real data examples are presented in Sections \ref{sec:sims} and \ref{sec:realdat}, respectively, and we conclude with a discussion in Section \ref{sec:disc}.

\section{Background}
\label{sec:netlit}

In this section we review the network modelling literature which forms the basis of our proposed hypergraph model. In Section \ref{sec:lsnet} and \ref{sec:rggs}, we discuss the latent space framework of \cite{hoff2002} random geometric graphs, respectively. Then, in Section \ref{sec:rghs} we describe how a random geometric graph can be extended to model a restricted class of hypergraphs.

\subsection{Latent Space Network Modelling}
\label{sec:lsnet}

Latent space models were introduced for network data in \cite{hoff2002}. The key assumption of this framework is that the nodes of a network can be represented in a low-dimensional latent space, and that the probability of an edge forming between each pair of nodes can be modelled as a function of their corresponding latent coordinates.

To describe this model for a $N$ node network, we let $\bm{Y} = \{y_{ij}\}_{i,j=1,2,\dots,N}$ denote the observed $(N \times N)$ adjacency matrix, where $y_{ij}$ represents the connection between nodes $i$ and $j$. For binary interactions, we take $y_{ij}=1$ if $i$ and $j$ share an edge and $y_{ij}=0$ otherwise. We also let $u_i \in \mathbb{R}^d$ denote the $d$-dimensional latent coordinate associated with the $i^{th}$ node, for $i= \{ 1,2,\dots,N \} = [N]$. The presence of an edge is then given by
\begin{align}
  \begin{split}
    Y_{ij} &\sim \mbox{Bernoulli}( p_{ij} ) \\
    p_{ij} &= P(y_{ij}=1 |u_i,u_j,\theta) = \dfrac{1}{1 + \exp\{- f( u_i, u_j, \theta ) \}}, \label{eq:model}    
  \end{split}
\end{align}
for $i,j \in [N]$, where $\theta$ represents additional model parameters and $f$ controls how the latent coordinates affect the propensity for ties to form. It is typical to chose $f$ to be monotonically decreasing in a measure of similarity between $u_i$ and $u_j$. For example, the distance model introduced in \cite{hoff2002} is obtained by choosing
\begin{align}
	f( u_i, u_j, \theta ) = \alpha - \| u_i - u_j \|, \label{eq:distmod}
\end{align}
where $\| \cdot \|$ is the Euclidean distance, and $\theta = (\alpha)$ represents the base-rate tendency for edges to form. The function $f$ may also be adapted to incorporate covariate information so that nodes which share certain characteristics are more likely to be connected. 
The likelihood, conditional on $\bm{U} = \{ u_i \}_{i=1}^N$ and $\theta$, is given by
\begin{align}
  \mathcal{L}(\bm{U}, \theta ; \bm{Y}) \propto \prod_{i<j} 	P( y_{ij}=1 | u_i, u_j, \theta )^{y_{ij}} \left[ 1 - 	P( y_{ij}=1 | u_i, u_j, \theta ) \right]^{1 - y_{ij}}, \label{eq:hofflike}
\end{align}
Properties of this model are well understood and we refer to \cite{rastelli2015} for details.

\subsection{Random Geometric Graphs}
\label{sec:rggs}

Random geometric graphs (RGGs) \citep{penrose2003} can be viewed as a special case of the latent space network model outlined in Section \ref{sec:lsnet}. To express the connection probabilities, we let $u_i = (u_{i1}, u_{i2}, \dots, u_{id}) \in \mathbb{R}^d$ and associate each node with a ball of radius $r$ that is centered on its latent coordinate, namely $B_r(u_i) = \{ u \in \mathbb{R}^d | \: \:\| u - u_i \| \leq r \} =$ $\left\{u \in \mathbb{R}^d \middle| \right. $ $ \left. \: \: \sqrt{ \sum_{j=1}^d (u_j - u_{ij})^2} \leq r \right\}$. The presence of an edge is then expressed deterministically as
\begin{align}
  p_{ij} = P( y_{ij} = 1 | u_i, u_j, \theta) = \mathbbm{1}\left( B_r(u_i) \cap B_r(u_j) \neq \emptyset \right), \label{eq:pij_rgg_ball}
\end{align}
where $\theta=r$, so that nodes $i$ and $j$ share a tie only if their corresponding sets have a nonempty intersection (see the left and middle panel of Figure \ref{fig:randomgraph}). Since this procedure is equivalent to connecting nodes $i$ and $j$ when $\| u_i - u_j\| \leq 2r$, we may also write
\begin{align}
  p_{ij} = P( y_{ij} = 1 | u_i, u_j, r) = \mathbbm{1}\left( \| u_i - u_j \| \leq 2r \right). \label{eq:pij_rgg_dist}
\end{align}
Maintaining the notation from Section \ref{sec:lsnet}, we can now express the likelihood of observing $\bm{Y}$ conditional on $\bm{U}$ and $r$ as
\begin{align}
  \mathcal{L}( \bm{U}, r ; \bm{Y} ) \propto \prod_{i<j} \mathbbm{1} ( \| u_i - u_j \| \leq 2r )^{y_{ij}} \left[ 1 - \mathbbm{1}( \| u_i - u_j \| \leq 2r) \right]^{1 - y_{ij}}. \label{eq:rgglike}
\end{align}
Since a RGG is deterministic conditional on $\bm{U}$ and $r$, \eqref{eq:rgglike} is equal to 1 only when there is a perfect correspondence between the observed connections $\bm{Y}$ and the connections induced by the latent positions and radius, and is otherwise equal to 0.


\subsection{Random Geometric Hypergraphs}
\label{sec:rghs}

We can generalise the graph generating procedure from Section \ref{sec:rggs} to model hypergraphs by considering the full intersection pattern of convex sets, and we refer to these hypergraphs as random geometric hypergraphs (RGHs). To begin, we introduce the concept of a nerve \citep[Section 3.2]{edelsbrunner2010} in the following definition. 
\begin{definition}{(Nerve)}
\label{def:nerve}
Let $\mathcal{A}=\{A_i\}_{i=1}^N$ represent a collection of non-empty sets. The nerve of $\mathcal{A}$ is given by $Nrv ( \mathcal{A} ) = \left\{ \sigma \subseteq \{ 1,2,\dots,N\} \middle| \underset{j \in \sigma}{\cap}  A_j \neq \emptyset \right\}$.

\end{definition} 

Intuitively, the nerve represents the set of indices whose corresponding regions have a non-empty intersection. Note also that the sets $\{1\}, \{2\}, \dots, \{N\}$ are included in $Nrv( \mathcal{A})$ and that $| \sigma | \leq N$ for $\sigma \in Nrv( \mathcal{A})$, where $| \sigma |$ is the order, or dimension, of the set. Clearly, the nerve defines a hypergraph where $\sigma \in Nrv(\mathcal{A})$ represents a hyperedge. Note that for sets $\sigma_1 \in Nrv (\mathcal{A})$ and $\sigma_2 \subset \sigma_1$, it follows immediately that $\sigma_2 \in Nrv(\mathcal{A})$. This property indicates that the hypergraph generated by a nerve is \emph{simplicial} \citep{kahle2016}.

 In Section \ref{sec:rggs}, we considered the choice $A_i = B_r(u_i)$ for generating a RGG. The nerve for this choice of $\mathcal{A}$ is well studied and it is referred to as the \v{C}ech complex \citep[Section 3.2]{edelsbrunner2010}, as given in Definition \ref{def:cech}. 

\begin{definition}{(\v{C}ech Complex)}
\label{def:cech}
  For a set of coordinates $\bm{U}=\{u_i\}_{i=1}^N$ and a radius $r$, the \v{C}ech complex $\mathcal{C}_r (\bm{U})$ is given by $\mathcal{C}_r (\bm{U}) = Nrv \left( \{ B_r(u_i) \}_{i=1}^N \right).$
\end{definition}


{
\begin{figure}[t!]
  \centering
  \includegraphics[width=0.32\textwidth]{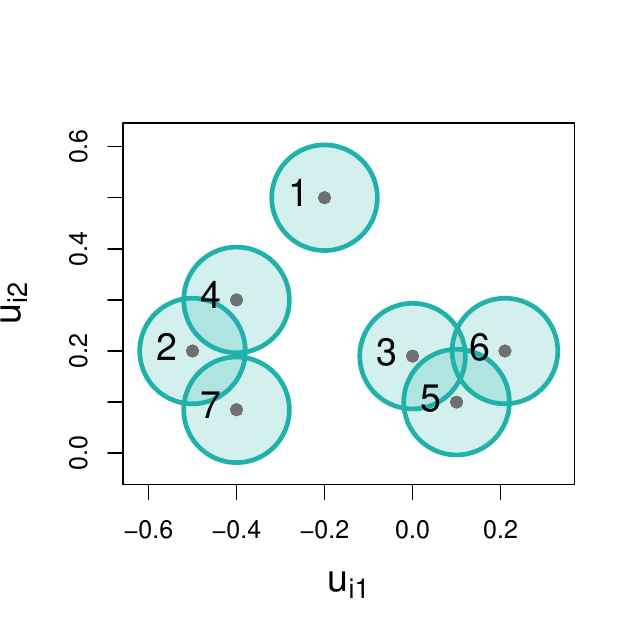}
  \includegraphics[width=0.32\textwidth]{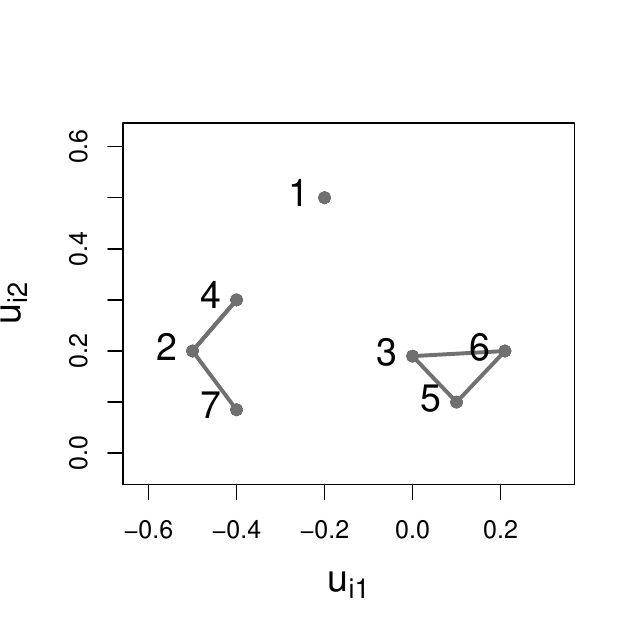}
  \includegraphics[width=0.32\textwidth]{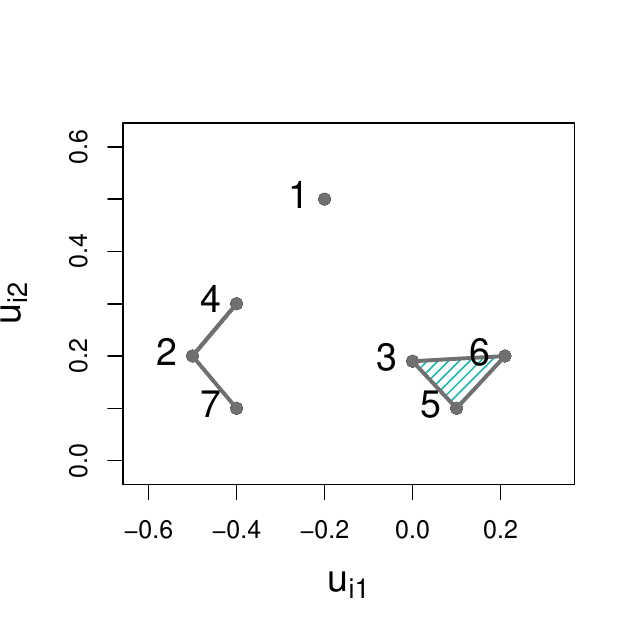}
  \vspace{-0.4cm}
  \caption{Example of a \v{C}ech complex. Left: $B_r(u_i)$ for $\{ u_i = (u_{i1}, u_{i2})\}_{i=1}^7$ in $\mathbb{R}^2$. Middle: the graph obtained by taking pairwise intersections. Right: the hypergraph obtained by taking intersections of arbitrary order. The shaded region indicates a order 3 hyperedge.}
  \label{fig:randomgraph}
\end{figure}
} 

We now introduce a subset of the \v{C}ech complex known as the $k$-skeleton which will be revisited in Section \ref{sec:unionsimp}. Below, we present a precise definition along with an example.
\begin{definition}{($k$-skeleton of the \v{C}ech complex) }
\label{def:jskel}
Let $\mathcal{C}_r (\bm{U})$ denote the \v{C}ech complex, as given in Definition \ref{def:cech}. The $k$-skeleton of $\mathcal{C}_r (\bm{U})$ is given by $ \mathcal{C}_r^{(k)} (\bm{U}) = \{ \sigma \in \mathcal{C}_r (\bm{U}) | | \sigma | \leq k\} $.
\end{definition}
 \begin{example}
   Figure \ref{fig:randomgraph} depicts an example of a \v{C}ech complex, where $\mathcal{C}_r(\bm{U}) = \{ \{1\},$ $\{2\},\{3\}, \{4\},$ $\{5\}, \{6\}, \{7\}, \{2,4\}, \{2,7\}, \{3,5\}, $ $\{3,6\}, \{5,6\} , \{3,5,6\}\}$. The $k$-skeletons are given by $\mathcal{C}_r^{(1)}(\bm{U}) = \{ \{1\},\{2\}, \{3\},$ $\{4\}, \{5\}, \{6\}, \{7\} \}$,  $\mathcal{C}^{(2)}_r(\bm{U}) = \mathcal{C}_r^{(1)} \cup \{ \{2,4\}, \{2,7\},$ $\{3,5\}, \{3,6\}, \{5,6\} \}$ and $\mathcal{C}^{(3)}_r(\bm{U})$ $= \mathcal{C}^{(2)}_r (\bm{U}) \cup \{3,5,6 \}$.
 \end{example}

\section{Latent space hypergraphs}
\label{sec:lshyps}

We now introduce a model for hypergraph data which builds upon the models discussed in the previous section. The motivation and notation for our approach are given in Section \ref{sec:motivation} and a preliminary model for non-simplicial hypergraphs is given in Section \ref{sec:unionsimp}. Finally, we present our generative model and likelihood in Section \ref{sec:like} along with a more flexible modification in Section \ref{sec:like}.

\subsection{Motivation and Notation}
\label{sec:motivation}

Recall the motivating examples of coauthorship and image co-tagging (see Figure \ref{fig:examples}) where we observe arbitrary interaction patterns. For these examples, the model discussed in Section \ref{sec:rghs} is likely too restrictive. This motivates us to develop a model which builds upon the ideas presented in Section \ref{sec:netlit} and is appropriate for non-simplicial hypergraphs. Crucially, we aim to develop a model which has (i) a convenient and computationally-tractable likelihood with (ii) a simple to describe support for general hypergraph data.

We consider hypergraphs on $N$ nodes with arbitrary interaction patterns and let $K$ denote the maximum hyperedge order, for $2 \leq K \leq N$. We denote the space of hypergraphs on $N$ and $K$ as $\mathcal{G}_{N,K}$ and write $\mathcal{G}_{N,K} = ( \mathcal{V}_N, \mathcal{E}_{N,K})$, where $\mathcal{V}_N = \{1,2,\dots,N\}$ and $\mathcal{E}_{N,K}$ denote the node labels and the set of possible hyperedges up to order $K$ on $N$ nodes, respectively. Additionally, we let $\mathcal{E}_{N,k}$ represent the possible hyperedges of exactly order $k$ on $N$ nodes, so that $\mathcal{E}_{N,K} = \cup_{k=2}^K \mathcal{E}_{N,k}$, and let $e_k \in \mathcal{E}_{N,k}$ denote an order $k$ interaction, where $|e_k| = k$ and $e_k$ contains no repeated elements. 

\subsection{Combining $k$-skeletons} 
\label{sec:unionsimp}

To extend the model in Section \ref{sec:rghs} to express non-simplicial hypergraphs we allow the radii to differ for each hyperedge order. More specifically, we generate a hypergraph by isolating the order $k$ hyperedges from the complex $\mathcal{C}_{r_k}(\bm{U})$ and combining these for $k=2,3,\dots,K$, where $r_k \neq r_{k'}$ for $k \neq k'$. As an example, consider radii $r_2$ and $r_3$ and their associated \v{C}ech complexes $\mathcal{C}_{r_2}(\bm{U})$ and $\mathcal{C}_{r_3}(\bm{U})$. Here we take the order 2 hyperedges in $\mathcal{C}_{r_2}(\bm{U})$ and the order 3 hyperedges in $\mathcal{C}_{r_3}(\bm{U})$ to obtain a non-simplicial interaction pattern, as demonstrated in Figure \ref{fig:nonsimpex}. We refer to a hypergraph constructed in this way as a non-simplicial random geometric hypergraph (nsRGH) and a formal description of this is given in Definition \ref{def:nsrgh}.

\begin{definition}{(nsRGH)} \label{def:nsrgh}
  Take $\bm{r}=(r_2,r_3,\dots,r_K)$ such that $r_k > r_{k-1} > 0$ for $k=3,4,\dots,K$. We define a nsRGH on $N$ nodes as the hypergraph with hyperedges given by $\cup_{k=2}^K \mathcal{D}_{r_k}^{(k)} (\bm{U})$, where $\mathcal{D}_{r_k}^{(k)} (\bm{U}) = \mathcal{C}_{r_k}^{(k)}(\bm{U}) \setminus \mathcal{C}_{r_k}^{(k-1)} (\bm{U})$ denotes the hyperedges of exactly order $k$ in the \v{C}ech complex with radius $r_k$ and $\mathcal{C}^{(k)}_{r_k} (\bm{U})$ is as in Definition \ref{def:jskel}. We denote this construction as $g(\bm{U}, \bm{r})$ throughout.
\end{definition}

\begin{example}
  For the nsRGH shown in the right panel of Figure \ref{fig:nonsimpex} we have $g(\bm{U}, \bm{r}) = \mathcal{D}_{r_2}^{(2)} (\bm{U}) \cup \mathcal{D}_{r_3}^{(3)} (\bm{U})$, where $\mathcal{D}_{r_2}^{(2)} (\bm{U}) = \{ \{2,4\}, \{2,7\}, \{3,5\}, \{5,6\} \}$ and $\mathcal{D}_{r_3}^{(3)} (\bm{U}) = \{ 3,5,6\}$.
\end{example}

{
\begin{figure}[t]
    \centering
        \includegraphics[width=0.32\textwidth]{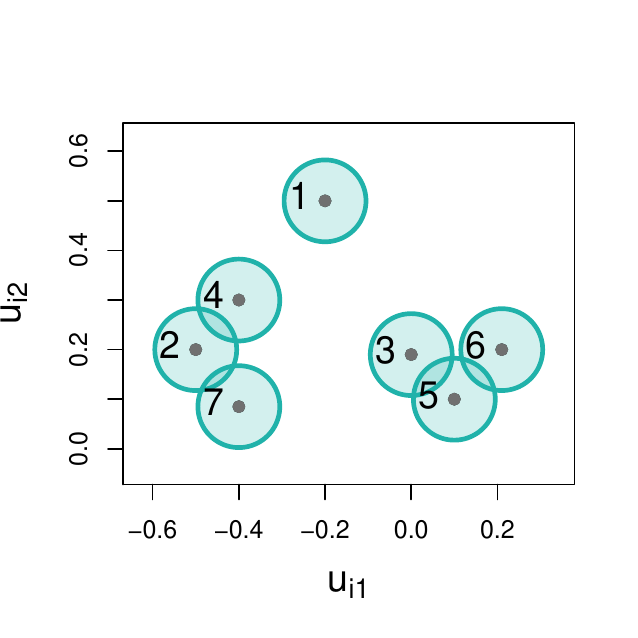}
        \includegraphics[width=0.32\textwidth]{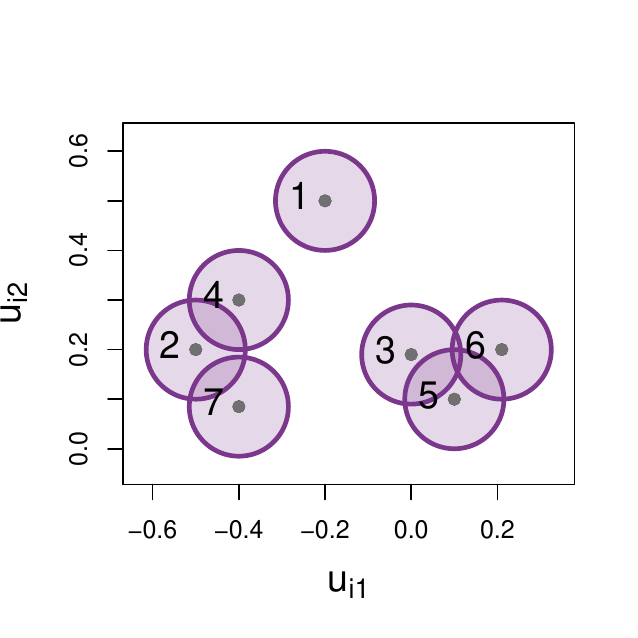}
        \includegraphics[width=0.32\textwidth]{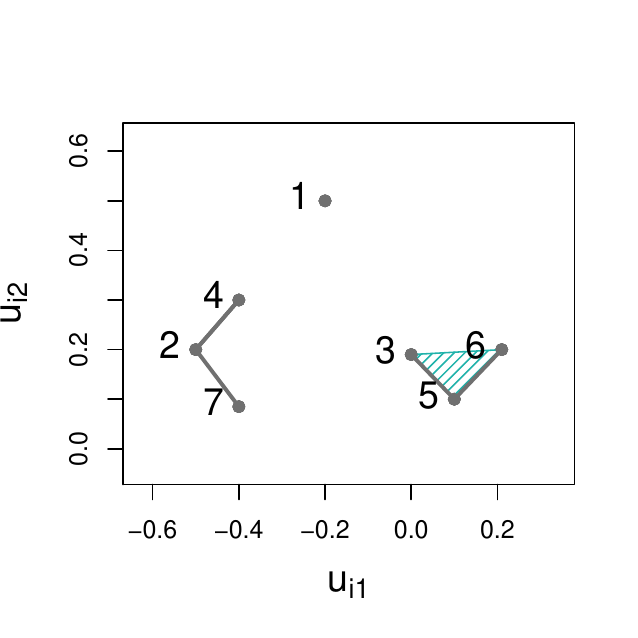}
        \vspace{-0.4cm}
    \caption{Example of a nsRGH (see Definition \ref{def:nsrgh}) with $\bm{U} = \{u_i\}_{i=1}^7 = \{(u_{i1}, u_{i2})\}_{i=1}^7$. Left: $\mathcal{C}_{r_2} (\bm{U})$. Middle: $\mathcal{C}_{r_3} (\bm{U})$. Right: $\cup_{k=2}^3 \mathcal{D}_{r_k}^{(k)}(\bm{U})$.}
    \label{fig:nonsimpex}
\end{figure}
} 

In Definition \ref{def:nsrgh}, constraints are imposed on the radii $\bm{r}=(r_2,r_3,\dots,r_K)$ to ensure that the generated hypergraphs are non-simplicial. To see this note that, if $r_3 < r_2$ and the hyperedge $\{i,j,k\}$ is present in the hypergraph, then it follows that the hyperedges $\{i,j\}, \{i,k\}$ and $\{j,k\}$ must also be in the hypergraph. 

\subsection{Generative Model and Likelihood}
\label{sec:like}

Since the procedure of generating a nsRGH $g( \bm{U}, \bm{r})$ (see Definition \ref{def:nsrgh}) is deterministic conditional on $\bm{U}$ and $\bm{r}$, there are two impediments to using this model for real data. Firstly, estimation via likelihood based methods is challenging since the probability of observing a hypergraph $h_{N,k} \in \mathcal{G}_{N,K}$ is binary and equal to one only if there is a perfect correspondence between hyperedges in $g( \bm{U}, \bm{r})$ and $h_{N,K}$. {Therefore, the likelihood cannot be used to guide an estimation procedure to probable values of $\bm{U}$ since a candidate estimate can only be evaluated as precisely correct or incorrect.} Secondly, it is not straightforward to characterise the space of hypergraphs that can be expressed as a nsRGH and so we do not know a priori whether a latent space representation can be found. 

To address these issues, we introduce a modification of a nsRGH whereby the state of the hyperedges in $g(\bm{U}, \bm{r})$ are perturbed according to a small amount of noise. More specifically, we modify the indicator denoting presence or absence for each hyperedge with probability $\upvarphi_k \in [0,1]$ as follows. Let $g^*(\bm{U}, \bm{r}, \bm{\upvarphi})$ denote the modified hypergraph, and let $y_{e_k}^{(g)} \in \{0,1\}$ and $y_{e_k}^{(g^*)} \in \{0,1\}$ denote the state of hyperedge $e_k$ in $g(\bm{U}, \bm{r})$ and $g^*(\bm{U}, \bm{r}, \bm{\upvarphi})$, respectively, where a state equal to one indicates presence. Then, we take
\begin{equation}
  y_{e_k}^{(g^*)}=
  \begin{cases}
    0, & \text{if}\ s_k=1 \ \text{and}\ y_{e_k}^{(g)}=1 \\
    1, & \text{if}\ s_k=1 \ \text{and}\ y_{e_k}^{(g)}=0 \\
    y_{e_k}^{(g)}, & \text{if}\ s_k=0  \\
  \end{cases} \label{eq:ygstr}
\end{equation}
where $S_k \sim \mbox{Bernoulli}( \upvarphi_k) $, so that $y_{e_k}^{(g)}$ is modified from either present to absent or absent to present with probability $\upvarphi_k$. The noise term $\upvarphi_k$ controls the amount of modification to the hyperedges of order $k$ and setting $\upvarphi_k$ close to $0$ obtains a hypergraph that is similar to a nsRGH. Crucially, this modification results in a model that has the desired properties highlighted in Section \ref{sec:motivation} since we extend the support of nsRGHs to the full space of hypergraphs and the likelihood is no longer binary for $\upvarphi_k \notin \{0, 1\}$.

To fully specify a generative model, we assign a probability distribution on the latent coordinates $\bm{U}$. We let $u_i \overset{iid}{\sim} \mathcal{N}(\mu, \Sigma)$, for $i \in [N]$ to reflect the intuition that nodes placed more centrally in the latent representation are likely to have high degree and neighbours with high degree. A hypergraph can now be generated by the procedure given in Algorithm \ref{alg:modhyp}, and details of efficient implementation of this are given in Appendix \ref{app:appnoise} and \ref{app:miniball}.

\begin{algorithm}[t]
  \caption{Sample a hypergraph $g^*_{N,K}$ given $N,K,\bm{r}, \bm{\upvarphi}, \mu$ and $\Sigma$.} \label{alg:modhyp}
  \begin{algorithmic}
\STATE Sample $\bm{U} = \{u_i\}_{i=1}^N$ such that $u_i \overset{iid}\sim \mathcal{N}(\mu, \Sigma)$, for $i=1,2,\dots,N$.
\STATE For $k=2,3,\dots,K$, \\
  \hspace{.5cm} a) Given $\bm{U}$ and $r_k$, check which $e_k = \{i_1, i_2, \dots, i_k \} \in \mathcal{E}_{N,k}$ satisfy $y_{e_k}^{(g)}=1$. \\
  \hspace{1cm} To determine if $y_{e_k}^{(g)}=1$, check that $\cap_{l=1}^k B_{r_k}(u_{i_l}) \neq \emptyset$. \\
  \hspace{.5cm} b) For all $e_k \in \mathcal{E}_{N,k}$, sample $S_k \sim \mbox{Bernoulli}(\upvarphi_k)$. \\
  \hspace{1cm} Let $y_{e_k}^{(g^*)} = \left(y_{e_k}^{(g)} + s_k \right) \mod 2$
  \end{algorithmic}
\end{algorithm}

We express the likelihood of an observed hypergraph $h_{N,K} \in \mathcal{G}_{N,K}$, conditional on $\bm{U}, \bm{r}$ and $\bm{\upvarphi}$, by considering the discrepancy between the hyperedge configurations in $h_{N,K}$ and $g_{N,K}(\bm{U}, \bm{r})$. For $k=2,3,\dots,K$, let
\begin{align}
  d_k( g_{N,K}(\bm{U}, \bm{r}), h_{N,K}) = \sum_{e_k \in \mathcal{E}_{N,k}} | y_{e_k}^{(g)} - y_{e_k}^{(h)} | \label{eq:l1}
\end{align}
denote the discrepancy between the order $k$ hyperedges in $g_{N,K}(\bm{U}, \bm{r})$ and $h_{N,K}$, where $y_{e_k}^{(h)} \in \{0,1\}$ indicates the state of $e_k$ in $h_{N,K}$. Intuitively, the distance \eqref{eq:l1} corresponds to the number of hyperedges whose state differs between $h_{N,K}$ and $g_{N,K}(\bm{U}, \bm{r})$. This is equivalent to the Hamming distance and is related to both the $l_1$ norm and the exclusive or (XOR) operator. Evaluating this distance does not require the $\sum_{k=2}^K {{N} \choose {k}}$ computations suggested by \eqref{eq:l1}, and we can instead evaluate the discrepancy by only considering hyperedges that are present in $g_{N,K}(\bm{U}, \bm{r})$ and $h_{N,K}$. In practice, this is likely to be far less than the number of possible hyperedges and details of this are discussed in Appendix \ref{app:evallikelihood}. 

Given this notion of hypergraph distance the likelihood of observing $h_{N,K}$, conditional on $\bm{U},\bm{r}$ and $\bm{\upvarphi}$, can be written as
\begin{align}
  \mathcal{L}( \bm{U}, \bm{r}, \bm{\upvarphi} ; h_{N,K} ) \propto \prod_{k=2}^K \upvarphi_k^{ d_k(g_{N,K}(\bm{U}, \bm{r}), h_{N,K}) } (1 - \upvarphi_k)^{{N \choose k} - d_k(g_{N,K}(\bm{U}, \bm{r}), h_{N,K})  }.  \label{eq:likelihood}
\end{align}

We obtain \eqref{eq:likelihood} by considering which hyperedges in $g_{N,K}(\bm{U}, \bm{r})$ must have their state modified to match the hyperedges in $h_{N,K}$, and which hyperedges are the same as in $h_{N,K}$. For order $k$ hyperedges which differ, the probability of switching the hyperedge state is given by $\upvarphi_k$. Since our likelihood is of the same form as \citet[][proof of Proposition 3.1]{lunagomez2020}, it follows that hypergraphs with a greater number of hyperedge modification are less likely for $0 < \upvarphi_k < 1/2$ and so \eqref{eq:likelihood} behaves in an intuitive way.

The model specification is complete with the following priors
\begin{align}
  \mu \sim \mathcal{N}(m_{\mu}, \Sigma_{\mu}) \hspace{.5cm} \Sigma_{\mu} \sim \mathcal{W}^{-1}(\Phi, \nu), \hspace{.5cm} r_k \sim \exp( \lambda_k ), \hspace{.25cm} \mbox{and} \hspace{.25cm} \upvarphi_k \sim \mbox{Beta}(a_k, b_k), \label{eq:modpri}
\end{align}
for $k=2,3,\dots,K$, where $\mathcal{W}^{-1}(\cdot, \cdot)$ denotes an Inverse-Wishart. 

\subsection{Can we improve model flexibility?}
\label{sec:ext}

In a nsRGH, the constraint $r_{k} > r_{k-1}$, for $k=3,\dots,K$, ensures that the interactions are non-simplicial. However, note that this implies $r_k$ will impact the higher-order hyperedges. To improve model flexibility, we can introduce an additional modification parameter for each hyperedge order. In Algorithm \ref{alg:modhyp} the noise $\upvarphi_k$ is applied independently across all hyperedges of order $k$ and, alternatively, we can modify each hyperedge depending on its state in $g_{N,K}(\bm{U}, \bm{r})$. For $k=2,3,\dots,K$, let $\bm{\psi}^{0} = ( \psi_2^{0}, \psi_3^{0}, \dots, \psi_K^{0}) \in [0,1]$ denote the probability of modifying the state of a hyperedge in $g_{N,K}(\bm{U}, \bm{r})$ from absent to present, and let $\bm{\psi}^{1} = ( \psi_2^{1}, \psi_3^{1}, \dots, \psi_K^{1}) \in [0,1]$ denote the probability of modifying the state of a hyperedge in $g_{N,K}(\bm{U}, \bm{r})$ from present to absent. This generative model is summarised in Algorithm \ref{alg:modhypext}, and as commented in Section \ref{sec:like}, we can implement this without the suggested $\sum_{k=2}^K {{N}\choose{k}}$ computations. Further details on efficient implementation can be found in the relevant sections of the supplement, as highlighted in Section \ref{sec:like}.

\begin{algorithm}[t]
  \caption{Sample a hypergraph $g^*_{N,K}$ given $N,K,\bm{r}, \bm{\psi}^{0}, \bm{\psi}^{1}, \mu$ and $\Sigma$.} \label{alg:modhypext}
  \begin{algorithmic}
\STATE Sample $\bm{U} = \{u_i\}_{i=1}^N$ such that $u_i \overset{iid}\sim \mathcal{N}(\mu, \Sigma)$, for $i=1,2,\dots,N$.
\STATE For $k=2,3,\dots,K$, \\
  \hspace{.5cm} a) Given $\bm{U}$ and $r_k$, check which $e_k = \{i_1, \i_2, \dots, i_k \} \in \mathcal{E}_{N,k}$ satisfy $y_{e_k}=1$. \\
  \hspace{1cm} To determine if $y_{e_k}=1$, that $\cap_{l=1}^k B_{r_k}(u_{i_l}) \neq \emptyset$. \\
  \hspace{.5cm} b) For all $e_k \in \mathcal{E}_{N,k}$\\
  \hspace{1cm} If $y_{e_k}^{(g)}=1$, set $y_{e_k}^{(g^*)}=0$ with probability $\psi_k^{1}$.\\
  \hspace{1cm} If $y_{e_k}^{(g)}=0$, set $y_{e_k}^{(g^*)}=1$ with probability $\psi_k^{0}$.
  \end{algorithmic}
\end{algorithm}

As in Section \ref{sec:like}, the likelihood of observing $h_{N,K}$ is based on a distance metric,
\begin{align}
  d_k^{(ab)}( g_{N,K}(\bm{U}, \bm{r}), h_{N,K}) = \# \{ e_k \in \mathcal{E}_{N,k} | y_{e_k}^{(g)} = a \: \cap \: y_{e_k}^{(h)} = b\},  \label{eq:subset}  
\end{align}
which records the number of hyperedges that have state $a \in \{0,1\}$ in $g_{N,K}(\bm{U}, \bm{r})$ and state $b \in \{0,1\}$ in $h_{N,K}$. For example, $d_k^{(01)}( g_{N,K}(\bm{U}, \bm{r}), h_{N,K})$ represents the number of hyperedges absent in $g_{N,K}(\bm{U}, \bm{r})$ and present in $h_{N,K}$. Efficient evaluation of \eqref{eq:subset} is discussed in Appendix \eqref{app:evallikelihood} and the likelihood conditional on $\bm{U}, \bm{r}, \bm{\psi}^{1}$ and $\bm{\psi}^{0}$ is given by  
\begin{align}
    \mathcal{L}\left( \bm{U}, \bm{r}, \bm{\psi}^{1}, \bm{\psi}^{0} ; h_{N,K} \right) &\propto   \prod_{k=2}^K \left[ \left( \psi_k^{1} \right)^{d_k^{(10)}(g_{N,K}(\bm{U}, \bm{r}), h_{N,K})}  \left( 1 - \psi_k^{1} \right)^{d_k^{(11)}(g_{N,K}(\bm{U}, \bm{r}), h_{N,K})}  \right.  \nonumber \\
    &\hspace{1cm}\times \left. \left(\psi_k^{0}\right)^{d_k^{(01)}(g_{N,K}(\bm{U}, \bm{r}), h_{N,K})} \left(1-\psi_k^{0}\right)^{d_k^{(00)}(g_{N,K}(\bm{U}, \bm{r}), h_{N,K})} \right] .  \label{eq:likelihoodext}
\end{align}

We obtain \eqref{eq:likelihoodext} in a similar way to \eqref{eq:likelihood}, where we distinguish between hyperedges that are present and absent in the induced hypergraph. Note that \eqref{eq:likelihoodext} is equivalent to \eqref{eq:likelihood} when $\psi_k^{1} = \psi_k^{0} = \upvarphi_k$, for $k=2,3,\dots,K$.

The model specification is complete with the following prior distributions
\begin{gather}
  \mu \sim \mathcal{N}(m_{\mu}, \Sigma_{\mu}), \hspace{0.2cm} \Sigma_{\mu} \sim \mathcal{W}^{-1}(\Phi, \nu), \hspace{0.2cm} r_k \sim \exp( \lambda_k ), \hspace{0.2cm} \psi_k^{0} \sim \mbox{Beta}\left(a_k^{0}, b_k^{0}\right), \hspace{0.2cm} \psi_k^{1} \sim \mbox{Beta}\left(a_k^{1}, b_k^{1}\right), \label{eq:extpriors}
\end{gather}
for $k=2,3,\dots,K$, where $\mathcal{W}^{-1}(\cdot, \cdot)$ denotes an Inverse-Wishart. 

{
\subsection{Identifiability}
\label{sec:ident}

We now consider identifiability for the model presented in Section \ref{sec:like} and note that analogous observations can be made for the modification given in Section \ref{sec:ext}. To begin, we note that the joint distribution of the hypergraph $g^*( \bm{U}, \bm{r}, \bm{\upvarphi})$ and its associated nsRGH (see Definition \ref{def:nsrgh}) $g(\bm{U}, \bm{r})$ can be decomposed as follows.
\begin{align}
  p( g^*(\bm{U}, \bm{r}, \bm{\upvarphi}) , g(\bm{U}, \bm{r}) | \mu, \Sigma, \bm{\upvarphi}, \bm{r} ) = p\left( g^*(\bm{U}, \bm{r}, \bm{\upvarphi}) | g(\bm{U}, \bm{r}), \bm{\upvarphi} \right) p\left( g(\bm{U}, \bm{r}) | \mu, \Sigma, \bm{r} \right) \label{eq:likedecomp} 
\end{align}

The conditional distribution $p( g(\bm{U},\bm{r}) | \mu, \Sigma, \bm{r} )$ in \eqref{eq:likedecomp} contains the hyperedges obtained via the geometric component of the model. These hyperedges are determined through their relative latent positions and therefore will be unchanged given distance-preserving transformations of $\bm{U}$ or, alternatively, scaling of $\bm{U}$ and $\bm{r}$ by the same factor. To make this clear, consider how the interaction patterns in Figure \ref{fig:randomgraph} vary according to translations, reflections, rotations of $\bm{U}$ or by jointly scaling $\bm{U}$ and $\bm{r}$. We remove these sources of non-identifiability by defining $\bm{U}$ on the Bookstein space of coordinates (see \cite{bookstein1986} and \cite[Section 2.3.3]{dryden1998}). Intuitively, Bookstein coordinates define a translation, rotation and re-scaling of the points $\bm{U}$ with respect to a set of anchor points which remain fixed throughout the posterior sampling procedure outlined in Section \ref{sec:postsmp}. Furthermore, by restricting the latent coordinates in this way, we also address the non-identifiability associated with rescaling $\bm{r}$. Details of this are given Appendix \ref{app:bookstein}.

Our approach to removing non-identifiability has not been previously considered in latent space network modelling literature, where it is typical to use Procrustes analysis \citep[Section 5]{dryden1998} in which an additional post-processing step is included to remove the effect of distance preserving transformations of $\bm{U}$. However, due to the additional source of non-identifiability associated with scaling $\bm{U}$ and $\bm{r}$, we note that this approach is not sufficient for our setting.

Finally, \eqref{eq:likedecomp} shows that a hyperedge in our model can either be attributed to the geometric or noise component. Therefore, to maintain the properties imposed on the hypergraph by $g(\bm{U}, \bm{r})$, we wish to keep the parameters $\bm{\upvarphi}$ small. In practice, when there are a small number of observed hyperedges, it will become increasingly difficult to distinguish between these competing sources. This implies an additional source of non-identifiability which we address by ensuring the noise terms $\bm{\upvarphi}$ are sufficiently small.
}
\section{Theoretical Results}
\label{sec:theory}

We now study the behaviour of the node degree in the hypergraph model detailed in Algorithm \ref{alg:modhyp}. Since the nodes in our hypergraph model are exchangeable, it is sufficient to study the degree properties of the $i^{th}$ node and we comment here that it is straightforward to extend the results in this section to the model detailed in Algorithm \ref{alg:modhypext}. To begin, we make explicit the following properties of our hypergraph model: (P1) A hypergraph generated from our model is a modification of a nsRGH (see Definition \ref{def:nsrgh}), and (P2) Conditional on $\bm{U}$ and $\bm{r}$, hyperedges of each order occur independently in our model.

To obtain our results, we also assume: (A1) The number of nodes $N$ and maximum hyperedge order $K$ are fixed, and (A2) The covariance of the latent coordinates $\Sigma$ is diagonal, where $\Sigma_{ll} = \sigma^2_{ll}$ for $l=1,2,$ $\dots,d$. Note that (A2)  is not restrictive since we can take a distance-preserving transformation of a Gaussian point cloud in $\mathbb{R}^d$ to map the covariance matrix onto a diagonal matrix.

Following Section \ref{sec:lshyps}, we denote a hypergraph obtained by modifying the nsRGH $g(\bm{U}, \bm{r})$ as $g^*(\bm{U}, \bm{r}, \bm{\upvarphi})$, and we let $y_{e_k}^{(g)}$ and $y_{e_k}^{(g^*)}$ indicate the presence or absence of the hyperedge $e_k$ in $g(\bm{U}, \bm{r})$ and $g^*(\bm{U}, \bm{r}, \bm{\upvarphi})$, respectively. The degree of node $i$ in a hypergraph is given by the number of hyperedges which contain $i$ and, for node $i$ in $g(\bm{U}, \bm{r})$, we denote the order $k$ degree as $\mbox{Deg}_{(i,k)}^g = \sum_{ \{e_k \in \mathcal{E}_{N,k} | i \in e_k \}} y_{e_k}^{(g)}$ and the full degree as $\mbox{Deg}_{(i)}^g = \sum_{k=2}^K \mbox{Deg}_{(i,k)}$. Analogous expressions for $g^*(\bm{U}, \bm{r}, \bm{\upvarphi})$ are obtained by replacing $g$ with $g^*$. {To organise our results, we separate the discussion into hyperedges with order $k=2$ and $k \geq 3$. For each case, we consider the probability of connections forming in our hypergraph model and how this translates into properties of the node degree. Proofs can be found in Appendix \ref{app:proofs}.}

{ 

\subsection{Degree properties for $k=2$}
\label{sec:degs_k2}

We begin by restating a result from \cite{rastelli2015} which allows us to express the degree distribution and average degree for order $k=2$ hyperedges in our model. Throughout, we express hyperedges of order 2 which contain node $i$ as $\{i,j\}$.

\begin{thm}[{Reproduced from \cite[Theorem 1]{rastelli2015}}] \label{thm:dd_k2}
  Let $p_{\{i,j\}}(u_i | r_2, \upvarphi_2, \mu, \Sigma)$ denote the probability of node $i$ with latent position $u_i$ belonging to an order $2$ hyperedge in the hypergraph model detailed in Section \ref{sec:like}. Conditional on $r_2, \upvarphi_2, \mu$ and $\Sigma$, the degree distribution of order 2 hyperedges for the $i^{th}$ node for our hypergraph model is given by
\begin{align}
  p \left( \mbox{Deg}_{i,2}^{(g^*)} = l | r_2, \upvarphi_2, \mu, \Sigma \right) &= \int p(u_i | \mu, \Sigma ) { N-1 \choose l } p_{\{ i,j \}}(u_i | r_2, \upvarphi_2, \mu, \Sigma)^l\nonumber \\
                                                                                &\hspace{3cm} \times (1 - p_{\{ i, j \}}(u_i | r_2, \upvarphi_2, \mu, \Sigma))^{N-1-l} \,d u_i, \label{eq:ddk2}
\end{align}
and the average degree is given by
\begin{align}
  \bar{d}_2^{(g^*)} (r_w, \upvarphi_2, \mu, \Sigma) = (N-1) \int p(u_i | \mu, \Sigma) p_{\{i,j\}}(u_i | r_2, \upvarphi_2, \mu, \Sigma) \,d u_i. \label{eq:avddk2}
\end{align}
  
\end{thm}

\begin{proof}
  See Appendix \ref{app:proofs_k2}.
\end{proof}

In order to apply this result, we must derive an expression for $p_{\{i,j\}}(u_i | r_2, \upvarphi_2, \mu, \Sigma)$. This is given in the following proposition.

\begin{prop}[Probability of an order $k=2$ hyperedge] \label{prop:prob_ij_give_ui}
  Let $e_2 = \{i,j\}$ denote an order $k=2$ hyperedge. The probability of $e_2$ occurring in $g^*$, given that node $i$ has latent position $u_i$, can be expressed as
\begin{align}
  p_{\{ i,j \}} \left( y_{ij}^{(g^*)} =1 | u_i, r_2, \upvarphi_2, \mu, \Sigma \right) &= p\left( y_{\{ i,j \}}^{(g)} = 1 | u_i, r_2, \mu, \Sigma \right) (1 - \upvarphi_2 ) \nonumber \\
  &\hspace{2.5cm} + \left(1 -  p \left( y_{\{ i,j \}}^{(g)} = 1 | u_i, r_2, \mu, \Sigma \right) \right) \upvarphi_2,
\end{align}
where $p\left( y_{\{i,j\}}^{(g)} = 1 | u_i, r_2, , \mu, \Sigma \right)$ denotes the probability of $e_2 = \{i,j\}$ being present in $g(\bm{U}, \bm{r})$ when node $i$ is positioned at $u_i$. We then have
\begin{align}
p \left( y_{\{ i,j \}}^{(g)} = 1 | u_i, r_2, \mu, \Sigma \right) = p( (U_j^*)^T U_j^* \leq 4 r_2^2), \label{eq:p_ustr_sqr}
\end{align}
where $U_j^* = U_j - u_i \sim \mathcal{N}( \mu - u_i, \Sigma)$. Then
\begin{align}
  (U_j^*)^T U_j^* \sim \sum_{l=1}^d \sigma_l^2 \chi^2_{1, ( (\mu - u_i)_l / \sigma_l)^2 }. \label{eq:ustr_sqr_dist}
\end{align}
where $(\mu - u_i)_l$ is the $l^{th}$ component of $(\mu - u_i)$, for $l=1,2,\dots,d$ and $\Sigma = \mbox{diag}( \sigma_1^2, \sigma_2^2, \dots, \sigma_d^2 )$.
\end{prop}

\begin{proof}
  See Appendix \ref{app:proofs_k2}.
\end{proof}

In combination with Theorem \ref{thm:dd_k2}, Proposition \ref{prop:prob_ij_give_ui} allows us to evaluate properties of the node degree of order $k=2$ hyperedges. Note however that the integrals \eqref{eq:ddk2} and \eqref{eq:avddk2} are intractable so we require numerical evaluation to explore these quantities. Furthermore, determining probabilities for the distribution \eqref{eq:ustr_sqr_dist} is straightforward when the variances constant but requires numerical methods for general diagonal $\Sigma$ where, in the latter case, we rely on the R packages \texttt{CompQuadForm} \citep{duchense2010}. The expression for the connection probabilities in Proposition \ref{prop:prob_ij_give_ui} is validated for an example in Figure \ref{fig:pek_given_ui_sig1}.

\subsection{Degree properties for $k \geq 3$}
\label{sec:degs_kg2}

We now present some results for the order $k$ degree distribution. First, we note that the expression for the degree distribution of order $k=2$ hyperedges given in Theorem \ref{thm:dd_k2} arises since hyperedges of the form $\{i,j\}$ occur independently given $u_i$. This is no longer true for higher-order hyperedges since, for example, if $k=3$ it is clear that hyperedges $\{i,j,l\}$ and $\{i,j,m\}$ are not independent given $i$. Deriving an exact expression for the order $k$ degree distribution is therefore challenging, and we consider an approximation of this in the following proposition where we denote an order $k$ hyperedge containing node $i$ as $e_{k,i}$.

\begin{prop}[Approximate order $k$ degree distribution] \label{prop:approx_dd_kg2}
    Let $p_{ e_{k,i} }\left( y_{e_{k,i}}^{(g^*)} =1 |u_i, r_k, \upvarphi_k, \right.$ $\left. \mu, \Sigma \right)$ denote the probability of node $i$ with latent position $u_i$ belonging to an order $k$ hyperedge, denoted as $e_{k,i}$, in the hypergraph model detailed in Section \ref{sec:like}. Conditional on $r_k, \upvarphi_k, \mu$ and $\Sigma$, the degree distribution of order $k \geq 3$ hyperedges for the $i^{th}$ node for our hypergraph model is approximately
\begin{align}
  p \left( \mbox{Deg}_{i,k}^{(g^*)} = l | r_2, \upvarphi_2, \mu, \Sigma \right) &\approx \int p(u_i | \mu, \Sigma ) \nonumber \\
  &\hspace{-4.5cm} \times \dfrac{ \left({N-1 \choose k-1} p(y_{e_{k,i}}^{(g^*)}=1 |u_i, r_k, \upvarphi_k, \mu, \Sigma) \right)^l \exp \left( -  {N-1 \choose k-1} p(y_{e_{k,i}}^{(g^*)} = 1 | u_i, r_k, \upvarphi_k, \mu, \Sigma) \right) }{l !} \,d u_i\label{eq:ddkh2}
\end{align}

\end{prop}

\begin{proof}
  See Appendix \ref{app:proofs_kg2}.
\end{proof}

This proposition relies on a Poisson approximation to the sum of dependent Bernoulli trials. Bounds for the quality of this approximation have been studied (for example, see \cite{teerapabolarn2014}) and it is well understood that this approximation will degrade as either the number of possible hyperedges, as controlled by $N$ and $k$, or the connection probability $p_{ e_{k,i} }(u_i | r_k, \upvarphi_k, \mu, \Sigma)$ grows, or both. 

As in the previous section, we require an expression for $p_{ e_{k,i} }(u_i | r_k, \upvarphi_k, \mu, \Sigma)$ to explore this distribution. Similarly to Proposition \ref{prop:prob_ij_give_ui}, an order $k$ hyperedge can be expressed either by the geometric or noise component of our model, where the probability of occurrence in $g(\bm{U}, \bm{r})$ is the more challenging to determine. Recall that this quantity is given by the probability that the balls of radius $r_k$ corresponding to the nodes in $e_{k,i}$ have a non empty intersection. More specifically, the hyperedge $e_{k,i} = \{i, i_2, i_3, \dots, i_k\}$ is present in $g(\bm{U}, \bm{r})$ if $\cap_{j \in e_{k,i}} B_{r_k}(u_j) \neq \emptyset$. This condition is equivalent to the coordinates $\{ u_j \}_{j \in e_{k,i}}$ lying within a ball of radius $r_k$ (see \cite[Section 3.2]{edelsbrunner2010} and Appendix \ref{app:miniball}). However, since the latent coordinates are assumed to follow a Normal distribution, evaluating this probability directly requires evaluation of an intractable integral (for example, see \cite{gilliland1962}). To address this, we consider an approximation of this quantity in which the region of integration is approximated by a square of side length $s_k$, denoted as $S_{s_k}$. We choose $s_k = \sqrt{\pi} r_k$ so that it has equal area to a disk of radius $r_k$. This choice greatly simplifies the region of integration by allowing each dimension to be considered independently and, using this, we obtain the expression in Lemma \ref{lemma:prob_ekg2}.

{
\begin{figure}[t!]
  \centering
  \includegraphics[width=.75\textwidth, trim={0cm 0cm 1cm 2cm}, clip]{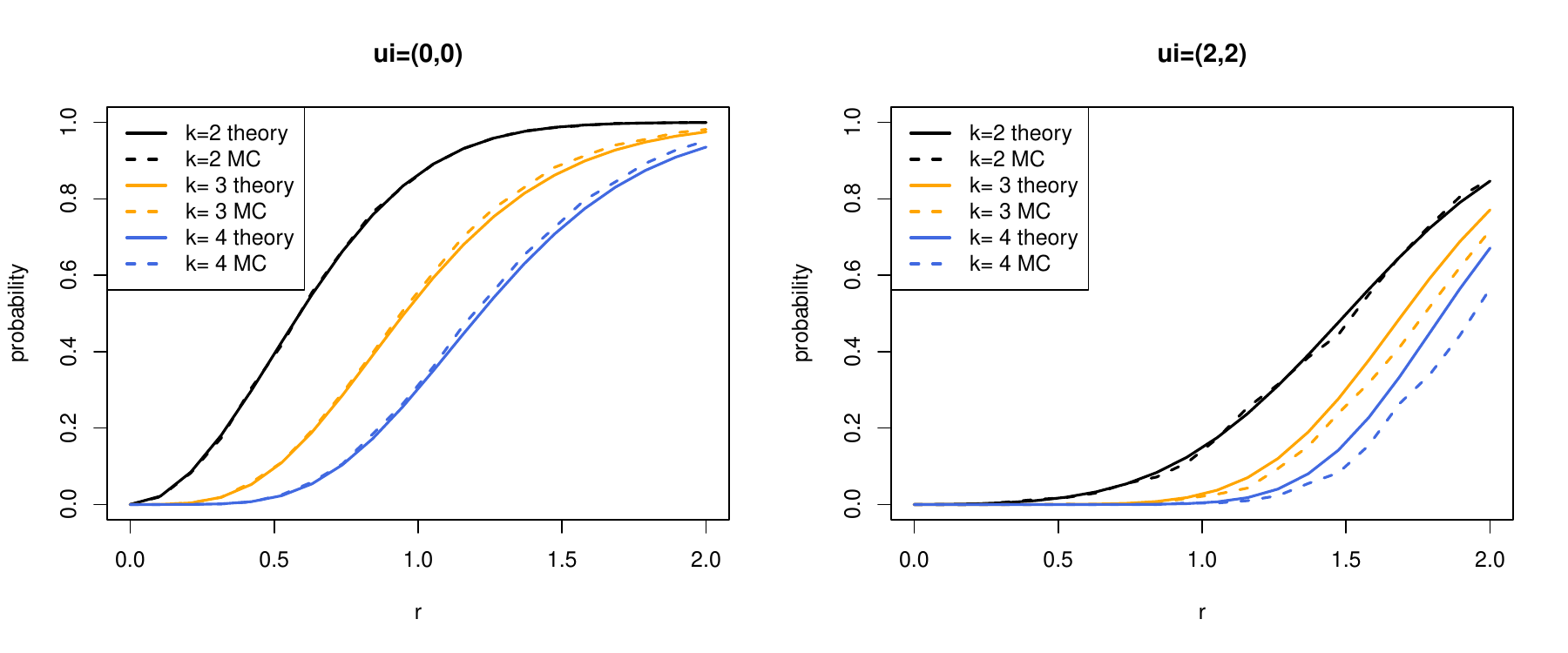}
  \caption{Comparison of theoretical (solid) and Monte Carlo (dashed) estimates of $p(y_{e_{k,i}}^{(g)}=1 | u_i, r_k, \mu, \Sigma)$ for varying $r_k$. We take $\Sigma = \mbox{diag}(1,1), \mu=(0,0)$ and consider connection probabilities for $k=2,3,4$. In the left plot $u_i = \mu$ and in the right plot $u_i = (1,2)$. The same study with $\Sigma = \mbox{diag}(1,2)$ is provided in Appendix \ref{app:theory_figures}.}  \label{fig:pek_given_ui_sig1}
\end{figure}
} 

\begin{lemma}[Probability of an order $k \geq 3$ hyperedges] \label{lemma:prob_ekg2}
    We can express the probability of an order $k$ hyperedge containing node $i$, denoted as $e_{k,i}$, being present in $g^*$ conditional on the latent coordinate $u_i$ as
\begin{align}
    p \left( y_{ e_{k,i} }^{(g^*)} = 1 | u_i, r_k, \upvarphi_k, \mu, \Sigma \right) &= p\left( y_{ e_{k,i} }^{(g)} = 1 | u_i, r_k, \mu, \Sigma \right) (1 - \upvarphi_k ) \nonumber \\
  &\hspace{3cm} + \left(1 -  p \left( y_{ e_{k,i} }^{(g)} = 1 | u_i, r_k, \mu, \Sigma \right) \right) \upvarphi_k,
\end{align}
where $p\left( y_{ e_{k,i} }^{(g)} = 1 | u_i, r_2, \mu, \Sigma \right)$ is the probability of the order $k$ hyperedge $e_{k,i} = \{i,i_2, i_3,$ $\dots, i_k\}$ being present in $g(\bm{U}, \bm{r})$ when node $i$ is positioned at $u_i$.

The probability of $e_{k,i}$ occurring in $g(\bm{U}, \bm{r})$ with $U_j \sim \mathcal{N}(\mu, \Sigma)$, conditional on $u_i$, can then be approximated as follows.
  \begin{align}
    p \left( y_{ e_{k,i} }^{(g)} = 1 | u_i, r_k, \mu, \Sigma \right) \approx \nonumber \\
    &\hspace{-4cm} \prod_{l=1}^d \left[ (k-1) \int f(u | \mu_l, \sigma_l ) \left[ F_m(u + \sqrt{\pi} r_k | \mu_l, \sigma_l) - F_m(u | \mu_l, \sigma_l) \right]^{k-2} \right.  \nonumber \\
                                                       &\hspace{-3cm} \left. \times \mathbbm{1}( u_i \in (u, u+ \sqrt{\pi} r_k) ) \, d u + \left[ F_m(u_i + \sqrt{\pi} r_k | \mu_l, \sigma_l) - F_m(u_i | \mu_l, \sigma_l) \right]^{k-1} \right], \label{eq:range_l_given_ui}
  \end{align}
  where $f(\cdot | \mu_l, \sigma_l)$ and $F(\cdot | \mu_l, \sigma_l)$ denote the pdf and cdf of the random variable with distribution $\mathcal{N}(\mu_l, \sigma_l^2)$, $\mu_l$ denotes the $l^{th}$ element of $\mu$ and $\Sigma = \mbox{diag}(\sigma_1^2, \sigma_2^2, \dots, \sigma_d^2)$.
  
\end{lemma}

\begin{proof}
  See Appendix \ref{app:proofs_kg2}.
\end{proof}

{
\begin{figure}[t!]
  \centering
  \includegraphics[width=.8\textwidth, trim={0cm 0cm 1cm 2cm}, clip]{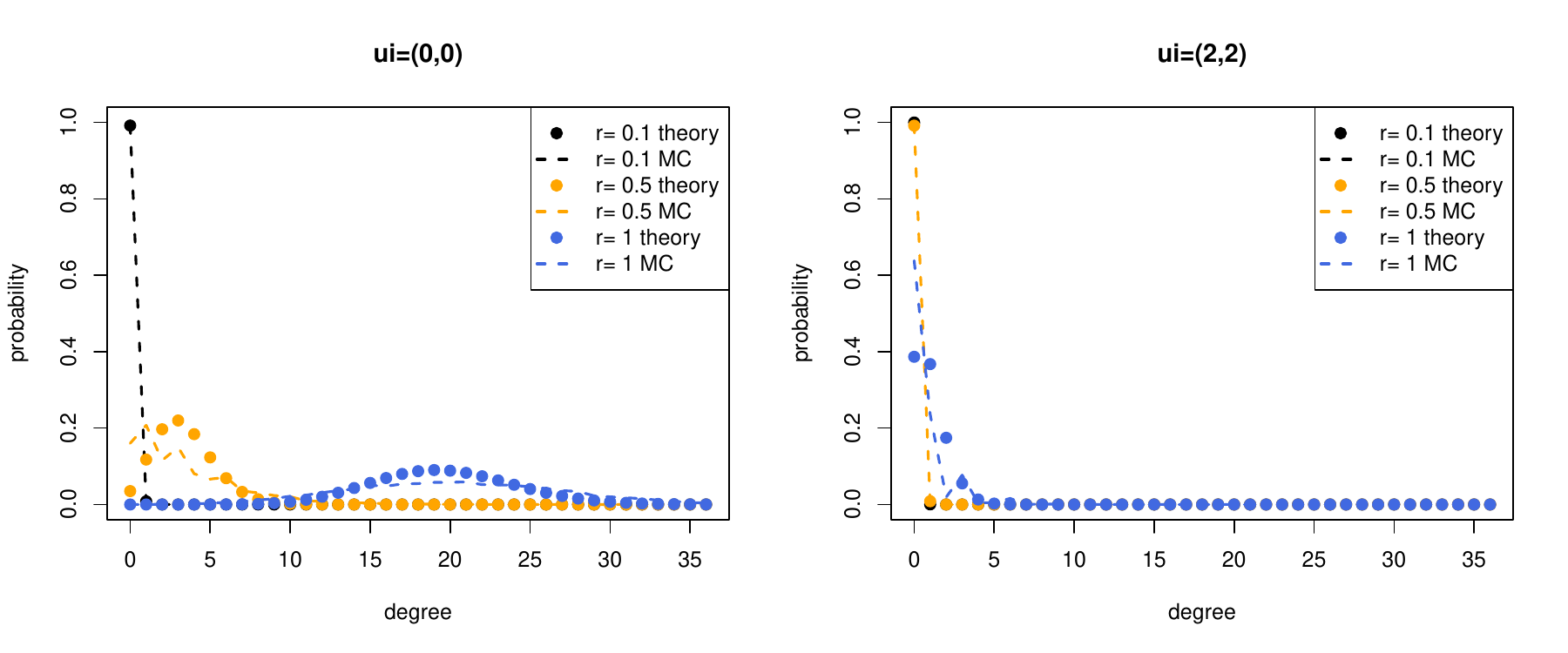}
  \caption{Comparison between empirical (dashed line) and Poisson approximation (points) of the order $k=3$ degree distribution conditional on the latent coordinate $u_i$. We take $N=10, \mu=(0,0), \Sigma = \mbox{diag}(1,1)$ and evaluate the distribution for $r_3 \in (0.1, 0.4, 1.0)$. The left plot shows $u_i = \mu$ and the right plot shows $u_i = (1,2)$. The equivalent Figures with $\Sigma = \mbox{diag}(1,2)$ and $N=20$ are given in Appendix \ref{app:theory_figures}.}  \label{fig:poisson_apprx_ui_sig1}
\end{figure}
} 

These results allow us to examine the behaviour of the degree distribution. We examine the quality of the $p(y_{e_{k,i}}^{(g)} = 1 | u_i, r_k, \mu, \Sigma)$ approximation and the Poisson approximation in Figures \ref{fig:pek_given_ui_sig1} and \ref{fig:poisson_apprx_ui_sig1}, respectively. We note that there is an interplay between the hyperedges which arise in our model from the latent coordinate and those which arise as random noise. Whilst we may obtain similar probabilities of connection through parameter combinations which result in the majority of hyperedges occurring due to either the geometric or random component, it is important to observe that the resulting properties of the hyperedges will differ considerably. Beyond the degree distribution, differences may also appear, for example, within motif counts and measures of transitivity. The above calculations suggest that examining such properties theoretically is likely to be challenging and, whilst techniques such as those employed for studying random simplicial complexes may be used here (for example see \cite{kahle2016, bobrowski2021}), our choice of underlying distribution on the latent coordinates may limit their applicability.
}

{ 
\section{Posterior Sampling}
\label{sec:postsmp}

We rely on Markov chain Monte Carlo (MCMC) \citep{gamerman2006} to obtain posterior samples for the models specified in Sections \ref{sec:like} and \ref{sec:ext}. Here we provide a high-level description of our MCMC scheme for the model detailed in Algorithm \ref{alg:modhypext} and note that this scheme can easily be modified for the model detailed in Algorithm \ref{alg:modhyp}. We refer the reader to the Appendix for further details.

We use a Metropolis-Hastings-within-Gibbs MCMC scheme \citep[Section 6.4.2]{gamerman2006} to sample from our posterior. The latent coordinates $\bm{U}$ and radii $\bm{r}$ are updated with Metropolis-Hastings (MH) steps and the remaining parameters are updated via Gibbs steps. As is typically the case, evaluation of the likelihood \eqref{eq:likelihoodext} presents a computational bottleneck due to the calculations involved in determining the hypergraph $g_{N,K}( \bm{U}, \bm{r})$. We rely on the efficient implementation available from the GUDHI library \citep{gudhi} to evaluate the \v{C}ech complex and find that the cost of calculating the order $k$ hyperedges in $g_{N,K}( \bm{U}, \bm{r})$ grows with both $N$ and $k$. This suggests that our target is a natural candidate for delayed-acceptance (DA) (see \cite{banterle2019}) in which terms of the likelihood are sequentially included in order of increasing computational cost, allowing for a faster and computationally cheaper rejection of poor proposals. 

\begin{algorithm}[t!]
  \caption{MCMC scheme to obtain posterior samples for the model in Section \ref{sec:ext}.} \label{alg:mcmc} 
  \begin{algorithmic}
    \STATE Input: observations $h_{N,K}$ and $N, K, j_{max} \in \mathbb{N}$.
    \STATE \underline{\textbf{Initialisation}}
    \STATE \hspace{.5cm} Determine initialisation $\bm{U}^{(0)}, \bm{r}^{(0)}, \bm{\psi}^{0(0)}, \bm{\psi}^{1(0)}, \Sigma^{(0)}$ and $\mu^{(0)}$ and specify Bookstein 
    \STATE \hspace{.5cm} anchor coordinates $\{B_1, B_2, \dots, B_d\}$ according to Algorithm \ref{alg:mcmcinit} (see Appendix \ref{app:mcmcinit}).
    \STATE \textbf{For $j$ in $1,2,\dots, j_{max}$}
    \STATE \hspace{.5cm} \underline{\textbf{Gibbs update steps for $\bm{\psi}^{(0)}, \bm{\psi}^{(0)}, \mu$ and $\Sigma$}}
    \STATE \hspace{.5cm} 1) \textbf{For $k = 2,3,\dots,K$} 
    \STATE \hspace{1.5cm}  Sample $\psi^{0(i)}_k$ from $p\left(\psi_k^{0} | \bm{U}^{(i)}, \bm{r}^{(i)}, h_{N,K}, a_k^{0}, b_k^{0}\right)$ (see Appendix \ref{app:psi0cond}). 
\STATE \hspace{1.5cm}  Sample $\psi^{1(i)}_k$ from $p\left(\psi_k^{1} | \bm{U}^{(i)}, \bm{r}^{(i)}, h_{N,K}, a_k^{1}, b_k^{1}\right)$ (see Appendix \ref{app:psi1cond}).
    \STATE \hspace{.5cm} 2) Sample $\mu^{(j)}$ from $p(\mu | \bm{U}^{(i-1)}, \Sigma^{(i-1)}, m_{\mu}, \Sigma_{\mu})$ (see Appendix \ref{app:mucond}).
    \STATE \hspace{.5cm} 3) Sample $\Sigma^{(j)}$ from $p(\Sigma | \bm{U}^{(i-1)}, \mu^{(i)}, \Phi, \nu)$ (see Appendix \ref{app:sigmacond}).
    \STATE \hspace{.5cm} \underline{\textbf{Delayed-acceptance update steps for $\bm{U}$}}
    \STATE \hspace{.5cm} 4) \textbf{For $i = \{ 1, 2, \dots, N\} \backslash \{B_1, B_2, \dots, B_d\} $ }
    \STATE \hspace{1.5cm} Propose $u_i^* = u_i + \epsilon_{u_i}$, where $\epsilon_{u_i} \sim \mathcal{N}(0, \sigma_{u_i} I_d)$
    \STATE \hspace{1.5cm} \textbf{For $l \in \{2,3,\dots,K\}$}
    \STATE \hspace{2cm} Calculate $AR_l = \min \left\{1 ,\prod_{k=2}^l \rho_k( \bm{U}, \bm{U}^*) \right\}$ according to \eqref{eq:rho_k}
    \STATE \hspace{2cm} Continue with probability $AR_l$, otherwise break and set $u_j^{(j)}=u_j$
    \STATE \hspace{2cm} If $l = K$, set $u_j^{(j)}=u_j^*$ with probability $AR_K$, otherwise set $u_j^{(j)}=u_j$
    \STATE \hspace{.5cm} \underline{\textbf{Random-walk update step for $\bm{r}$}}
    \STATE \hspace{.5cm} 5) \textbf{For $k=2,3,\dots,K$}
    \STATE \hspace{1.5cm} Propose ${r_k}^* = {r_k} + \epsilon_{r_k}$ and $\epsilon_{r_k} \sim \mathcal{N}(0, \sigma_{r_k})$
\STATE \hspace{1.5cm} With probability \eqref{eq:radiiAR} set ${r_k}^{(j)} = {r_k}$, otherwise set ${r_k}^{(j)} = {r_k}$
  \end{algorithmic}
\end{algorithm}

To specify a random-walk MH update for the latent coordinates $\bm{U}$, first recall that we define $\bm{U}$ on the Bookstein space of coordinates (Section \ref{sec:ident}) so that a set of anchor points remain fixed throughout estimation. For $u_i \in \mathbb{R}^d$, let the $d$ anchor points be denoted by $\{ u_{B_1}, u_{B_2}, \dots, u_{B_d} \}$. Then, for $i \in \{ 1, 2, \dots, N\} \backslash \{B_1, B_2, \dots, B_d\}$ we propose $\bm{U}^*$ with $i^{th}$ entry $u_i^* = u_i + \epsilon_u$ where $\epsilon_u \sim \mathcal{N}(0, \sigma_u I_d)$ and all other entries $u_i^* = u_i$. This proposal is then accepted as a sample from $p(\bm{U} | \mu, \Sigma, \bm{r}, \bm{\psi}^{0}, \bm{\psi}^{1}, h_{N,K})$ with probability
\begin{align}
  \min  \left\{ 1, \dfrac{\mathcal{L}( \bm{U}^*, \bm{r}, \bm{\psi}^{1}, \bm{\psi}^{0} ; h_{N,K} ) p( u_i^* | \mu, \Sigma ) }{\mathcal{L}( \bm{U}, \bm{r}, \bm{\psi}^{1}, \bm{\psi}^{0} ; h_{N,K} ) p( u_i | \mu, \Sigma ) } \right\}.  \label{eq:uAR}
\end{align}
Note that the term associated with the proposal is symmetric and so does not appear in \eqref{eq:uAR}. Let $h_{N,k}$ denote the observed hyperedges of order $k \in \{2,3,\dots,K\}$, then we can equivalently express \eqref{eq:uAR} as
\begin{align}
  &\min  \left\{ 1, \dfrac{ p( h_{N,2} | \bm{U}^*, r_2, \psi_2^{1}, \psi_2^{0}) p( u_i^* | \mu, \Sigma ) }{ p( h_{N,2} | \bm{U}, r_2, \psi_2^{1}, \psi_2^{0}) p( u_i | \mu, \Sigma ) } \times \prod_{k=3}^K \dfrac{ p( h_{N,k} | \bm{U}^*, r_k, \psi_k^{1}, \psi_k^{0})}{p( h_{N,k} | \bm{U}, r_k, \psi_k^{1}, \psi_k^{0}) }\right\} \\
  &\hspace{2cm} = \min \left\{ 1, \rho_2(\bm{U}, \bm{U}^*) \prod_{k=3}^K \rho_k(\bm{U}, \bm{U}^*) \right\} \label{eq:rho_k}
\end{align}
where $p( h_{N,k} | \bm{U}, r_k, \psi_k^{1}, \psi_k^{0})$ denotes the contribution of the order $k$ terms in \eqref{eq:likelihoodext}. Following \cite{banterle2019}, we can then implement DA in which we first accept according to the order $k=2$ terms, $\rho_2(\bm{U}, \bm{U}^*)$, and for $k=3,4,\dots,K$ sequentially include the terms $\rho_k(\bm{U}, \bm{U}^*)$. We detail this in Algorithm \ref{alg:mcmc} and note that a proposal can only be accepted according to the full likelihood. Therefore asymptotic exactness is unaffected, but poor proposals can be quickly rejected according to the terms involving lower-order hyperedges.

The $k^{th}$ radii is updated according to a standard random-walk MH step, where the proposal $r_k^* = r_k + \epsilon_r$ with $\epsilon_r \sim \mathcal{N}(0, \sigma_r)$, is accepted with probability
\begin{align}
  \min  \left\{ 1, \dfrac{ p(h_{N,k} | \bm{U}, r_k^*, \upvarphi_k ) p( r_k^* | \lambda_k ) }{p(h_{N,k} | \bm{U}, r_k, \upvarphi_k ) p( r_k | \lambda_k ) } \right\}.  \label{eq:radiiAR}
\end{align}

The remaining parameters can be sampled directly from their full conditionals and an outline of our MCMC scheme for $j_{max}$ iterations is presented in Algorithm \ref{alg:mcmc}. Further details for implementation of our scheme, including conditional distributions, initialisation and efficient evaluation of \eqref{eq:likelihoodext}, can be found in the supplement. Finally, we comment that Algorithm \ref{alg:mcmc} outlines a basic random walk and our implementation instead uses adaptive updates for $\bm{U}$ and $\bm{r}$ as detailed in \cite{vihola2012} and implemented in the R library \texttt{ramcmc} (see \cite{ramcmcpackage}). In this context, it is the distributions of $\epsilon_{u_i}$ and $\epsilon_{r_k}$ that are adapted to reflect the optimal acceptance rates for a random-walk MCMC scheme.

}

\section{Simulations}
\label{sec:sims}

In this section we describe three simulation studies. We begin in Section \ref{sec:modeldepth} with an investigation of the flexibility of our modelling approach in comparison with two other hypergraph models from the literature. Then, in Section \ref{sec:privspost}, we examine efficacy of our estimation procedure via the predictive degree distribution and explore the scalability of our approach for simulated hypergraphs in Section \ref{sec:scalability}. An additional simulation study is presented in Appendix \ref{sec:misspec} where we consider the robustness of our model under misspecification. 

\subsection{Model depth comparisons}
\label{sec:modeldepth}

Here we explore the range of hypergraphs that can be expressed via our model and provide a comparison with two competing models from the literature. To ensure our comparison is fair, we specify several cases which are designed to highlight important aspects of each model. For each case, we sample hypergraphs from the generative models and compare summary statistics which characterise certain properties of the simulated data.

{
\begin{figure}[t!]
  \begin{minipage}{\linewidth}
    \renewcommand{\arraystretch}{1}
\centering
  \begin{tabular}{ p{4.5cm} | p{.75cm} | p{7cm} }
    Model & Case & Description\\
    \hline  \hline
    \multirow{2}{4.5cm}{ \cite{stasi2014} } & 1 & All nodes equally likely to form connections \\
    & 2 & Some nodes more likely to form connections \\
    \hline
    \multirow{4}{4.5cm}{ \cite{ng2018}  } & 1 & Hyperedges in a single cluster \\
          & 2 & Distinct topic clusters only \\
          & 3 & Distinct size clusters only \\
          & 4 & Fuzzy topic clusters \\
    \hline
    \multirow{5}{4.5cm}{ Latent Space Hypergraph } & 1 & Strongly correlated $\Sigma$ \\
          & 2 & No correlation in $\Sigma$ \\
          & 3 & Dense in $e_2$, sparse in $e_3, e_4$ \\
          & 4 & Sparse $e_2, e_4$, dense in $e_3$ \\
          & 5 & Increase latent dimension from $d=2$ to $d=3$ \\
  \end{tabular}
\captionof{table}{Summary of cases for each hypergraph model considered in the model depth comparison study. The case numbers correspond to the labels in Figure \ref{fig:modeldepth}. } \label{table:cases_summary}
\end{minipage}
\end{figure}
} 

We will focus on the hypergraph model described in Algorithm \ref{alg:modhypext} in addition to the uniform hypergraph model of \cite{stasi2014} and the extended latent class analysis hypergraph model of \cite{ng2018} (details of these models are given in Appendix \ref{sec:altern-hypergr-model}). The uniform hypergraph model allows control over the degree distribution via node-specific parameters and the extended latent class analysis hypergraph model is designed to express hyperedges which exhibit a clustering structure according to topic and size. In contrast, our model utilises a latent geometry to model the connections and so we expect transitive connections in which ``friends of friends are likely to be friends''. Furthermore, our model also encourages a higher-order `transitivity' since, for example, the presence of hyperedges $\{i,j\}, \{i,k\}$ and $\{j,k\}$ suggests that the hyperedge $\{i,j,k\}$ is also likely to be present. Table \ref{table:cases_summary} summarises the cases we consider for each model, and details of specific parameter choices are provided in Table \ref{table:cases} in Appendix \ref{sec:sims-set-up}. 

\begin{figure}[t!]
    \centering
\begin{subfigure}{.6\textwidth}
    \begin{subfigure}[t]{0.18\textwidth}
      \renewcommand{\thesubfigure}{\alph{subfigure}1}
        \centering
        \includegraphics[trim={0 1.5cm 0 1.5cm}, clip, width=.89\textwidth]{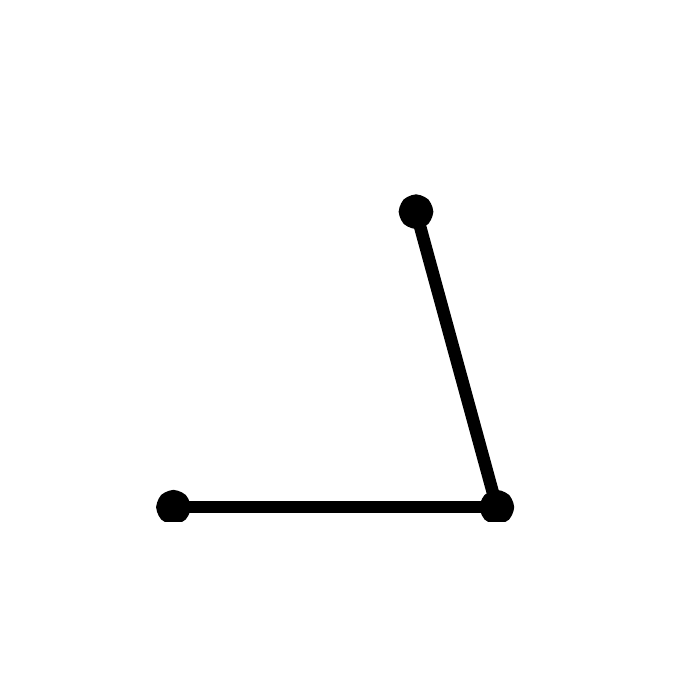}
        \caption{3 star}
    \end{subfigure}%
    ~ 
    \begin{subfigure}[t]{0.18\textwidth}
      \addtocounter{subfigure}{-1}
      \renewcommand{\thesubfigure}{\alph{subfigure}2}
        \centering
        \includegraphics[trim={0 1.5cm 0 1.5cm}, clip, width=.89\textwidth]{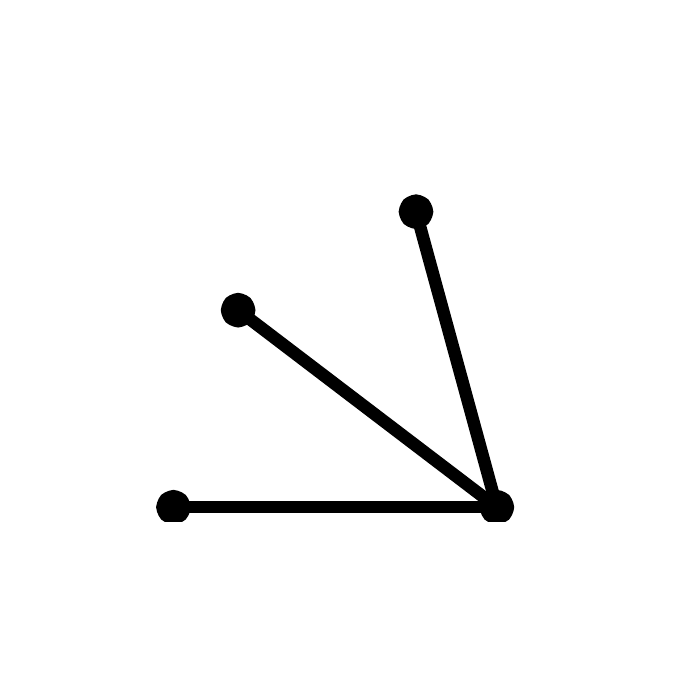}
        \caption{4 star}
    \end{subfigure}
    ~ 
    \begin{subfigure}[t]{0.16\textwidth}
      \addtocounter{subfigure}{-1}
      \renewcommand{\thesubfigure}{\alph{subfigure}3}
        \centering
        \includegraphics[trim={0 1.5cm 0 1.5cm}, clip, width=\textwidth]{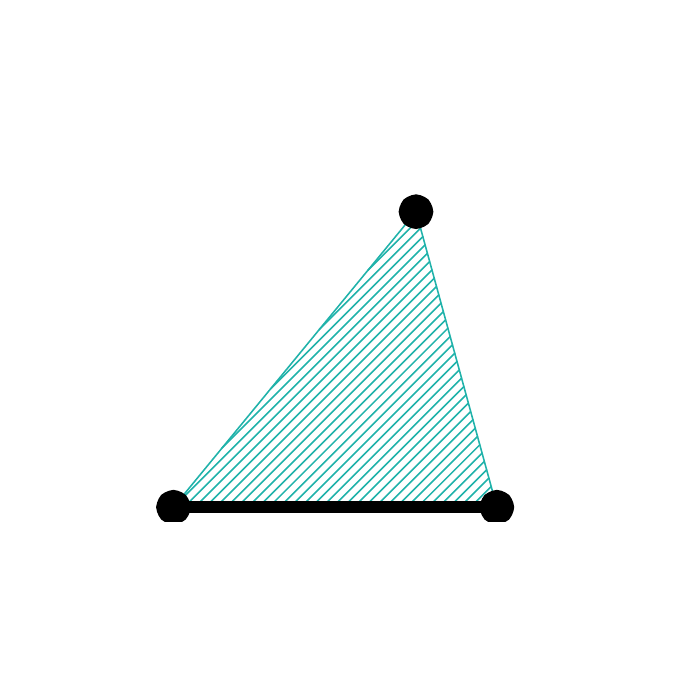}
        \caption{$h_1$} \label{fig:h1}
    \end{subfigure}
    ~ 
    \begin{subfigure}[t]{0.16\textwidth}
      \addtocounter{subfigure}{-1}
      \renewcommand{\thesubfigure}{\alph{subfigure}4}
        \centering
        \includegraphics[trim={0 1.5cm 0 1.5cm}, clip, width=\textwidth]{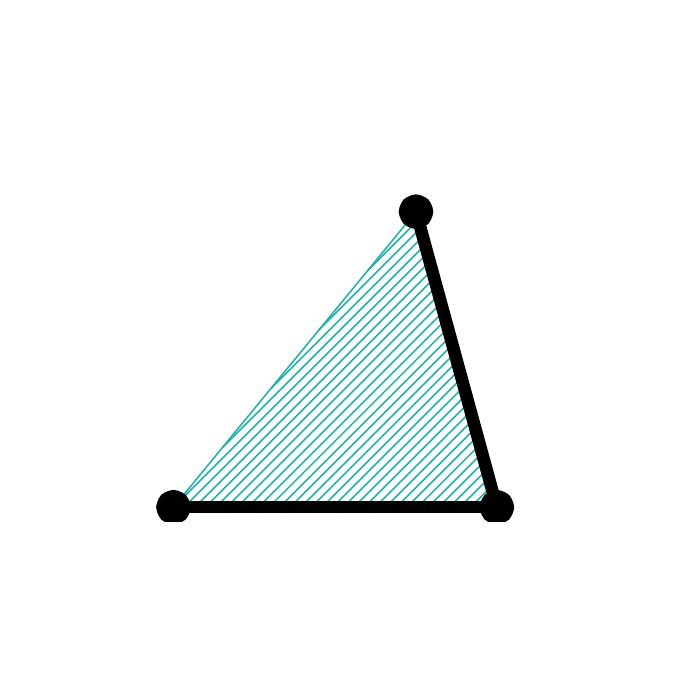}
        \caption{$h_2$} \label{fig:h2}
    \end{subfigure}
    ~ 
    \begin{subfigure}[t]{0.16\textwidth}
      \addtocounter{subfigure}{-1}
      \renewcommand{\thesubfigure}{\alph{subfigure}5}
        \centering
        \includegraphics[trim={0 1.5cm 0 1.5cm}, clip, width=\textwidth]{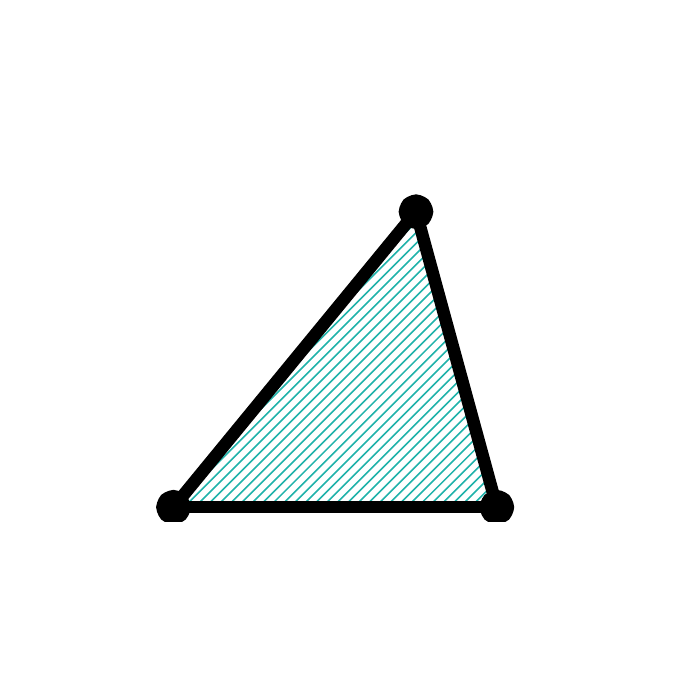}
        \caption{$h_3$} \label{fig:h3}
    \end{subfigure}
    \addtocounter{subfigure}{-1}
    \caption{Motifs counted in model simulations.}
    \label{fig:motifs}
\end{subfigure}
\begin{subfigure}{.38\textwidth}
    \centering
    \begin{subfigure}[t]{0.25\textwidth}
      \renewcommand{\thesubfigure}{\alph{subfigure}1}
        \centering
        \includegraphics[trim={0 1.5cm 0 1.5cm}, clip, width=1\textwidth]{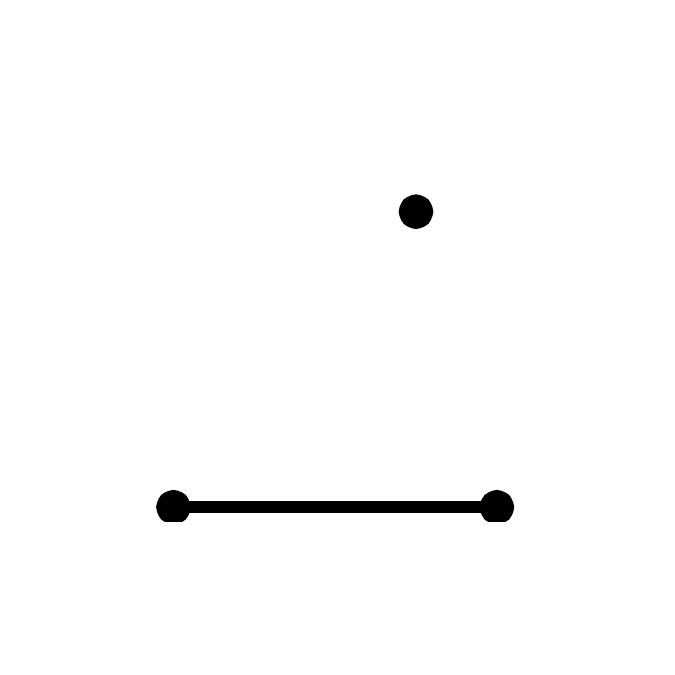}
        \caption{$m_1$} \label{fig:m1}
    \end{subfigure}%
    ~ 
    \begin{subfigure}[t]{0.25\textwidth}
      \addtocounter{subfigure}{-1}
      \renewcommand{\thesubfigure}{\alph{subfigure}2}
        \centering
        \includegraphics[trim={0 1.5cm 0 1.5cm}, clip, width=\textwidth]{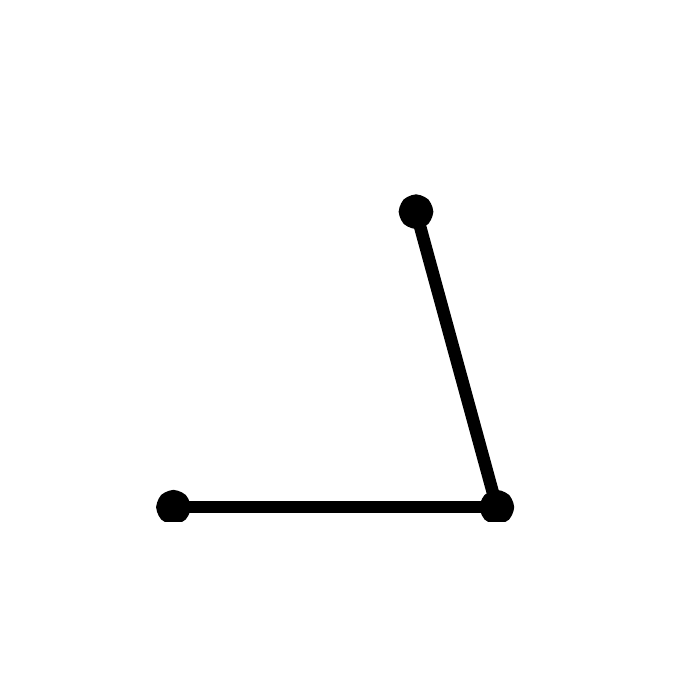}
        \caption{$m_2$} \label{fig:m2}
    \end{subfigure}
~
    \begin{subfigure}[t]{0.25\textwidth}
      \addtocounter{subfigure}{-1}
      \renewcommand{\thesubfigure}{\alph{subfigure}3}
        \centering
        \includegraphics[trim={0 1.5cm 0 1.5cm}, clip, width=\textwidth]{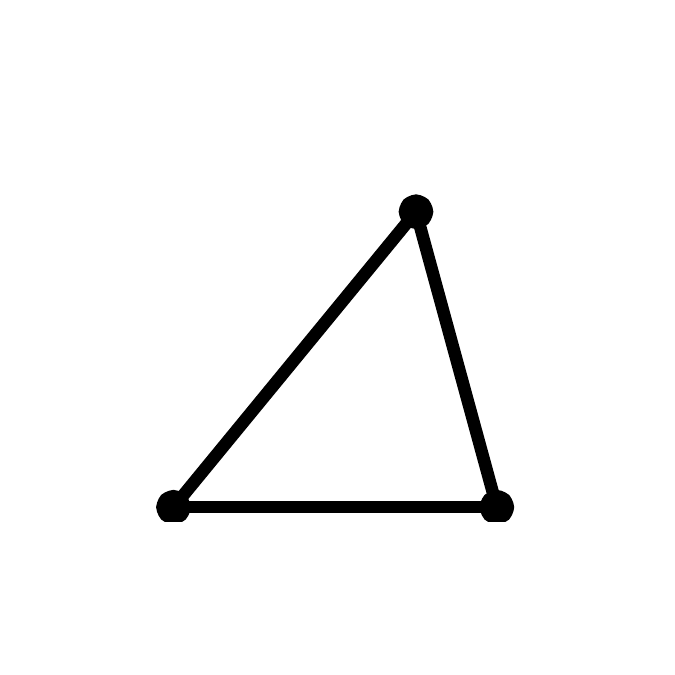}
        \caption{$m_3$} \label{fig:m3}
    \end{subfigure}
    \addtocounter{subfigure}{-1}
    \caption{Additional motifs.}
    \label{fig:motifs_dataex}
\end{subfigure}
\caption{Depiction of motifs considered in Sections \ref{sec:sims} and \ref{sec:realdat}.} \label{fig:motif_summs}
\end{figure}

To characterise the simulated hypergraphs we record subgraph counts, properties of the degree distribution and a measure of clustering. The particular subgraphs considered are depicted in Figure \ref{fig:motifs}, and the degree distribution and spread of hyperedge orders are recorded using the percentiles of the node degrees and edge sizes, respectively. in addition to the density of the hyperedges of order $k=2$ and $k=3$. Finally, to measure clustering, we project the hypergraph onto a pairwise graph, such that the edge $\{i,j\}$ exists if $i$ and $j$ are present in the same hyperedge, and calculate the community structure using the leading eigenvector with the function \texttt{cluster\_leading\_eigen()} in the \textit{igraph} package in R \citep{csardi2006}. 

\begin{sidewaysfigure}
  \begin{subfigure}{\textwidth}
    \hspace{-2cm} \includegraphics[width=1.1\textwidth]{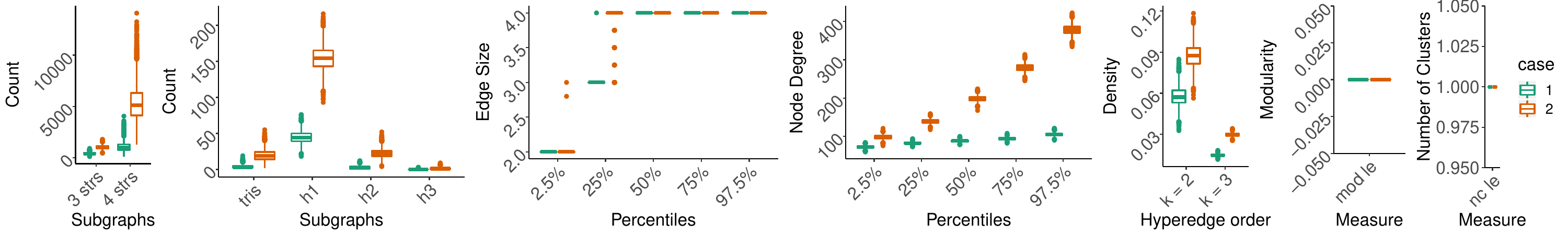}
      \caption{Uniform hypergraph model of \cite{stasi2014}.} \label{fig:beta}
  \end{subfigure}
  \begin{subfigure}{\textwidth}
    \hspace{-2cm} \includegraphics[width=1.1\textwidth]{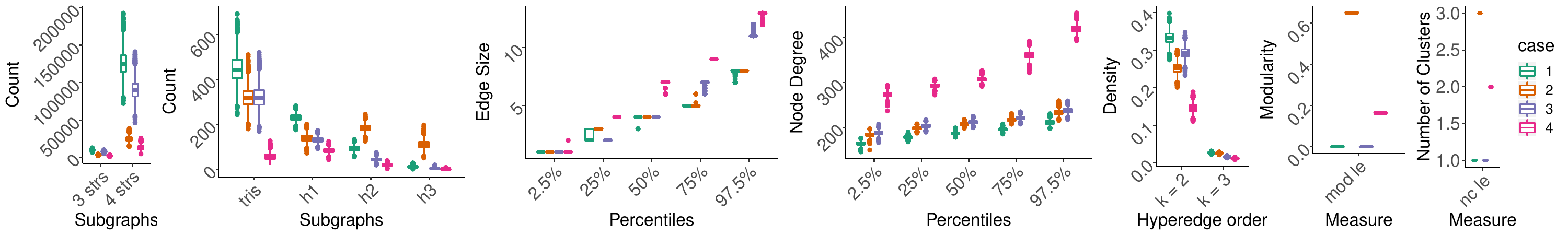}
      \caption{Extended latent class analysis hypergraph model of \cite{ng2018}.} \label{fig:murphy}
  \end{subfigure}
  \begin{subfigure}{\textwidth}
    \hspace{-2cm} \includegraphics[width=1.1\textwidth]{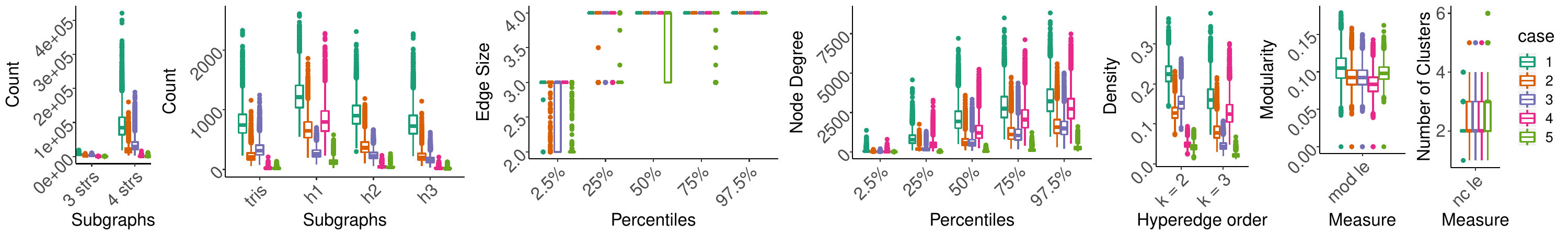}
      \caption{Latent space hypergraph model detailed in Algorithm \ref{alg:modhypext}.} \label{fig:lsm}
  \end{subfigure}
\caption{Summary of hypergraphs simulated from each of the models considered in Section \ref{sec:modeldepth}. The cases considered are summarised in Table \ref{table:cases_summary} and Table \ref{table:cases} in the Appendix.} \label{fig:modeldepth}
\end{sidewaysfigure}

Figure \ref{fig:modeldepth} summarises the results of our study. For the model of \cite{stasi2014} (Figure \ref{fig:beta}) we see that, by comparing Cases 1 and 2, this model can express hypergraphs with very different degree distributions. We also observe a larger number of motif counts for Case 2 since this case generates denser hypergraphs. Note that the maximum hyperedge order was set to 4, and so no hyperedges for $k \geq 5$ are generated. For the model of \cite{ng2018} (Figure \ref{fig:murphy}), we see that the topic clustering is consistently captured for all simulated hypergraphs. Generally, the degree distributions are similar, however by comparing Cases 1 and 4 we note that this model can express different levels of connectivity. We also observe little variation in the motif counts and, for most cases, find that triangles are more prevalent than the hypergraph motifs considered. For our latent space hypergraph model (Figure \ref{fig:lsm}), we observe a greater control over the motif counts by comparing Cases 3 and 4 where the order $k=2$ and $k=3$ hyperedges are denser. The counts for triangles and $h_1$ subgraphs clearly reflect the number of hyperedges of each order, and we also observe some control over the degree distribution and density. When we increase the latent dimension to $d=3$ and fix all other parameters, we obtain a sparser hypergraph as is demonstrated by comparing Cases 1 and 5. Finally, since our model is not designed to capture clustering, we do not observe consistent estimates for the number of clusters. 

Here we have explored the advantages of each modelling approach and shown that our model presents a flexible framework that is well-suited to hypergraph data which exhibit large motif counts. We note our approach could be modified to express different hypergraph characteristics by, for example, changing the distribution on $\bm{U}$ or the choice of complex.

\subsection{Assessing Model Fit}
\label{sec:privspost}

In this section we examine the performance of our MCMC scheme through inspection of the predictive degree distribution. For this study, we simulate a hypergraph from the model outlined in Algorithm \ref{alg:modhyp} with $N=50, d=2, K=3, \bm{r}=(0.35,0.45), \bm{\upvarphi}_0=\bm{\upvarphi}_1=(0.001,0.001), \mu = (0,0)$ and $\Sigma = \left(\begin{smallmatrix} 1 & 0 \\ 0 & 1\end{smallmatrix} \right)$. After $5000$ post burn-in iterations we obtain posterior mean estimates to (2 significant figures) $\hat{\bm{U}}$, $\hat{\bm{r}}= (0.15, 0.19),\hat{\bm{\psi}}^{0}=\hat{\bm{\psi}}^{1}=(0.000904,0.00021), \hat{\mu}=(-0.029,-0.0072)$ and $\hat{\Sigma} = \left(\begin{smallmatrix} 0.087 & 0.019 \\ 0.019 & 0.18 \end{smallmatrix} \right)$. Recalling that the anchor coordinates force a scaling on the estimated latent positions, we note that we cannot make a direct comparison between the true and estimated model parameters. This is further demonstrated by comparing the known and estimated latent coordinates in Appendix \ref{app:privpost}.

\begin{figure}[t!] 
  \centering
  \includegraphics[width=.45\textwidth]{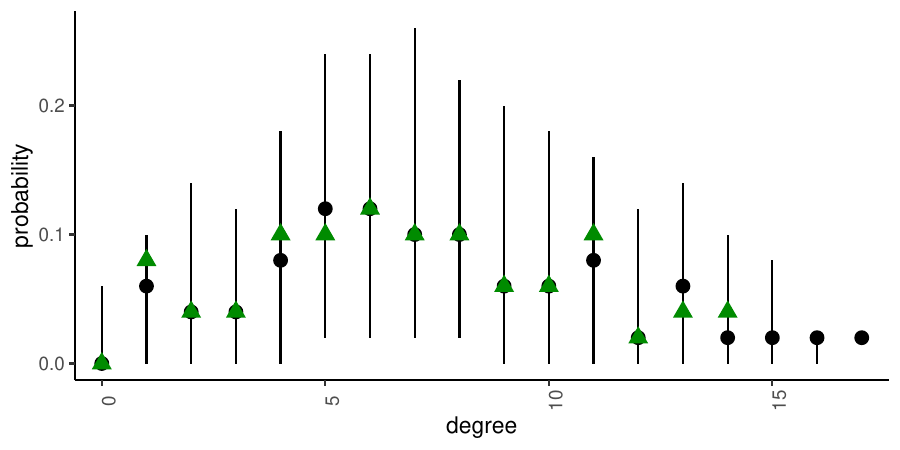}
  \includegraphics[width=.45\textwidth]{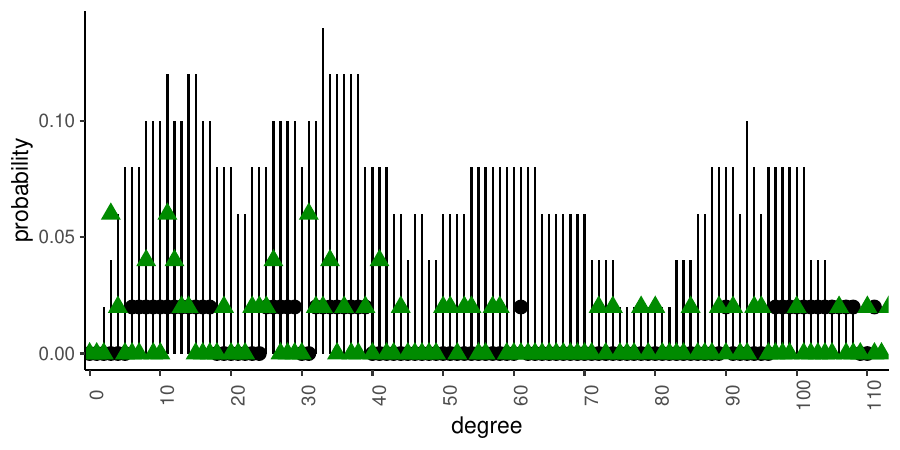}
    \caption{Comparison of true and posterior predictive degree distributions for hyperedges of order $k=2$ (left) and $k=3$ (right). For each figure, the vertical lines and black dots correspond to the range and median of the probabilities of observing each degree as calculated via the posterior predictive. The observed degree is shown as green triangles.}
    \label{fig:ddprivpost}
  \end{figure}

Alternatively, we can examine the posterior predictive and compare this to the simulated hypergraph. If the sampling procedure has produced meaningful estimates, we should find that there is a reasonable correspondence between these two quantities. Note however that a hypergraph is a complex object and so, in order to make a reasonable comparison, we summarise each hypergraph using its degree distribution. Figure \ref{fig:ddprivpost} provides a visual comparison by showing the posterior predictive distribution for the probability of each degree alongside the observed degree distribution. To allow a more nuanced comparison we present the degree distributions for hyperedges with order $k=2$ and $k=3$ separately. Overall we see that there is reasonable overlap between these quantities. For completeness, we also summarise the output of the MCMC scheme in Appendix \ref{app:privpost} where we see that the sampling procedure appears to have converged and the estimated latent coordinates correspond reasonably to the truth.

\subsection{Scalability study}
\label{sec:scalability}

In our MCMC scheme (see Section \ref{sec:postsmp}) the latent coordinates are updated via a delayed-acceptance (DA) step. This allows a poor proposal to be quickly rejected on a subset of the likelihood and we commented that DA offers computational advantages over a standard Metropolis-Hastings sampler. Here, we substantiate this claim on data simulated according to Algorithm \ref{alg:modhyp} by comparing average per-iteration cost times for an MCMC scheme implemented with and without DA. Since the computational cost is closely connected to the number of nodes $N$ and the density of the hypergraphs, we consider timings for hypergraphs with increasing $N$ under two different density regimes, referred to as (R1) and (R2). We specify the parameters for each regime to allow (R1) to be sparser than (R2), and we control this through $\bm{r}$ and $\Sigma$. Details of the parameters used to simulate the data are given in Appendix \ref{app:scalability} alongside plots which show the density of the simulated data.

\begin{figure}[t!]
  \centering
  \includegraphics[trim = {0cm, 0.5cm, 1cm, 1cm}, clip, width=.6\textwidth]{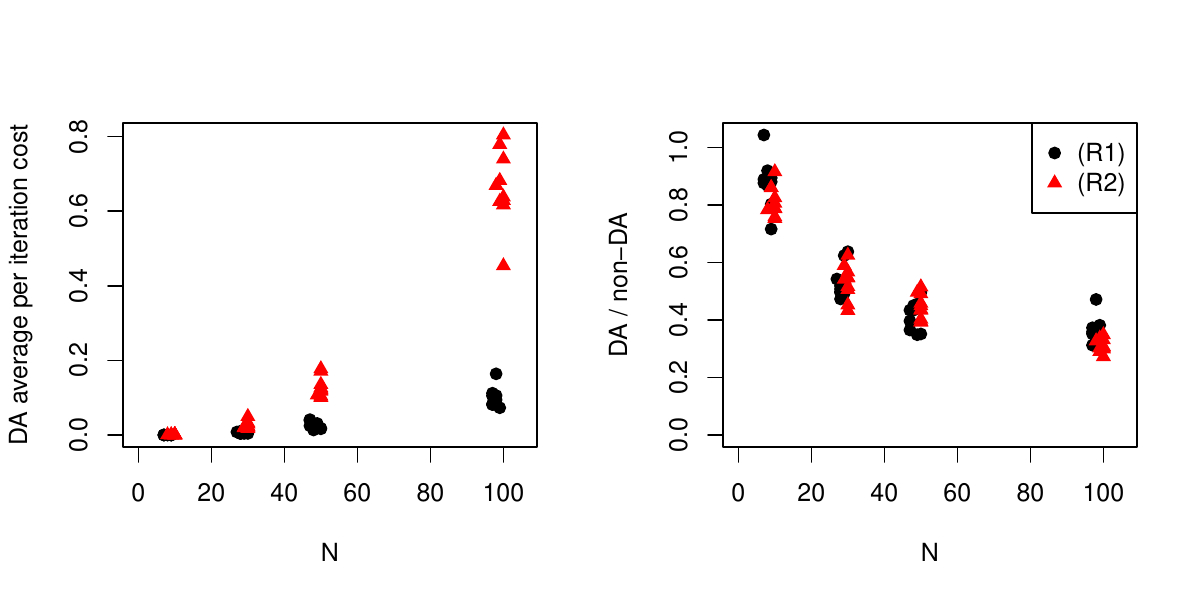}
  \caption{Summary of average per-iteration costs after $500$ iterations for an MCMC implemented with and without delayed-acceptance (DA) for regimes (R1) (black, circle) and (R2) (red, triangle). Left: average per-iteration cost in seconds for a DA scheme. Right: ratio of average per-iteration cost for a DA scheme and a scheme without DA. A number less than 1 implies that the DA offers a computational speed up.}
  \label{fig:scalability_timings}
\end{figure} 

This study allows us to examine the scalability of our approach as $N$ grows and to demonstrate the reduction in computational cost offered by DA. Figure \ref{fig:scalability_timings} reports average per-iteration times for the DA MCMC scheme for both regimes in addition to the ratio of average per-iteration times for a DA and non-DA MCMC scheme. This figure confirms our intuitions that hypergraphs with a larger $N$ are more computationally challenging, and this cost is further increased as the density of the hypergraphs grows. By inspecting the right-hand plot of Figure \ref{fig:scalability_timings}, we observe that incorporating DA into our MCMC offers a significant improvement in average per-iteration cost. We note that this relative improvement increases with $N$ and is largely unaffected by the density of the hypergraphs. Whilst $K$ remains fixed throughout this section, we do expect the improvement offered by DA to increase with $K$.

\section{Real data examples}
\label{sec:realdat}

In this section we analyse two real world hypergraph datasets using the model detailed in Algorithm \ref{alg:modhypext}. The first of these considers a dataset describing co-purchases of grocery items and the second returns to the motivating example of academic coauthorship.

\subsection{Grocery Items}
\label{sec:grocery}

\begin{figure}[t!]
  \centering
\begin{subfigure}[t]{.47\textwidth}
  \centering
  \includegraphics[width=.6\textwidth]{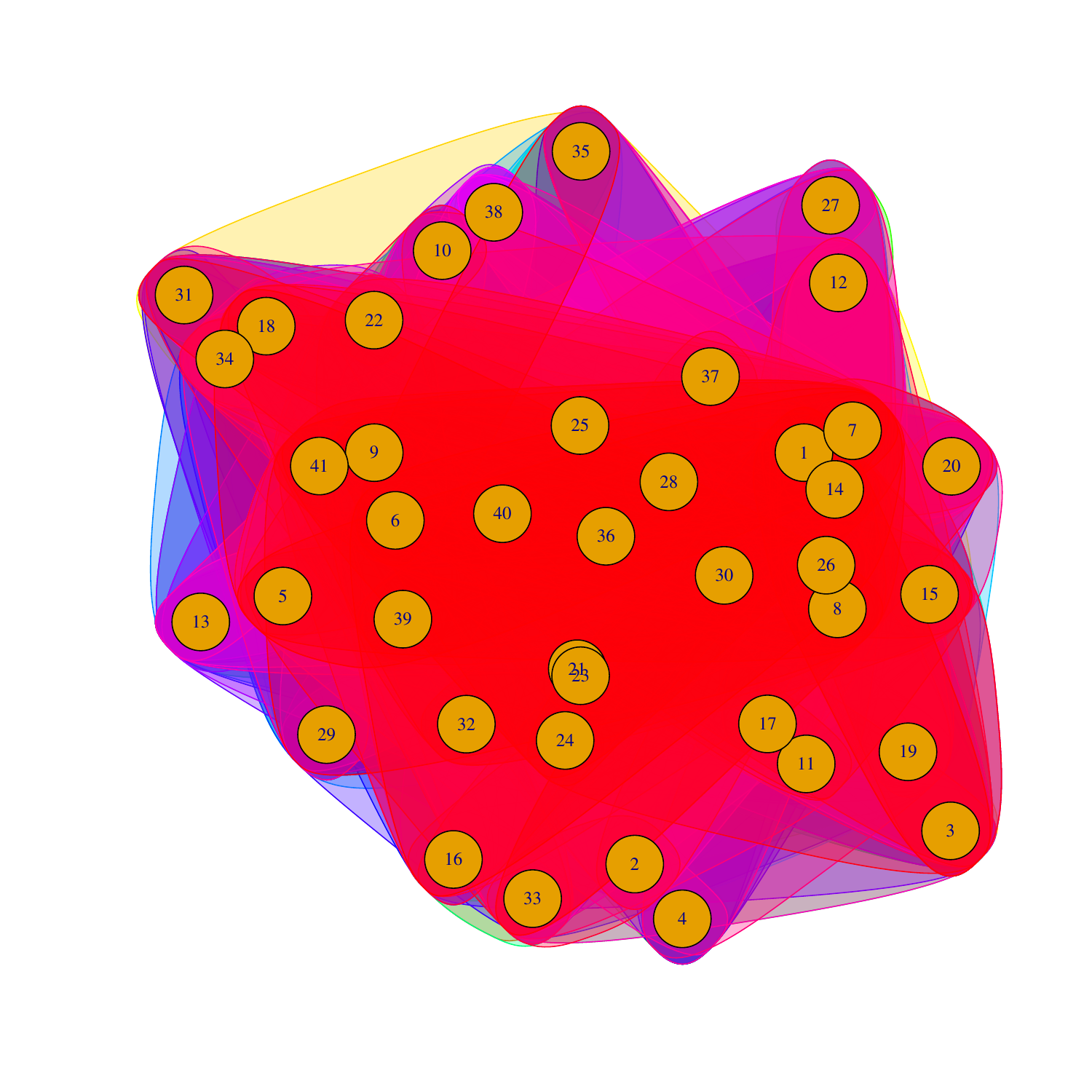}
\caption{Grocery dataset (Section \ref{sec:grocery}).} \label{fig:umn_grocery}
\end{subfigure}
~
\begin{subfigure}[t]{.47\textwidth}
  \centering
  \includegraphics[width=.6\textwidth]{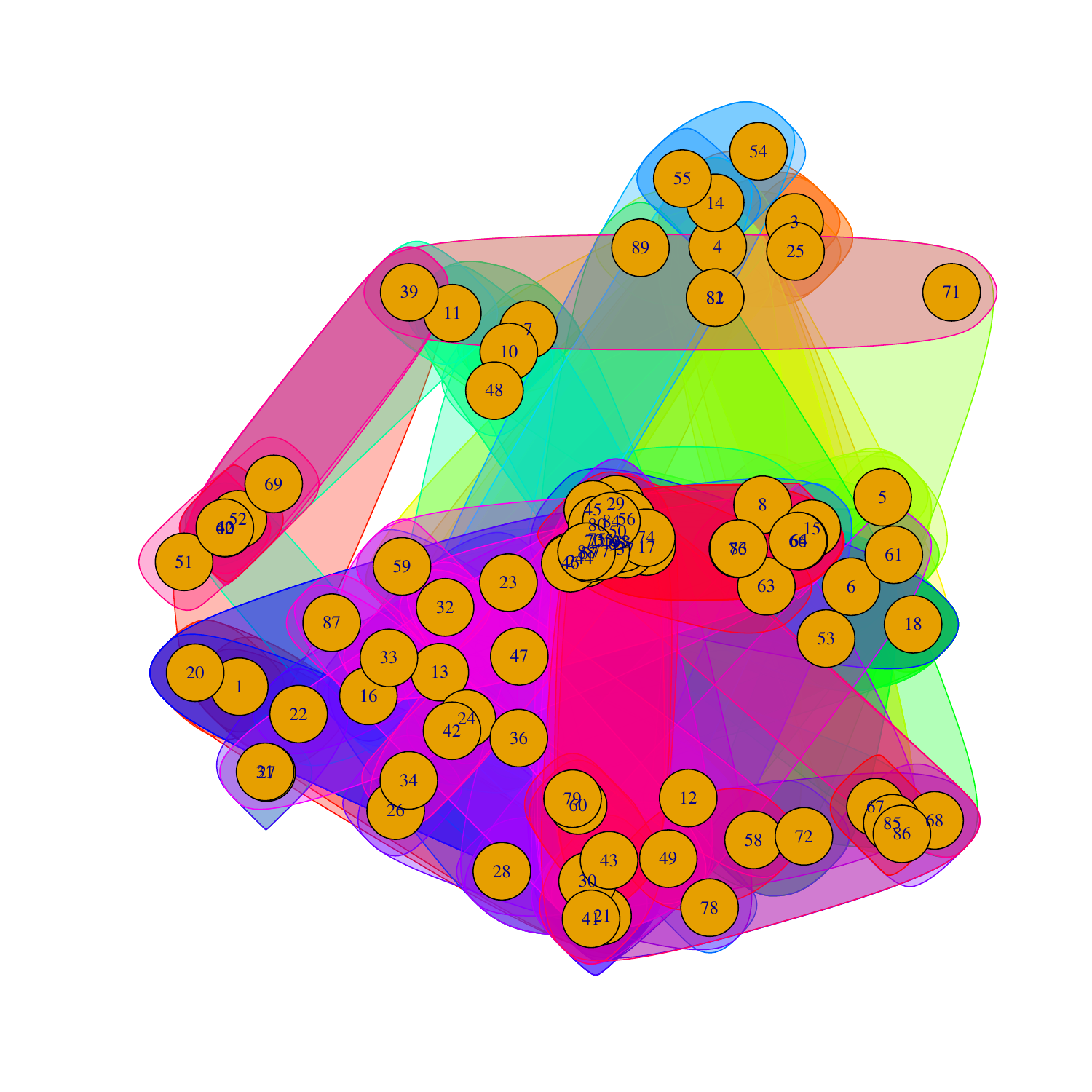}
\caption{Coauthorship dataset (Section \ref{sec:coauthorship}).} \label{fig:umn_dblp}
\end{subfigure}
\caption{Visualisation of each of the datasets considered in Section \ref{sec:realdat}. The hyperedges correspond to the observations for each dataset and the node positions are the posterior mean of the latent coordinates $\hat{\bm{U}}$.} \label{fig:data_upst}
\end{figure} 

Here we consider a subset of the grocery transaction data available as part of the R package \texttt{arules} \citep{arules}. In this example, each node corresponds to a category of items and each hyperedge indicates which set of items were bought together. To analyse this data we take a random subset of size $N=40$ with $K=6$ and hold back $N^* = 4$ for validation of the predictive distribution for additional nodes. Since the hyperedges can be expressed through the geometric or noise component of our model, we impose constraints on the noise terms to ensure that the set of hyperedges explained by the latent positions is non-empty. More specifically, we limit the order $k$ noise terms so that they cannot exceed the density of the order $k$ hyperedges in the observed hypergraph. With these constraints, we implemented our MCMC scheme for $40,000$ iterations and discarded the first half as burn-in. Figure \ref{fig:umn_grocery} depicts the observed hypergraph with the nodes positioned at posterior mean of the latent coordinates $\hat{\bm{U}}$ and we obtain the posterior mean estimate of the radii $\hat{\bm{r}} = (0.141, 0.147, 0.149, 0.152, 0.166)$. Traceplots for each parameter are reported in Appendix \ref{app:data_examples} and the average per-iteration cost for our scheme was $0.0079$ minutes. 

\begin{figure}[t!]
    \centering
\begin{subfigure}[t]{\textwidth}
  \centering
  \includegraphics[width=.6\textwidth, trim={.1cm 0.2cm 0.5cm 0.5cm}, clip]{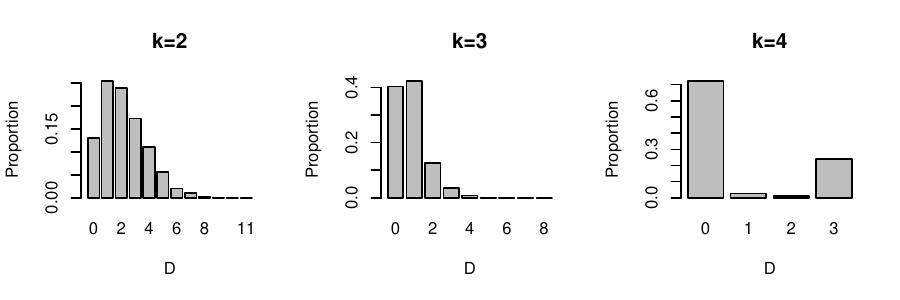}
  \caption{Node degree. Left to right: $k=2, k=3$ and $k=4$.}
  \end{subfigure}
\begin{subfigure}[t]{\textwidth}
  \centering
  \includegraphics[width=.7\textwidth, trim={.1cm 0.2cm 0.5cm 0.5cm}, clip]{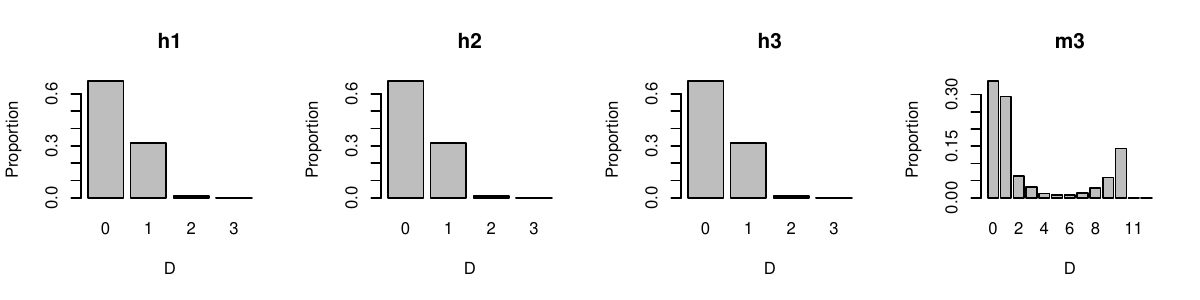}
  \caption{Motif counts. Left to right: motif in Figure \ref{fig:h1}, \ref{fig:h2}, \ref{fig:h3} and \ref{fig:m3}.}
\end{subfigure}
  \caption{Summary of predictive distributions for $N^*$ additional nodes for the Grocery dataset. For each measure we report the proportion of predictions which are distance $D$ from the truth, where $D$ is the absolute difference between the predictive and the truth. } \label{fig:pred_grocery} \end{figure}

To select the $N^*$ nodes for which we evaluate the predictive distribution, we take the nodes with the smallest eigen-centralities in the graph obtained by connecting all node pairs who share a hyperedge. To reflect how these nodes were selected, we then sample additional latent coordinates from the posterior predictive by first obtaining $N + N^*$ samples from $\mathcal{N}\left( \mu^{(i)}, \Sigma^{(i)} \right)$ and then selecting the $N^*$ coordinates that are furthest from the mode $\mu^{(i)}$ in terms of Euclidean distance. Similarly to Section \ref{sec:privspost}, we summarise the predictive hypergraphs by considering node degree and motif counts and Figure \ref{fig:pred_grocery} reports the results of this. In particular, for each measure, we report the proportion of predictive samples which have discrepancy $D$ from the true observed value where $D$ denotes the absolute difference between the prediction and the truth. Overall, we see that the majority of samples have either zero or close to zero discrepancy from the observed values, suggesting that we reasonably capture the unobserved interaction patterns.

\subsection{Coauthorship for Statisticians}
\label{sec:coauthorship}

\begin{figure}[t!]
    \centering
\begin{subfigure}[t]{\textwidth}
  \centering
  \includegraphics[width=.6\textwidth, trim={.1cm 0.2cm 0.5cm 0.5cm}, clip]{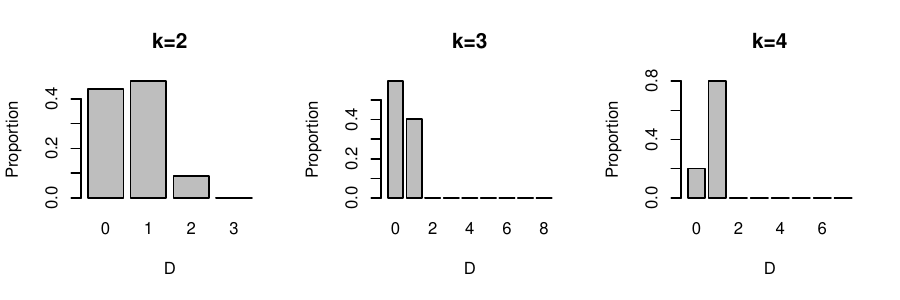}
  \caption{Node degree. Left to right: $k=2, k=3$ and $k=4$.}
  \end{subfigure}
\begin{subfigure}[t]{\textwidth}
  \centering
  \includegraphics[width=.7\textwidth, trim={.1cm 0.2cm 0.5cm 0.5cm}, clip]{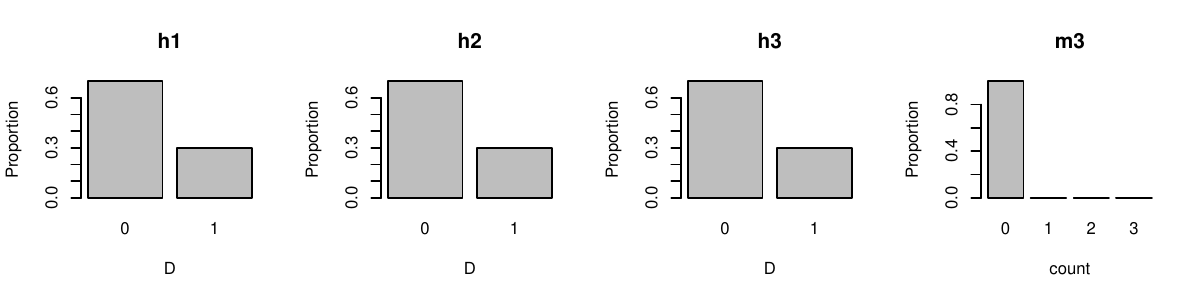}
  \caption{Motif counts. Left to right: motif in Figure \ref{fig:h1}, \ref{fig:h2}, \ref{fig:h3} and \ref{fig:m3}.}
\end{subfigure}
\caption{Summary of predictive distributions for $N^*$ additional nodes for the coauthorship dataset. For each measure we report the proportion of predictions which are distance $D$ from the truth, where $D$ is the absolute difference between the predictive and the truth. } \label{fig:pred_coauth}
\end{figure}

We now repeat the analysis from the previous section for a hypergraph with larger $N$. In this section we consider a coauthorship hypergraph constructed from the DBLP dataset\footnote{\url{https://dblp.uni-trier.de/xml/}}, presented in \cite{mao2020} and available online here\footnote{{\url{https://xueyumao.github.io/coauthorship.html}}}. From this dataset we construct a subset of size $N=99$ with $K=4$ via a random walk and, similarly to Section \ref{sec:grocery}, we hold out $N^*=10$ of these nodes according to the graph eigen-centrality and use these to validate predictive inference. We fit the model outlined in Section \ref{sec:ext} by implementing our MCMC for $80,000$ iterations and taking the initial $3/5^{th}$ of these as burn-in. Given equivalent restrictions on the noise parameters as described for the previous example, we obtain the posterior mean of the radii $\hat{\bm{r}} = (0.086, 0.096, 0.103)$ and the posterior means of the latent position as shown in Figure \ref{fig:umn_dblp}. The average per-iteration cost for our scheme was $0.0106$ minutes and traceplots showing convergence for the model parameters are given in Appendix \ref{app:data_examples}. A summary of the predictive distribution of the $N^*$ additional nodes is given in Figure \ref{fig:pred_coauth} and, for all summaries of the predicted interactions, we find that the majority of the discrepancies between the predicted and true values are either zero or close to zero. This suggests that our predicted interaction patterns are close to the truth. 

\begin{figure}[t]
  \centering
  \includegraphics[width=.8\textwidth, trim={1cm 0cm 0cm 1cm}, clip]{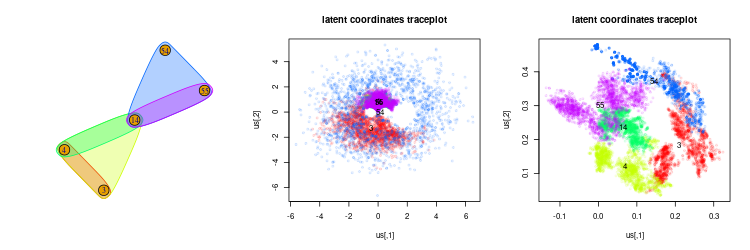}
  \caption{Comparison between latent positions for subset of the coauthorship network on nodes $\{3,4,14,54,55\}$. Left: observed interactions. Middle: traceplot of latent positions for model with $r_k=r$. Right: traceplot of latent positions for our model with $r_{k} > r_{k-1}$.} \label{fig:dblp_simp_nonsimp_comp}
\end{figure}

To provide further justification for our assumption of increasing radii we compare our model to a restriction in which $r_k = r$ for $k=2,3,\dots,K$. In the restricted model the geometric component is simplicial and, since this data example is highly non-simplicial, we expect fewer of the hyperedges to be explained by this component. To verify this, we present the latent coordinate traceplots for a subset of the data in Figure \ref{fig:dblp_simp_nonsimp_comp} for each of these models. By comparing the middle and right panels of this figure, we see that the latent coordinates are more constrained for our model with a non-simplicial geometric component. Indeed, by inspecting the hyperedges generated from each of the samples of the latent positions, we find that our model explains all but one of the observed interactions through the geometric component and the restricted model only explains two hyperedges through the geometric component. This matches our intuition and helps strengthen our case for the utility of our model.

\section{Discussion}
\label{sec:disc}

In this article we have introduced a model for hypergraph data in which each node is associated with a coordinate in $\mathbb{R}^d$ that influences the interaction patterns. Our approach extends the framework in \cite{hoff2002} to the hypergraph setting via a modification of a nerve construction that allows us to express non-simplicial hypergraphs. This application of a nerve draws a connection between stochastic geometry and latent space network models, and allows us to avoid the prohibitively expensive likelihood associated with the direct hypergraph analogue of \cite{hoff2002}. Similarly to the case of pairwise interactions, the latent representation imposes appealing properties on the hypergraphs generated from our model. Most notably, we observe high levels of transitivity and motif counts which suggest interactions among subsets of a group of nodes often occur jointly, and this feature is highlighted and contrasted with existing hypergraph models in the simulation in Section \ref{sec:modeldepth}. We stress that the geometric component of our model may however be unable to express certain hypergraph relationships. For example, the maximum number of leaves in a star will be limited by the dimension of the latent space. This may be mediated by either choosing an alternative convex set to generate the nerve, increasing the probability of hyperedge modification, or adopting a different specification for the latent positions.  

The modification of the indicators for the hyperedges in our model has two main motivations. Firstly, without this modification, the likelihood of observing a given hypergraph is binary and equal to one only when there is a perfect correspondence between the observed and estimated hyperedges. Secondly, the modification extends the support of the model to the space of all hypergraphs. Whilst techniques such as tempering could be used to resolve the model fitting issues associated with a binary likelihood, the challenge of characterising the support of the model would remain and it would be unclear a priori whether a suitable geometric representation of an observed hypergraph could be found. From \eqref{eq:likedecomp} we observed that a hypergraph generated from our model is a modification of a nsRGH and, as the probability of modification grows, the generated hypergraphs will behave more like the hypergraph analogue to an Erd\H{o}s-R\'{e}nyi random graph. In the context of networks, \cite{le2018} investigate the recovery of an underlying true network given a set of noisy observations. Whilst this differs from our setting, we can view our model in a similar way in which the nsRGH represents the truth. Additionally, we can draw a parallel with our approach to a regression comprised of a trend and noise term. In the most interesting case, the trend would explain as much of the data as possible and this helps motivate our observations on the magnitude of the probability of hyperedge modification. 

Our proposed approach provides a framework for modelling hypergraph data that could be extended to a number of different settings, and examples include modelling non-binary hyperedges, the addition of community structure analogously to \citep{handcock2007}, or the introduction of covariate information. Additional avenues for future work follow from considering alternative choices for the nerve construction, latent geometry and restrictions on the radii. For example, we could instead utilise the more scalable Vietoris-Rips complex \citep{vietoris1927}, consider non-Euclidean embeddings \citep{krioukov2010} or explore models with node-specific radii. Whilst there exist a myriad of possible extensions, it is important to stress that estimation for a given modification of our model may be challenging. Even for our proposed model, scalability remains a key limitation and this may be exacerbated by certain extensions. In this article we found that delayed-acceptance offered improvements in the computational efficiency of our MCMC sampler, and it remains an open line of enquiry to consider how methodology to improve scalability for latent space networks, such as likelihood approximations \citep{raftery2012, rastelli2018} or variational methods \citep{saltertownshend2013}, may be adapted to this setting.

\section*{Acknowledgements}
We acknowledge the support of the EPSRC funded  EP/L015692/1 STOR-i CDT, and Christopher Nemeth acknowledges the support of EPSRC grants EP/S00159X/1, EP/R01860X/1 and EP/V022636/1. This work was also supported, in part, by NSF awards CAREER IIS-1149662 and IIS-1409177, by ONR awards YIP N00014-14-1-0485 and N00014-17-1-2131. Thank you to Tin Lok James Ng for providing the Star Wars dataset analysed in a previous version of this work and Matthew Kahle, Sandipan Roy, Brendan Murphy and Marco Battiston for helpful comments. We also acknowledge the High End Computing facility at Lancaster University which was used to implement our simulations and data exmples.

\bibliographystyle{apalike}
\bibliography{biblio}

\appendix

\section{Bookstein coordinates}
\label{app:bookstein}

In Bookstein coordinates a set of points are chosen as the anchor points. These points are fixed in the space and all other points are translated, rotated and scaled accordingly. In Appendix \ref{sec:bk2d} we describe the Bookstein coordinates in $\mathbb{R}^2$, and in Appendix \ref{sec:bk3d} we describe the Bookstein coordinates in $\mathbb{R}^3$.

\subsection{Bookstein coordinates in $\mathbb{R}^2$}
\label{sec:bk2d}
Section 2.3.2 of \cite{dryden1998} details the Bookstein coordinates in $\mathbb{R}^2$. In this case, we set the anchor points $u_1^B = (u_{11}^B, u_{12}^B)$ and $u_2^B = (u_{21}^B, u_{22}^B)$ to be $\left( - 1/2,0 \right)$ and $\left( 1/2, 0 \right)$, respectively. Let $\bm{U}^B$ denote the Bookstein coordinates and $\bm{U}$ denote the untransformed coordinates. Then $\bm{U}^B$ is given by
\begin{align}
  \bm{U}^B &= cR( \bm{U} - \bm{b} ) \nonumber \\ 
&= \dfrac{1}{\sqrt{ (u_{21}^B - u_{11}^B)^2 + (u_{22}^B - u_{12}^B)^2 }}  
  \begin{bmatrix}
    \cos (a) & \sin (a) \\
    - \sin (a) & \cos(a)
  \end{bmatrix}
 \left( \bm{U} - \dfrac{1}{2}
   \begin{bmatrix}
     u_{11}^B + u_{21}^B \\
     u_{12}^B + u_{22}^B
   \end{bmatrix}
 \right), \label{eq:bookstein}
\end{align}
\begin{figure}[t]
  \centering
    \begin{subfigure}[t]{0.45\textwidth}
        \centering
        \includegraphics[width=.7\textwidth, trim={0cm 0cm 1cm 1cm}, clip]{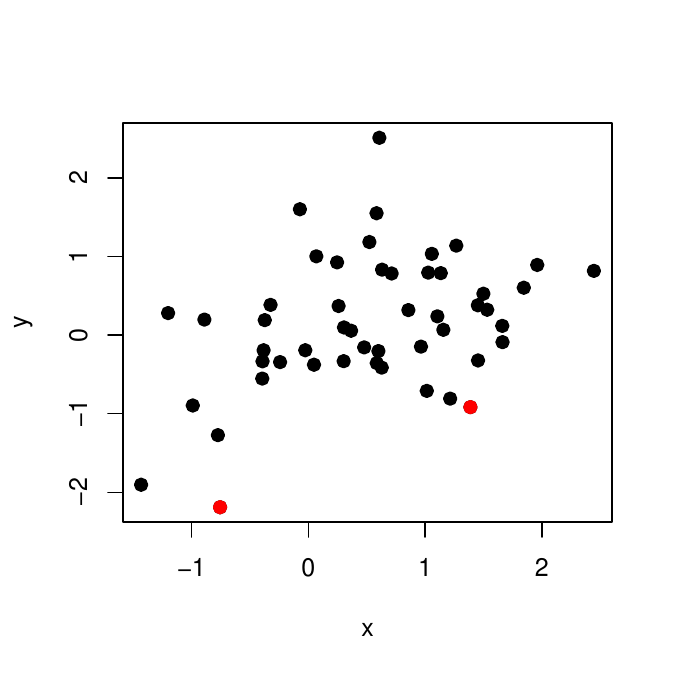}
    \end{subfigure}
~
    \begin{subfigure}[t]{0.45\textwidth}
        \centering
        \includegraphics[width=.7\textwidth, trim={0cm 0cm 1cm 1cm}, clip]{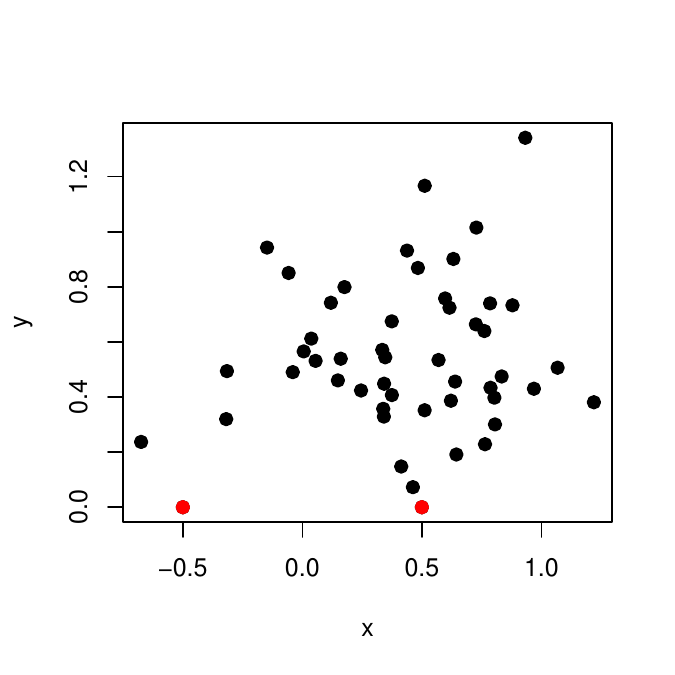}
    \end{subfigure}
    \caption{Bookstein transformation in $\mathbb{R}^2$. Left: original coordinates. Right: transformed Bookstein coordinates. The points highlighted in red are mapped to $(-1/2,0)$ and $(1/2,0)$.}
    \label{fig:bkstn}
\end{figure}

where $a = \mbox{arctan}\{ (u_{22}^B-u_{12}^B)/(u_{21}^B - u_{11}^B)\}$. The Bookstein transformation can hence be seen as a translation, rotation and re-scaling of the coordinates $\bm{U}$. Figure \ref{fig:bkstn} shows an example of the Bookstein transformation in $\mathbb{R}^2$.  

Furthermore, if $\bm{U} \sim \mathcal{N}(\mu, \Sigma)$, then we know that $\bm{U}^B \sim \mathcal{N}(\mu^B, \Sigma^B)$ where
\begin{align}
  \mu^B &= cR( \mu - b), \\
  \Sigma^B &= c^2 R \Sigma R^T.
\end{align}

\subsection{Bookstein coordinates in $\mathbb{R}^3$}
\label{sec:bk3d}
Section 4.3.3 of \cite{dryden1998} gives the Bookstein transformation for $\bm{U}$ where $u_i \in \mathbb{R}^3$. Following from \eqref{eq:bookstein} we first set $u_1^B = (-1/2, 0,0), u_2^B = (1/2,0,0) $ and $u_3^B = (u_{31}^B, u_{32}^B, 0) $. Then for $i=4,5,\dots,N$ and $l=1,2,3$ we calculate
\begin{align}
  w_{il} = u_{il} - \dfrac{1}{2} (u_{1l}^B + u_{2l}^B).
\end{align}
The Bookstein coordinate $u_i^B$ for $i=4,5,\dots,N$ is then given by
\begin{align}
  u_i^B = R_1 R_2 R_3 (w_{i1}, w_{i2}, w_{i3} ) / D_{12}
\end{align}
where
\begin{align}
R_1 &= 
  \begin{bmatrix}
    1 & 0 & 0 \\
    0 & \cos (\phi) & \sin( \phi ) \\
    0 & - \sin (\phi) & \cos (\phi) \\
  \end{bmatrix}
,  &&R_2 =  
  \begin{bmatrix}
    \cos ( \omega ) & 0 & \sin (\omega) \\
    0 & 1 & 0 \\
    - \sin (\omega ) & 0 & \cos( \omega )
  \end{bmatrix}
\\
  R_3 &= 
        \begin{bmatrix}
          \cos( \theta) & \sin(\theta) & 0 \\
          - \sin (\theta) & \cos (\theta) & 0 \\
          0 & 0 & 1
        \end{bmatrix}
, &&D_{12} = 2( w_{21}^2 + w_{22}^2 + w_{23}^2)^{1/2}.
\end{align}
Furthermore, we have
\begin{align}
  \theta = &\mbox{arctan}( w_{22} / w_{21} ) \\
  \omega = &\mbox{arctan}( w_{23} / (w_{21}^2 + w_{22}^2)^{1/2} ) \\
  \phi = &\mbox{arctan} \left( \dfrac{(w_{21}^2 + w_{22}^2)w_{33} - (w_{21}w_{31} + w_{22}w_{32}) w_{23}}{ (w_{21}^2 + w_{22}^2 + w_{23}^2)^{1/2} (w_{21}w_{32} - w_{31}w_{22}) } \right).
\end{align}

We see that the transformation in $\mathbb{R}^3$ is more involved than in $\mathbb{R}^2$ since we need to consider the effect of rotations over three different axes. Note that $R_1, R_2$ and $R_3$ correspond to a rotation around the $x,y$ and $z$ axes, respectively. 

\section{Modifying the Hyperedge Indicators}
\label{app:appnoise}

In the generative model detailed in Algorithm \ref{alg:modhyp}, the indicators for all order $k$ hyperedges are modified with probability $\upvarphi_k$. Since the probability $\upvarphi_k$ is small, we expect very few of the ${N \choose k}$ possible order $k$ hyperedges to be modified and so naively simulating a $\mbox{Bernoulli}(\upvarphi_k)$ random variable for each hyperedge is needlessly computationally expensive. Alternately, we can instead sample the number of order $k$ hyperedges whose indicator we modify, $n_k$, and then uniformly sample $n_k$ hyperedges from $\mathcal{E}_{N,k}$.

To sample an order $k$ hyperedge, we require a sample from the ordered set $\{i_1 < i_2 < \dots < i_k | i_1,i_2,\dots,i_k \in \{1,2,\dots,N\} \}$. Since the hyperedge indices are strictly increasing, it is sufficient to simply obtain $k$ samples from $\{1,2,\dots,N\}$ without replacement and order this sample as follows. 
\begin{enumerate}
\item Sample indices $ j_{1:k}=\{j_1, j_2, \dots, j_k\}$ from $\{1,2,\dots,N\}$ without replacement.
\item Set $e_k = \{i_1, i_2, \dots, i_k\} = \{j_{\sigma(1)}, j_{\sigma(2)}, \dots, j_{\sigma(k)} \}$ where $\sigma(l)$ denotes the index of the $l^{th}$ largest element of $j_{1:k}$ so that the elements of $e_k$ are increasing.
\end{enumerate}

We note that this procedure ignores the dependence between samples. However, we expect this effect to be negligible since the majority of hyperedges are not modified. Furthermore, we can also adapt this procedure for the model detailed in Algorithm \ref{alg:modhypext}.

\section{Conditional Posterior Distributions}
\label{app:condpost}

\subsection{Conditional posterior for $\mu$}
\label{app:mucond}

The conditional posterior for $\mu$ is given by
\begin{align}
  p(\mu | \bm{U}, \Sigma, m_{\mu}, \Sigma_{\mu} ) \propto p( \bm{U} | \mu, \Sigma ) p( \mu | m_{\mu}, \Sigma_{\mu} )
\end{align}
where $ p(\mu| m_{\mu}, \Sigma_{\mu}) = \mathcal{N}( m_{\mu} , \Sigma_{\mu} )$ and $p( \bm{U} | \mu, \Sigma) = \prod_{i=1}^N \mathcal{N}( u_i | \mu, \Sigma)$.

We have
\begin{align}
  p(\mu | \bm{U}, \Sigma, m_{\mu}, \Sigma_{\mu}) \propto \exp \left\{ -\dfrac{1}{2} \sum_{i=1}^N ( u_i - \mu)^T \Sigma^{-1} ( u_i - \mu) - \dfrac{1}{2} ( \mu - m_{\mu} )^T \Sigma_{\mu}^{-1} ( \mu - m_{\mu} )\right\}
\end{align}

By applying the result in Section of 8.1.7 \cite{peterson2012} we obtain
\begin{align}
  p(\mu | \bm{U}, \Sigma, m_{\mu}, \Sigma_{\mu}) = \mathcal{N}\left( ( N \Sigma^{-1} + \Sigma_{\mu}^{-1})^{-1} \left( \Sigma^{-1} \sum_{i=1}^N u_i + \Sigma_{\mu}^{-1} m_{\mu} \right), ( N \Sigma^{-1} + \Sigma_{\mu}^{-1})^{-1}  \right).
\end{align}

\subsection{Conditional posterior for $\Sigma$}
\label{app:sigmacond}

The conditional posterior for $\Sigma$ is given by
\begin{align}
  p(\Sigma | \bm{U}, \mu, \Phi, \nu ) \propto p( \bm{U} | \mu, \Sigma ) p( \Sigma | \Phi, \nu )
\end{align}
where $ p( \Sigma | \Phi, \nu  ) = \mathcal{W}^{-1}( \Phi, \nu)$ and $p( \bm{U} | \mu, \Sigma) = \prod_{i=1}^N \mathcal{N}( u_i | \mu, \Sigma)$.

We have
\begin{align}
  p( \Sigma | \bm{U}, \mu, \Phi, \nu ) &\propto \prod_{i=1}^N \dfrac{1}{ | \Sigma |^{1/2} } \exp \left\{ - \dfrac{1}{2} ( u_i - \mu)^t \Sigma^{-1} (u_i - \mu) \right\} | \Sigma|^{- (\nu + d + 1)/2} \exp \left\{ - \dfrac{1}{2} tr \left( \Phi \Sigma^{-1} \right)\right\} \\
  &\propto | \Sigma |^{- (\nu + d + N + 1)/2} \exp \left\{ - \dfrac{1}{2} \left[ tr \left( \Sigma^{-1} \sum_{i=1}^N (u_i - \mu)(u_i - \mu)^T \right)  + tr( \Phi \Sigma^{-1}) \right]\right\} \\
  &\propto  | \Sigma |^{- (\nu + d + N + 1)/2} \exp \left\{ - \dfrac{1}{2}  tr \left( \left[ \sum_{i=1}^N (u_i - \mu)(u_i - \mu)^T + \Phi \right]  \Sigma^{-1}  \right)  \right\}. \label{eq:postforsig}
\end{align}
Line \eqref{eq:postforsig} follows from the symmetry of $\Sigma$ and $\sum_{i=1}^N( v_i - \mu) (v_i - \mu)^T$, and properties of the trace operator. Hence, we obtain
\begin{align}
  p(\Sigma | \bm{U}, \mu, \Phi, \nu ) = \mathcal{W}^{-1} \left( \Phi + \sum_{i=1}^N (u_i - \mu)(u_i - \mu)^T, \nu + N \right).
\end{align}

\subsection{Conditional posterior for $\psi_k^{0}$}
\label{app:psi0cond}

The conditional posterior for $\psi_k^{0}$ is given by
\begin{align}
  p\left(\psi_k^{0} | \bm{U}, \bm{r}, h_{N,K}, a_k^{0}, b_k^{0} \right) \propto \mathcal{L}\left( \bm{U}, \bm{r}, \bm{\psi}^{1}, \bm{\psi}^{0} ; h_{N,K} \right) p\left( \psi_k^{0} | a_k^{0}, b_k^{0} \right)
\end{align}
where $ \mathcal{L}\left( \bm{U}, \bm{r}, \bm{\psi}^{1}, \bm{\psi}^{0} ; h_{N,K} \right)$ is as in \eqref{eq:likelihoodext} and $p\left( \psi_k^{0} | a_k^{0}, b_k^{0}\right) = \mbox{Beta}\left(a_k^{0}, b_k^{0}\right)$.

We have
\begin{align}
  p ( &\psi_k^{0} | \bm{U}, \bm{r}, h_{N,K}, a_k^{0}, b_k^{0} ) \nonumber \\
  &\propto \left(\psi_k^{0}\right)^{d_k^{(01)}(g_{N,K}(\bm{U}, \bm{r}), h_{N,K})} \left(1-\psi_k^{0}\right)^{d_k^{(00)}(g_{N,K}(\bm{U}, \bm{r}), h_{N,K})} \left(\psi_k^{0}\right)^{a_k^{0}-1}\left(1 - \psi_k^{0}\right)^{b_k^{0}-1} \\
  &\propto \left(\psi_k^{0}\right)^{d_k^{(01)}(g_{N,K}(\bm{U}, \bm{r}), h_{N,K}) + a_k^{0} - 1} \left(1-\psi_k^{0}\right)^{d_k^{(00)}(g_{N,K}(\bm{U}, \bm{r}), h_{N,K}) + b_k^{0} - 1}.
\end{align}

Hence, we obtain
\begin{align}
  p ( \psi_k^{0} | &\bm{U}, \bm{r}, h_{N,K}, a_k^{0}, b_k^{0} )  \nonumber \\
&= \mbox{Beta} \left(d_k^{(01)}(g_{N,K}(\bm{U}, \bm{r}), h_{N,K}) + a_k^{0}, d_k^{(00)}(g_{N,K}(\bm{U}, \bm{r}), h_{N,K}) + b_k^{0} \right).
\end{align}

\subsection{Conditional posterior for $\psi_k^{1}$}
\label{app:psi1cond}

The conditional posterior for $\psi_k^{1}$ is given by
\begin{align}
  p\left(\psi_k^{1} | \bm{U}, \bm{r}, h_{N,K}, a_k^{1}, b_k^{1} \right) \propto \mathcal{L}\left( \bm{U}, \bm{r}, \bm{\psi}^{1}, \bm{\psi}^{0} ; h_{N,K} \right) p\left( \psi_k^{1} | a_k^{1}, b_k^{1} \right)
\end{align}
where $ \mathcal{L}\left( \bm{U}, \bm{r}, \bm{\psi}^{1}, \bm{\psi}^{0} ; h_{N,K} \right)$ is as in \eqref{eq:likelihoodext} and $p\left( \psi_k^{1} | a_k^{1}, b_k^{1}\right) = \mbox{Beta}\left(a_k^{1}, b_k^{1}\right)$.

We have
\begin{align}
  p ( &\psi_k^{1} |  \bm{U}, \bm{r}, h_{N,K}, a_k^{1}, b_k^{1} ) \nonumber \\
      &\propto \left(\psi_k^{1}\right)^{d_k^{(10)}(g_{N,K}(\bm{U}, \bm{r}), h_{N,K})} \left(1-\psi_k^{1}\right)^{d_k^{(11)}(g_{N,K}(\bm{U}, \bm{r}), h_{N,K})} \left(\psi_k^{1}\right)^{a_k^{1}-1}\left(1 - \psi_k^{1}\right)^{b_k^{1}-1} \\
  &\propto \left(\psi_k^{1}\right)^{d_k^{(10)}(g_{N,K}(\bm{U}, \bm{r}), h_{N,K}) + a_k^{1} - 1} \left(1-\psi_k^{1}\right)^{d_k^{(11)}(g_{N,K}(\bm{U}, \bm{r}), h_{N,K}) + b_k^{1} - 1}.
\end{align}

Hence, we obtain
\begin{align}
  p ( \psi_k^{1} |  &\bm{U}, \bm{r}, h_{N,K}, a_k^{1}, b_k^{1} )  \nonumber \\
  &= \mbox{Beta} \left(d_k^{(10)}(g_{N,K}(\bm{U}, \bm{r}), h_{N,K}) + a_k^{1}, d_k^{(11)}(g_{N,K}(\bm{U}, \bm{r}), h_{N,K}) + b_k^{1} \right).
\end{align}

\section{MCMC initialisation}
\label{app:mcmcinit}

To initialise the latent coordinates $\bm{U}$ we rely on generalised multidimensional scaling (GMDS) (similarly to \cite{sarkar2006}) with the shortest path length as the distance measure. We calculate this on a weighted adjacency matrix which incorporates the intuition than nodes which appear in a hyperedge are likely closer than nodes which only share a pairwise edge. Given an initial value of $\bm{U}$ we then transform these coordinates onto the Bookstein space of coordinates, where we specify our anchor points so that they lie close to the smallest axis of variation as indicated by PCA applied to $\bm{U}_0$. Our initialisation procedure for $\bm{U}$ is given in Algorithm \ref{alg:uinit}. 

\begin{algorithm}[t]
  \caption{Initialise $\bm{U}$.} \label{alg:uinit}
  \begin{algorithmic}
    \STATE \textbf{Input}: Observed hypergraph $h_{N,K}$
    \STATE 1) Let $A \in \mathbb{R}^{N \times N}$ denote a weighted adjacency matrix. 
    \STATE \hspace{.35cm} For $i,j \in \{ 1,2,\dots,N\}$, if $\{i,j\}$ are connected by a hyperedge 
    \STATE \hspace{.75cm} - let $A_{(i,j)} = 1$ if $\{i,j\}$ are only connected by a hyperedge of order $k=2$,
    \STATE \hspace{.75cm} - let $A_{(i,j)} = \lambda$ if $\{i,j\}$ are connected by a hyperedge of order $k>2$.
    \STATE 2) Find the distance matrix $D \in \mathbb{R}^{N \times N}$, where $D_{(i,j)}$ is the shortest path between 
    \STATE \hspace{.35cm} nodes $\{i,j\}$ in the weighted graph determined by $A$. For $i=j$, let $D_{(i,j)}=0$.
    \STATE 3) Apply MDS to $D$ to obtain coordinates $\bm{U}_0 \in \mathbb{R}^{N \times d}$.
    \STATE 4) Apply PCA to $\bm{U}_0$ and determine the axis of smallest variation. Specify anchor 
    \STATE \hspace{.35cm} coordinates by finding points in $\bm{U}_0$ along this axis.
    \STATE 5) Given the index of the anchor points, transform $\bm{U}_{0}$ onto Bookstein coordinates
    \STATE \hspace{.35cm}  (see Appendix \ref{app:bookstein}). 
  \end{algorithmic}
\end{algorithm}

\begin{algorithm}[h]
  \caption{Initialise $\mu$ and $\Sigma$.} \label{alg:musiginit}
  \begin{algorithmic}
    \STATE \textbf{Input}: Hypergraph $h_{N,K}$, $\bm{r}_0$, $\bm{\psi}^{0}_0$, $\bm{\psi}^{1}_0$, $\mu \sim \mathcal{N}(m_{\mu}, \Sigma_{\mu})$, $\Sigma \sim \mathcal{W}^{-1}( \Phi, \nu)$, $N_{smp}$ and $\epsilon$.
    \STATE 1) Calculate $T(h_{N,K})$, where $T(\cdot)$ is a vector of hypergraph summary statistics. \STATE \hspace{.35cm} Let $n=0$.
    \STATE 2) While $n < N_{smp}$
    \STATE \hspace{.75cm} -Sample $\mu^* \sim \mathcal{N}( m_{\mu}, \Sigma_{\mu})$ and $\Sigma^* \sim \mathcal{W}^{-1}( \Phi,\nu)$. 
    \STATE \hspace{.75cm} -Sample $u_i^* \sim \mathcal{N}( \mu^*, \Sigma^*)$ for $i=1,2,\dots,N$.
    \STATE \hspace{.9cm} Let $\bm{U}^*$ be the $N \times d$ matrix whose $i^{th}$ row is $u_i^*$.
    \STATE \hspace{.75cm} -Given initial $\bm{r}_0$, determine the hypergraph $g_{N,K}(\bm{U}^*, \bm{r}_0)$.
    \STATE \hspace{.75cm} -Let $g_{N,K}^*$ by the hypergraph obtained by modifying $g_{N,K}(\bm{U}^*, \bm{r}_0)$ with noise
    \STATE \hspace{.9cm} $\bm{\psi}^{0}_0$ and $\bm{\psi}^{1}_0$
    \STATE \hspace{.75cm} -Calculate $T \left( g_{N,K}^* \right)$.
    \STATE \hspace{.75cm} -If $|T(h_{N,K}) -  T ( g_{N,K}^* )| < \epsilon$
    \STATE \hspace{1.45cm} Accept samples $\mu^*$ and $\Sigma^*$.
    \STATE 3) Let $\mu_0$ and $\Sigma_0$ be the average of $N_{smp}$ samples.
  \end{algorithmic}
\end{algorithm}
The radii $\bm{r}$ depend on the scale of $\bm{U}$, and so they are initialised in terms of $\bm{U}_{0}$. Given the initial latent coordinates, $\bm{r}_0$ is chosen to be the minimum radius which induces all edges that are present in $h_{N,K}$. The noise parameters $\bm{\psi}^{0}$ and $\bm{\psi}^{1}$ are initialised by sampling from their prior, where the prior values suggest that the perturbations are small. 

To initialise the parameters $\mu$ and $\Sigma$ we use an ABC scheme (see \cite{marin2012} for an overview). In this scheme we first sample $\mu$ and $\Sigma$ from their priors and then sample a hypergraph given $\bm{r}_0$. By comparing summary statistics of the sampled and observed hypergraphs, we determine whether or not our sampled hypergraph is similar enough to the observed hypergraph. If so, we accept the sampled $\mu$ and $\Sigma$. We choose the number of hyperedges of order $k=2$, the number of hyperedges of order $k=3$ and the number of triangles as our summary statistics. This initialisation scheme is detailed in Algorithm \ref{alg:musiginit} and the full initialisation scheme is given in Algorithm \ref{alg:mcmcinit}.
\begin{algorithm}[t]
  \caption{Procedure for initialising MCMC scheme in Algorithm \ref{alg:mcmc}.} \label{alg:mcmcinit}
  \begin{algorithmic}
    \STATE \textbf{Input} Observed hypergraph $h_{N,K}$
    \STATE 1) Determine $\bm{U}_0$ by applying Algorithm \ref{alg:uinit}. 
    \STATE 2) Let initial radii $\bm{r}_0$ be the smallest radii which induce all hyperedges observed in 
    \STATE \hspace{.2cm} $h_{N,K}$, conditional on $\bm{U}_0$.
    \STATE 3) Sample $\bm{\psi}_0^{0}$ and $\bm{\psi}_0^{1}$ from their prior distributions.
    \STATE 4) Sample $\mu_0$ and $\Sigma_0$ by applying Algorithm \ref{alg:musiginit}.
  \end{algorithmic}
\end{algorithm}

\section{Practicalities}
\label{app:pracs} 

To implement the MCMC scheme given in Algorithm \ref{alg:mcmc} there are a number of practical considerations we must address. In this section we comment on these where, in \ref{app:miniball} we discuss an approach for determining the presence of a hyperedge of arbitrary order and in \ref{app:evallikelihood} we discuss efficient evaluation of the likelihood.

\subsection{Smallest Enclosing Ball}
\label{app:miniball}
 
\begin{figure}[t]
  \centering
  \includegraphics[trim={0 2.25cm 0 2.25cm}, clip, width=.3\textwidth]{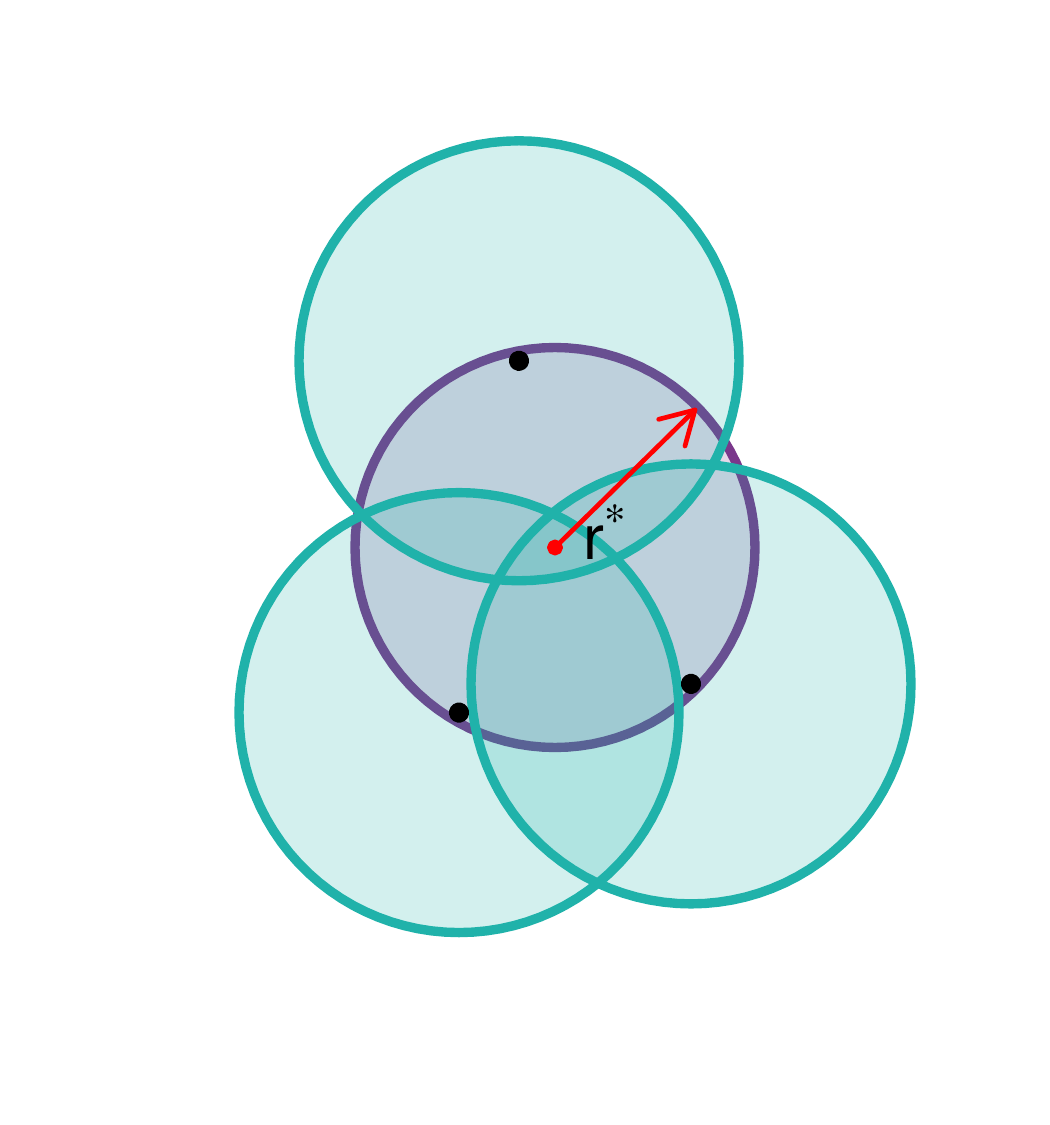}
  \caption{The blue shaded regions correspond to $B_r(u_i)$, for $i=1,2,3$, and the purple shaded region is the smallest enclosing ball of the points. The statements $r^* < r$ and $B_r(u_1) \cap B_r(u_2) \cap B_r(u_3) \neq \emptyset$ are equivalent.}
  \label{fig:mb}
\end{figure}

To determine whether a hyperedge $e_k = \{ i_1, i_2, \dots, i_k\}$ is present in $g(\bm{U}, \bm{r})$, we need to check whether $B_{r_k}(u_{i_1}) \cap B_{r_k}(u_{i_2}) \cap \dots \cap B_{r_k}(u_{i_k}) \neq \emptyset$. This is equivalent to checking whether the coordinates $\{ u_{i_1}, u_{i_2}, $ $\dots, u_{i_k} \}$ are contained within a ball of radius $r_k$ \citep[Section 3.2]{edelsbrunner2010} (see Figure \ref{fig:mb}). Therefore, we require
\begin{enumerate}
\item Determine the smallest enclosing ball $B$ for the coordinates $\{ u_{i_1}, u_{i_2}, \dots, u_{i_k} \}$.
\item If the radius of $B$ is less than $r_k$ take $y_{e_k}^{(g)}=1$, otherwise $y_{e_k}^{(g)}=0$.
\end{enumerate}

To compute the smallest enclosing ball we can rely on the the miniball algorithm , which may be also be referred to as the minidisk algorithm [Section 4.7]\citep{berg2008}. Details of this are given in Algorithm \ref{alg:miniball} and we refer the interested reader to \citep[Section 3.2]{edelsbrunner2010} for the intuition behind this. An alternative description of this algorithm can be found in \citep[Section 4.7]{berg2008}, and for efficient implementation of the \v{C}ech complex we rely on the GUDHI C++ library \citep{gudhi}.

\begin{algorithm}[t]
  \caption{Miniball} \label{alg:miniball}
  \begin{algorithmic}
    \STATE 1) Set $\sigma_1 = e_k$ and $\sigma_2 = \emptyset$
    \STATE 2) \textbf{if} $\sigma_1 = \emptyset$, compute the miniball $B$ of $\sigma_2$
    \STATE \hspace{.3cm} \textbf{else} choose $u \in \sigma_1$
    \STATE \hspace{.8cm} -Calculate the miniball $B$ which contains the points $\sigma_1 \setminus u$ in its interior and the
    \STATE \hspace{.95cm} points $\sigma_2$ on its boundary
    \STATE \hspace{.8cm} -\textbf{if} $u \notin B$, then set $B$ to be the miniball $B$ which contains the points $\sigma_1 \setminus u$ 
    \STATE \hspace{.95cm} in its interior and the points $\sigma_2 \cup u$ on its boundary
  \end{algorithmic}
\end{algorithm}

\subsection{Evaluating $\mathcal{L}(\bm{U},\bm{r}, \bm{\psi}^{1},\bm{\psi}^{0}; h_{N,K})$}
\label{app:evallikelihood}

The likelihood given in \eqref{eq:likelihoodext} requires the enumeration of hyperedge discrepancies between the observed hypergraph $h_{N,K}$ and the induced hypergraph $g_{N,K}(\bm{U}, \bm{r})$. In this section we note that this does not require a summation over all possible hyperedges, and so can be computed far more efficiently than \eqref{eq:subset} suggests. We first discuss evaluation of \eqref{eq:likelihoodext}, and then observe that \eqref{eq:likelihood} can be evaluated in a similar way.

To evaluate the likelihood we have the hyperedges present in $h_{N,K}$ and the hyperedges present in $g_{N,K}(\bm{U}, \bm{r})$. In practice, as the data examples from Section \ref{sec:realdat} suggest, the number of hyperedges in each of these hypergraphs is much smaller than the number of possible hyperedges $\sum_{k=2}^K {{N} \choose {k}}$. Let $n_{k}^{(h)}$ and $n_{k}^{(g)}$ denote the number of order $k$ hyperedges in $h_{N,K}$  and $g_{N,K}(\bm{U}, \bm{r})$, respectively. To evaluate the likelihood, we first enumerate the number of order $k$ hyperedges which are present in both hypergraphs to obtain $d_k^{(11)}(g_{N,K}(\bm{U}, \bm{r}), h_{N,K})$. This can easily be computed by evaluating the number of intersection between the hyperedges in $h_{N,K}$ and $g_{N,K}(\bm{U}, \bm{r})$. Then, for $k=2,3,\dots,K$, we have
\begin{align}
  d_k^{(10)}(g_{N,K}(\bm{U}, \bm{r}), h_{N,K}) &= n_k^{(g)} - d_k^{(11)}(g_{N,K}(\bm{U}, \bm{r}), h_{N,K}),\label{eq:d10} \\
  d_k^{(01)}(g_{N,K}(\bm{U}, \bm{r}), h_{N,K}) &= n_k^{(h)} - d_k^{(11)}(g_{N,K}(\bm{U}, \bm{r}), h_{N,K}),\label{eq:d01} \\
  d_k^{(00)}(g_{N,K}(\bm{U}, \bm{r}), h_{N,K}) &= {{N} \choose {k}} -  \left[ d_k^{(11)}(g_{N,K}(\bm{U}, \bm{r}), h_{N,K})  \right. \nonumber\\
  &\hspace{.75cm} \left. + d_k^{(10)}(g_{N,K}(\bm{U}, \bm{r}), h_{N,K})  + d_k^{(01)}(g_{N,K}(\bm{U}, \bm{r}), h_{N,K}) \right].
\end{align}

Hence, we are able to avoid summation over all possible hyperedges. We can easily calculate the distance specified in \eqref{eq:l1} from the above, by observing that it is given by the sum of \eqref{eq:d10} and \eqref{eq:d01}.

\section{Theoretical results in Section \ref{sec:theory}}
\label{app:proofs}

\subsection{Proofs for Section \ref{sec:degs_k2}}
\label{app:proofs_k2}

\begin{proof}[Theorem \ref{thm:dd_k2}]
  This follows immediately from \cite[Theorem 1]{rastelli2015}.
\end{proof}

\begin{proof}[Proposition \ref{prop:prob_ij_give_ui}] \label{proof:prob_ij_give_ui}
Since a hyperedges can form due to the geometric or noise component of our model, it immediately follows that
\begin{align}
  p_{\{ i,j \}}\left( y_{\{i,j\}}^{(g^*)} = 1 | u_i, r_2, \upvarphi_2, \mu, \Sigma \right) &= p\left( y_{\{ i,j \}}^{(g)} = 1 | u_i, r_2, , \mu, \Sigma \right) (1 - \upvarphi_2 ) \nonumber \\
  &\hspace{3cm} + \left(1 -  p \left( y_{\{ i,j \}}^{(g)} = 1 | u_i, r_2, , \mu, \Sigma \right) \right) \upvarphi_2,
\end{align}
where $p( y_{\{i,j\}}^{(g)} = 1 | u_i, r_2, \mu, \Sigma)$ denotes the probability of the hyperedge $e_2 = \{i,j\}$ being present in $g(\bm{U}, \bm{r})$, where node $i$ is positioned at $u_i$.

Recall that the hyperedge $\{i,j\}$ occurs in $g(\bm{U}, \bm{r})$ if $B_{r_2}(u_i) \cap B_{r_2}(u_j) \neq \emptyset$. Given $u_i$, this can equivalently be expressed as $p( \| U_j - u_i \| \leq 2 r_2 | u_i, r_2, \mu, \Sigma)$ where $U_j \sim \mathcal{N}(\mu, \Sigma)$. Therefore we require an expression for
\begin{align}
  p_i( y_{e_2}^{(g)} = 1 | u_i, r_2, \mu, \Sigma) = p( \| U_j^* \| \leq 2 r_2) =  p \left( (U_j^*)^T U_j^* \leq 4 r_2^2 \right), \label{eq:pe2_given_ui_r2}
\end{align}
where $U_j^* \sim \mathcal{N}( \mu - u_i, \Sigma)$.

Since we assume $\Sigma$ is diagonal, we may write
\begin{align}
  (U_j^*)^T U_j^* = \sum_{l=1}^d (U_{j,l}^*)^2,
\end{align}
where $U_{j,l}^* \sim \mathcal{N}( (\mu - u_i)_{l}, \sigma_l^2)$ denotes the $l^{th}$ dimension of $U_j^*$. We then obtain the distribution of $(U_{j,l}^*)^2$ by noting
\begin{align}
  \left( \dfrac{U_{j,l}^*}{\sigma_l} \right)^2 \sim \chi^2_{1, ((\mu - u_i)_{l} / \sigma_l)^2 } \hspace{.5cm} \Rightarrow \hspace{.5cm} \left( U_{j,l}^* \right)^2 \sim \sigma_l^2 \chi^2_{1, ((\mu - u_i)_{l} / \sigma_l)^2 }. 
\end{align}
Hence, we have
\begin{align}
  (U_j^*)^T U_j^* = \sum_{l=1}^d (U_{j,l}^*)^2 \sim \sum_{l=1}^d \sigma_l^2 \chi^2_{1, ( (\mu - u_i)_{l} / \sigma_l)^2 }. \label{eq:dist_ustrsqr}
\end{align}

For general diagonal $\Sigma$ we can evaluate \eqref{eq:pe2_given_ui_r2} using the R package \texttt{CompQuadForm} (\cite{duchense2010}). We note here that, if $\sigma_l = \sigma$ for $l=1,2,\dots,d$, we can simplify \eqref{eq:dist_ustrsqr} to the following
\begin{align}
  (U_j^*)^T U_j^* \sim  \sigma^2 \chi^2_{d, \lambda},
\end{align}
where $\lambda = \sum_{l=1}^d (\mu - u_i)_l ^2/ \sigma^2$. This implies that
\begin{align}
    p_i( y_{e_2}^{(g)} = 1 | u_i, r_2, \mu, \Sigma) = p( (U_j^*)^T U_j^* \leq 4 r_2^2) = p( \sigma^2 Z \leq 4 r_2^2) = p(  Z \leq 4 r_2^2 / \sigma^2),
\end{align}
where $Z \sim \chi^2_{d, \lambda}$. 

\end{proof}

\subsection{Proofs for Section \ref{sec:degs_kg2}}
\label{app:proofs_kg2}

\begin{proof}[Proposition \ref{prop:approx_dd_kg2}]

  Conditional on the latent coordinate $u_i$, we can view the order $k$ degree of the $i^{th}$ node as a sum of Bernoulli trials with success probability $p\left( y_{e_{k,i}}^{(g^*)}=1 | u_i, r_k, \mu, \Sigma\right)$ over all possible ${N-1 \choose k-1}$ hyperedges. Note however, that these Bernoulli trials are not dependent since there exist combinations of hyperedges containing node $i$ which share additional indices. For example, when $k=3$, the hyperedges $\{i,j,l\}$ and $\{i,j,m\}$ are not independent.

  Writing an exact expression for the order $k$ degree distribution is therefore challenging. To address the dependence, we consider a Poisson approximation of the sum of dependent Bernoulli trials (discussed, for example, in \cite{teerapabolarn2014}). The result then follows by extending the expression for the order $k=2$ degree distribution in Theorem \ref{thm:dd_k2} and substituting a Poisson approximation for the sum of dependent Bernoulli trials.
  
\end{proof}

\begin{proof}[Lemma \ref{lemma:prob_ekg2}]

  Firstly, the decomposition
  \begin{align}
    p \left( y_{ e_{k,i} }^{(g^*)} = 1 | u_i, r_k, \upvarphi_k, \mu, \Sigma \right) &= p\left( y_{ e_{k,i} }^{(g)} = 1 | u_i, r_k, \mu, \Sigma \right) (1 - \upvarphi_k ) \nonumber \\
  &\hspace{3cm} + \left(1 -  p \left( y_{ e_{k,i} }^{(g)} = 1 | u_i, r_k, \mu, \Sigma \right) \right) \upvarphi_k
\end{align}
follows immediately given the explanation in Proof \ref{proof:prob_ij_give_ui}. Additionally, we note that, since $U_j \sim \mathcal{N}(\mu, \Sigma)$ and $\Sigma = \mbox{diag}( \sigma_1^2, \sigma_2^2, \dots, \sigma_d^2)$, we have $(U_j)_l \sim \mathcal{N}(\mu_l, \sigma_l^2)$ where subscript $l$ refers to the $l^{th}$ dimension, for $l=1,2,\dots,d$.

  We approximate $p \left( y_{ e_{k,i} }^{(g)} = 1 | u_i, r_k, \mu, \Sigma \right)$ by considering the probability that coordinates $\{ u_j \}_{j \in e_{k,i}}$ fall within a square of side length $s_k$, denoted as $S_{s_k}$, instead of a ball of radius $r_k$. This greatly simplifies the region of integration and allows us to consider the integral independently for each dimension. We choose $s_k$ so that $S_{s_k}$ has area equal to that of a ball of radius $r_k$, and this gives $s_k = \sqrt{\pi} r_k$.

  Our goal is to evaluate the expression $p\left( \{ u_j \}_{j \in e_{k,i}} \in S_{s_k} | u_i, r_k, \mu, \Sigma \right)$, and we first note that
  \begin{align}
    p\left( \{ u_j \}_{j \in e_{k,i}} \in S_{s_k} | u_i, r_k, \mu, \Sigma \right) = \prod_{l=1}^d p \left( R_l \leq s_k | u_{il}, r_k, \mu, \Sigma \right),
  \end{align}
  where $u_{il}$ is the $l^{th}$ element of $u_i$ and $R_l$ denotes the range of $\{ u_{jl} \}_{j \in e_{k,i}}$.

  To evaluate this, we rely on well-known results from order statistics \citep{david2004}. In particular, let $X_1, X_2, \dots, X_m$ be independent but non-identically distributed, where $X_j$ has pdf $f_j(x)$ and cdf $F_j(x)$. Then, it is well known that
  \begin{align}
    F(t) = \sum_{j=1}^M \int f_j(x) \prod_{v \neq j} \left[ F_v(x+t) - F_v(x) \right] \,d x
  \end{align}
  is the cdf of the range of $X_1, X_2, \dots, X_m$. By applying this result to our setting and noting that the $i^{th}$ node has a fixed value, we obtain the result.
  
\end{proof}

\subsection{Additional figures}
\label{app:theory_figures}

\begin{figure}[t!]
  \centering
  \includegraphics[width=.7\textwidth, trim={0cm 0cm 1cm 2cm}, clip]{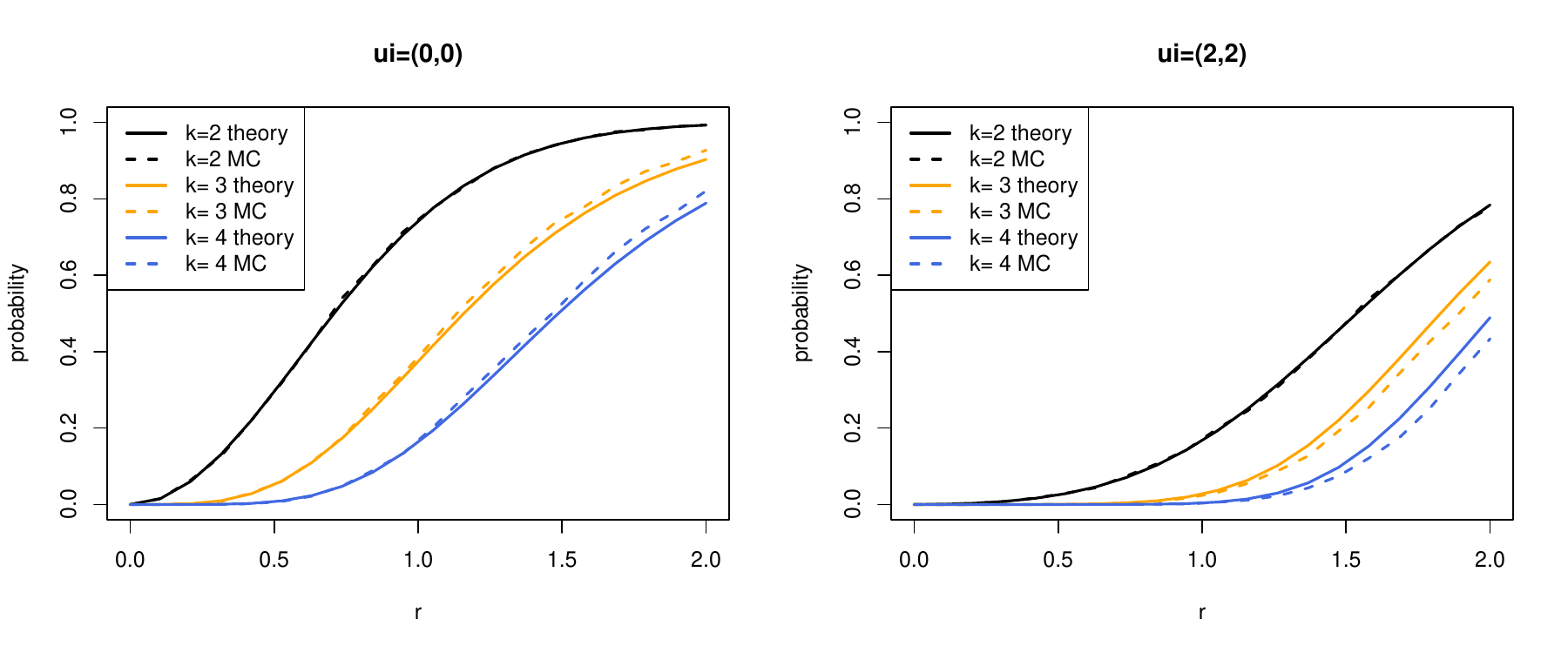}
  \caption{Comparison of theoretical (solid) and Monte Carlo (dashed) estimates of $p(y_{e_{k,i}}^{(g)}=1 | u_i, r_k, \mu, \Sigma)$ for varying $r_k$. We take $\Sigma = \mbox{diag}(1,2), \mu=(0,0)$ and consider connection probabilities for $k=2,3,4$. In the left plot $u_i = \mu$ and in the right plot $u_i = (1,2)$.}  \label{fig:pek_given_ui_sig2}
\end{figure}

\begin{figure}[t!]
  \centering
  \includegraphics[width=.7\textwidth, trim={0cm 0cm 1cm 2cm}, clip]{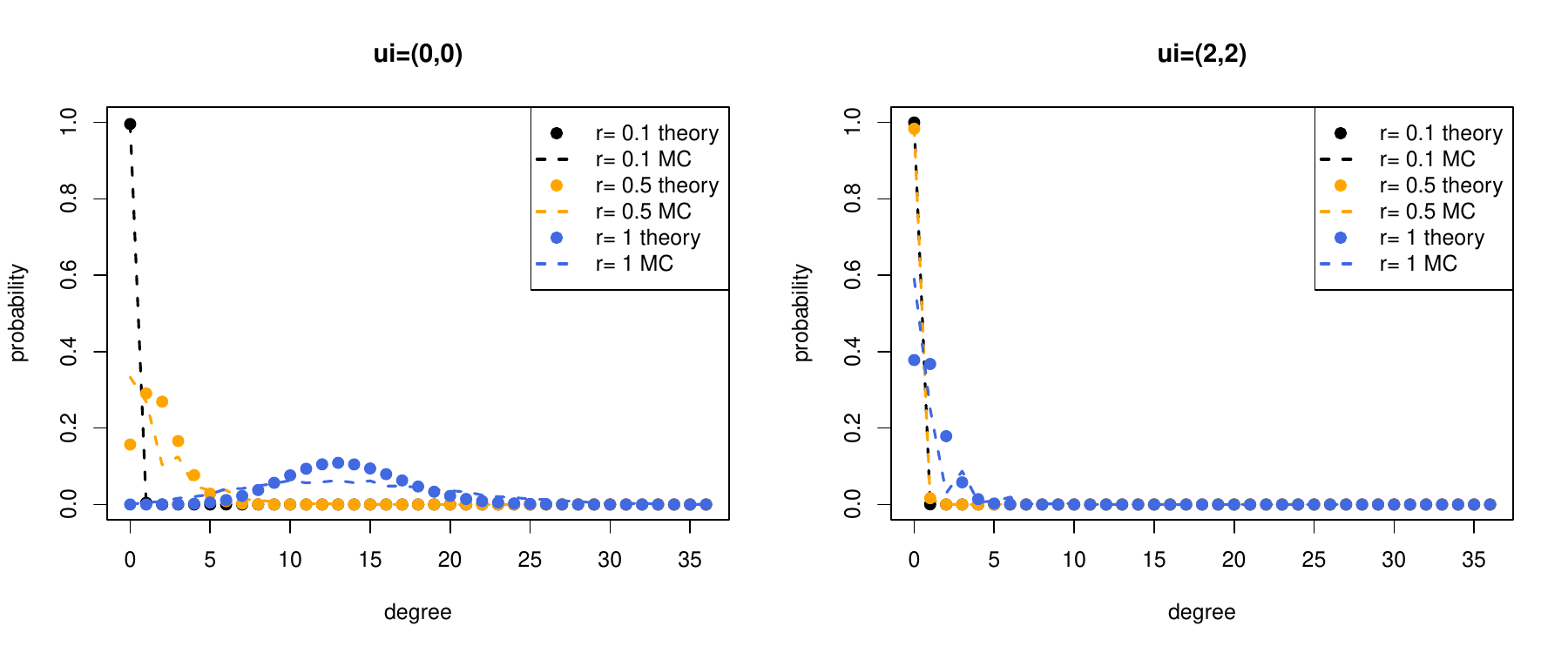}
  \caption{Comparison between empirical (dashed line) and Poisson approximation (points) of the order $k=3$ degree distribution. We take $N=10, \mu=(0,0), \Sigma = \mbox{diag}(1,2)$ and evaluate the distribution for $r_3 \in (0.1, 0.4, 1.0)$. The left plot shows $u_i = \mu$ and the right plot shows $u_i = (1,2)$.}  \label{fig:poisson_apprx_ui_sig2}
\end{figure}

\begin{figure}[t!]
  \centering
  \includegraphics[width=.7\textwidth, trim={0cm 0cm 1cm 2cm}, clip]{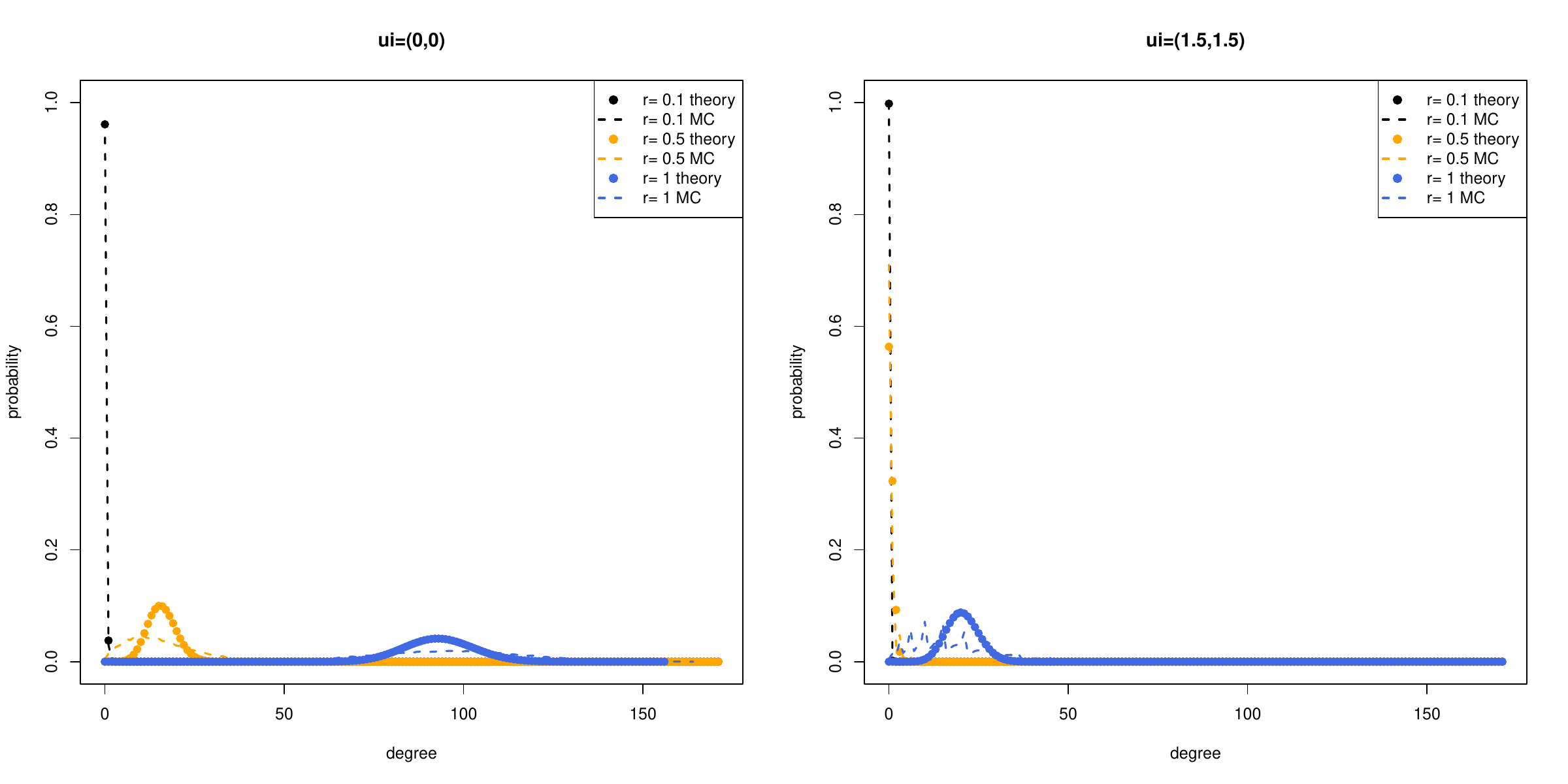}
  \caption{Comparison between empirical (dashed line) and Poisson approximation (points) of the order $k=3$ degree distribution conditional on the latent coordinate $u_i$. We take $N=20, \mu=(0,0), \Sigma = \mbox{diag}(1,1)$ and evaluate the distribution for $r_3 \in (0.1, 0.4, 1.0)$. The left plot shows $u_i = \mu$ and the right plot shows $u_i = (1,2)$.}  \label{fig:poisson_apprx_ui_sig1_N=20}
\end{figure}

Figure \ref{fig:pek_given_ui_sig2} evaluates the approximation for the intersection probabilities when $\Sigma = \mbox{diag}(1,2)$. Figure \ref{fig:poisson_apprx_ui_sig2} evaluates the Poisson approximation for the order $k=3$ degree distribution when $N=10$ and $\Sigma = \mbox{diag}(1,2)$. Figure \ref{fig:poisson_apprx_ui_sig1_N=20} evaluates the Poisson approximation for the order $k=3$ degree distribution when $N=20$ and $\Sigma = \mbox{diag}(1,1)$.

\section{Alternative hypergraph models}
\label{sec:altern-hypergr-model}

\subsection{\cite{stasi2014} model}
\label{sec:beta-model}

In this model each node in the hypergraph is assigned a parameter which controls its tendency to form edges, and we denote this parameter by $\beta_i$, for $i=1,2,\dots,N$. Let $y_{e_k} = 1 $ denote the presence of the hyperedge $e_k = \{ i_1, i_2, \dots, i_k\} \subseteq \{1,2,\dots,N\}$ for $k \geq 2$. The probability of the hyperedge $e_k$ occurring is then given by
\begin{align}
  p( y_{i_1 i_2 \dots i_k} = 1 ) = \dfrac{\exp\{ \beta_{i_1} + \beta_{i_2} + \dots \beta_{i_k} \} }{ 1 + \exp\{ \beta_{i_1} + \beta_{i_2} + \dots \beta_{i_k} \}}.
\end{align}
Since the hyperedges are assumed to occur independently conditional on $\bm{\beta}=(\beta_1, \beta_2, \dots,$ $\beta_N)$, the likelihood is obtained by taking the product of Bernoulli likelihoods over all possible hyperedges $\mathcal{E}_{N,K}$. This likelihood can be shown to belong to the exponential family. \cite{stasi2014} introduce several variants of this model, but we only consider the above model for the study in Section \ref{sec:modeldepth}.

\subsection{\cite{ng2018} model}
\label{sec:citeng2018-model}

The model introduced by \cite{ng2018} assumes that hyperedges can be clustered according to their topic and size. In this context, the topic clustering implies that the hyperedges can be partitioned into latent classes and the probability of a node belonging to a hyperedge depends on its latent class. As an example, consider a coauthorship network where papers are represented as hyperedges. We may classify papers according to their academic discipline and impose that certain authors are more likely to contribute to papers within different disciplines. The size clustering is with respect to the hyperedge order, and this allows the model to capture variation in the size of hyperedges. To specify this model, we assume $T$ topic clusters and $S$ size clusters. It is assumed that the $i^{th}$ node belongs to a hyperedge with size label $s$ and topic label $t$ with probability $\alpha_s \phi_{it}$, so that $\bm{\alpha}=(\alpha_1,\alpha_2,\dots,\alpha_S)$ controls the size clusters and $\bm{\phi} = \{ \phi_{it} \}_{i=1,2,\dots,N, t=1,2,\dots,T}$ controls the topic clusters. Additionally, we let $\pi = (\pi_1, \pi_2, \dots, \pi_T)$ and $\tau = (\tau_1, \tau_2, \dots, \tau_S)$ denote the prior topic and size assignment probabilities, respectively. To write down the likelihood, we let $x_{ij}=1$ indicate that the $i^{th}$ node belongs to the $j^{th}$ hyperedge, $z_{jt}^{(1)}=1$ indicate that the $j^{th}$ hyperedge has topic label $t$, and $z_{js}^{(2)}=1$ indicate that the $j^{th}$ hyperedge has size label $s$. The likelihood is then given by
\begin{align}
  \mathcal{L}( x, z^{(1)}, z^{(2)}; \theta ) = \prod_{j=1}^M \prod_{t=1}^T \prod_{s=1}^S \left[ \pi_t \tau_s \prod_{i=1}^N (\alpha_s \phi_{it})^{x_{ij}} (1 - \alpha_s \phi_{it})^{1 - x_{ij}} \right]^{z_{jt}^{(1)} z_{js}^{(2)}}.
\end{align}
Finally, to ensure the model is identifiable, we set $\alpha_S=1$. \cite{ng2018} also introduce a version of this model which only assumes a topic clustering, but we do not use this for our study in Section \ref{sec:modeldepth}.

\section{Simulation set-up for Section \ref{sec:modeldepth}}
\label{sec:sims-set-up}

The model parameters for the cases described in Table \ref{table:cases_summary} in Section \ref{sec:modeldepth} are provided in Table \ref{table:cases}.

\begin{figure}[t]
\begin{minipage}[t]{\linewidth}
\def\arraystretch{.9}
  \begin{tabular}{ p{3cm} | p{1cm} | p{9cm} }
    Model & Case & Parameters   \\
    \hline  \hline
    \multirow{2}{3cm}{ \cite{stasi2014} } & 1 & $\beta_{i}=-1.4$ for $i=1,2,\dots,N$ \\
    \cline{2-3}
    & 2 & $\bm{\beta}=(-0.5, -0.53, \dots, -1.97, -2)$ \\
    \hline
    \multirow{7}{3cm}{ \cite{ng2018}  } & \multirow{1}{1cm}{1} & $G=K=1$, $a=1, \phi_{i1}=0.075, \pi=1, \tau=1$  \\
    \cline{2-3}
          & \multirow{2}{1cm}{2} & $G=3, K=1$, $a=1, \phi_{i1}=0.25$ for $i \in \mathcal{A}$, $\phi_{i2}=0.25$ for $i \in \mathcal{B}$, $\phi_{i3}=0.25$ for $i \in \mathcal{C}$,   $\pi = (1/3, 1/3, 1/3), \tau=1$\\
    \cline{2-3} 
          & \multirow{2}{1cm}{3} & $G=1, K=3$, $a=(0.2, 0.5, 1)$,   \\
          & & $\phi_{i1}=0.15, \pi = 1, \tau=(1/3, 1/3, 1/3)$\\
    \cline{2-3}
          & \multirow{3}{1cm}{4} & $G=2, K=3$, $a=(0.4,1), \phi_{i1}=0.3$ for $i \in \mathcal{A}$, $\phi_{i2}=0.3$ \\
    & & for $i \in \mathcal{B}$, $\phi_{i1}=\phi_{i2}=0.2$ for $i \in \mathcal{C}$, $\pi = (1/2, 1/2), \tau=(1/3,1/3,1/3)$ \\
    \hline
    \multirow{10}{3cm}{ LSH } & \multirow{2}{1cm}{1} & $\bm{r}=(0.18,0.3,0.35)$, $\mu=(0,0)$, $\Sigma = 0.25 \left( \begin{smallmatrix} 1 & 0.9 \\ 0.9 & 1\end{smallmatrix} \right)$, \\
    && $\bm{\psi}_0=(0.01, 0.01, 0.01)$, $\bm{\psi}_1=(0.01, 0.01, 0.01)$ \\
    \cline{2-3}
          & \multirow{2}{1cm}{2} & $\bm{r}=(0.18,0.3,0.35), \mu=(0,0), \Sigma = 0.25 \left( \begin{smallmatrix} 1 & 0 \\ 0 & 1\end{smallmatrix} \right),$\\
    && $\bm{\psi}_0=(0.01, 0.01, 0.01)$, $\bm{\psi}_1=(0.01, 0.01, 0.01)$  \\
    \cline{2-3}
          & \multirow{2}{1cm}{2} & $\bm{r}=(0.2,0.3,0.35), \mu=(0,0), $ $\Sigma = 0.25 \left( \begin{smallmatrix} 1 & 0 \\ 0 & 1\end{smallmatrix} \right),$\\
&&  $\bm{\psi}_0=(0.01, 0.01, 0.01)$, $\bm{\psi}_1=(0.01, 0.5, 0.01)$ \\
    \cline{2-3}
          & \multirow{2}{1cm}{2} & $\bm{r}=(0.1,0.35,0.4), \mu=(0,0),$ $\Sigma = 0.25 \left( \begin{smallmatrix} 1 & 0 \\ 0 & 1\end{smallmatrix} \right),$  \\
&& $\bm{\psi}_0=(0.01, 0.01, 0.01)$, $\bm{\psi}_1=(0.01, 0.01, 0.01)$ \\
    \cline{2-3}
          & \multirow{2}{1cm}{2} & $\bm{r}=(0.18,0.3,0.35), \mu=(0,0), $ $\Sigma = 0.25 \left( \begin{smallmatrix} 1 & 0 & 0 \\ 0 & 1 & 0 \\ 0 & 0 & 1\end{smallmatrix} \right)$,\\
          &&   $ \bm{\psi}_0=(0.01, 0.01, 0.01)$, $\bm{\psi}_1=(0.01, 0.01, 0.01)$ \\
  \end{tabular}
\captionof{table}{Cases for each hypergraph model considered in the model depth comparison study. The case numbers correspond to the labels in Figures \ref{fig:beta}, \ref{fig:murphy} and \ref{fig:lsm}, and are described in Table \ref{table:cases_summary}. Throughout we take $N=50$ and, where appropriate, $K=4$.} \label{table:cases}
\vspace{.5cm}
\end{minipage}
\end{figure}

\section{Assessing Model Fit}
\label{app:privpost}

\begin{figure}[t]
  \centering
  \includegraphics[trim={0 0cm 0 1.5cm}, clip,width=.4\textwidth]{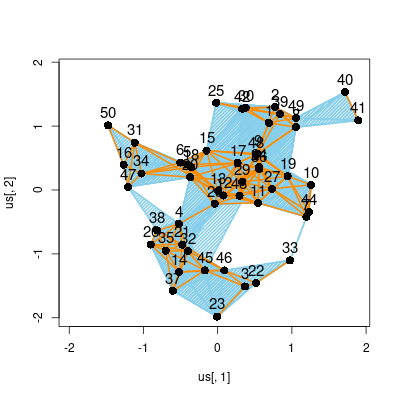}
  \caption{Visualisation of simulated data with $N=50, K=3, \bm{r}=(0.35,0.45), \bm{\upvarphi}^0=\bm{\upvarphi}^1=(0.001,0.001), \mu = (0,0)$ and $\Sigma = I_2$. }
  \label{fig:simdat_S61}
\end{figure}

\begin{figure}[t]
  \centering
  \begin{subfigure}[t]{\textwidth}
    \centering
    \includegraphics[trim={0 0cm 0 1.5cm}, clip, width=.6\textwidth]{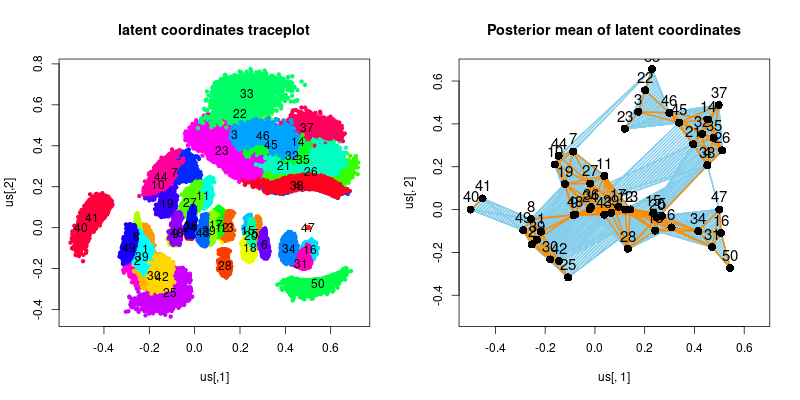} 
    \caption{Left: posterior samples of $\bm{U}$, Right: posterior mean of latent coordinates and observed hyperedges.}
  \end{subfigure}
  \begin{subfigure}[t]{\textwidth}
    \centering
    \includegraphics[trim={0 0cm 0 0cm}, clip,width=\textwidth]{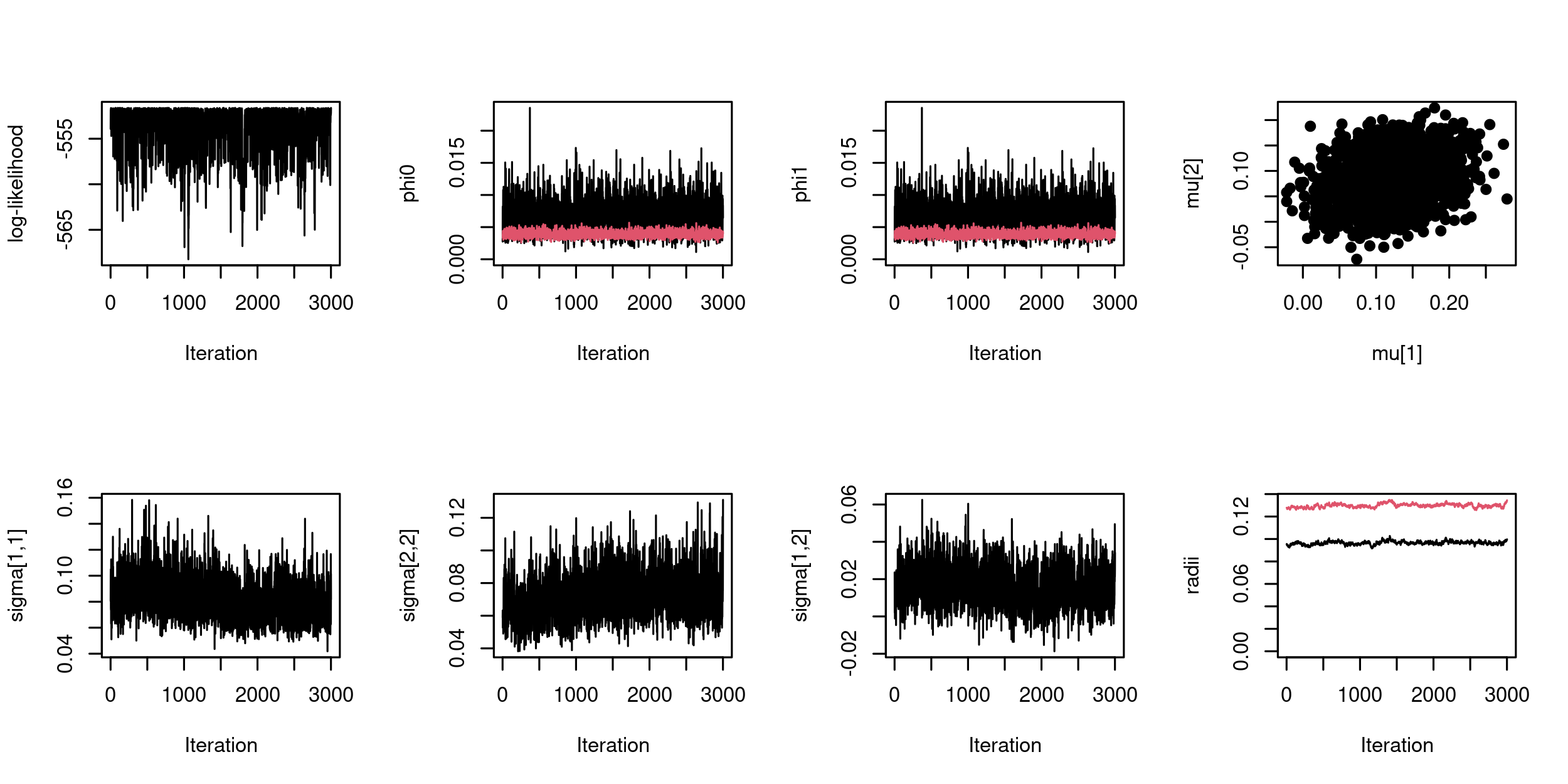}
    \caption{Top, left to right: traceplots of $\mathcal{L}$, $\bm{\upvarphi}^0$, $\bm{\upvarphi}^1$, $\mu$. Bottom, left to right: traceplots of $\Sigma_{1,1}$, $\Sigma_{2,2}$, $\Sigma_{1,2}$ and $\bm{r}$, where $\Sigma_{i,j}$ denotes the $(i,j)^{th}$ entry of $\Sigma$. Where appropriate, $k=2$ and $k=3$ parameters are plotted in black and red, respectively.}
  \end{subfigure}
  \caption{Summary of MCMC output for simulated data with $N=50$ and $K=3$, as shown in Figure \ref{fig:simdat_S61}.}
  \label{fig:simdat_S61_fit}
\end{figure}

We provide a visualisation of the simulated data example used in Section \ref{sec:privspost} in Figure \ref{fig:simdat_S61} and a summary of the MCMC output in Figure \ref{fig:simdat_S61_fit}.

\section{Details for Scalability study}
\label{app:scalability}

\begin{figure}[t]
  \centering
  \includegraphics[width=.7\textwidth]{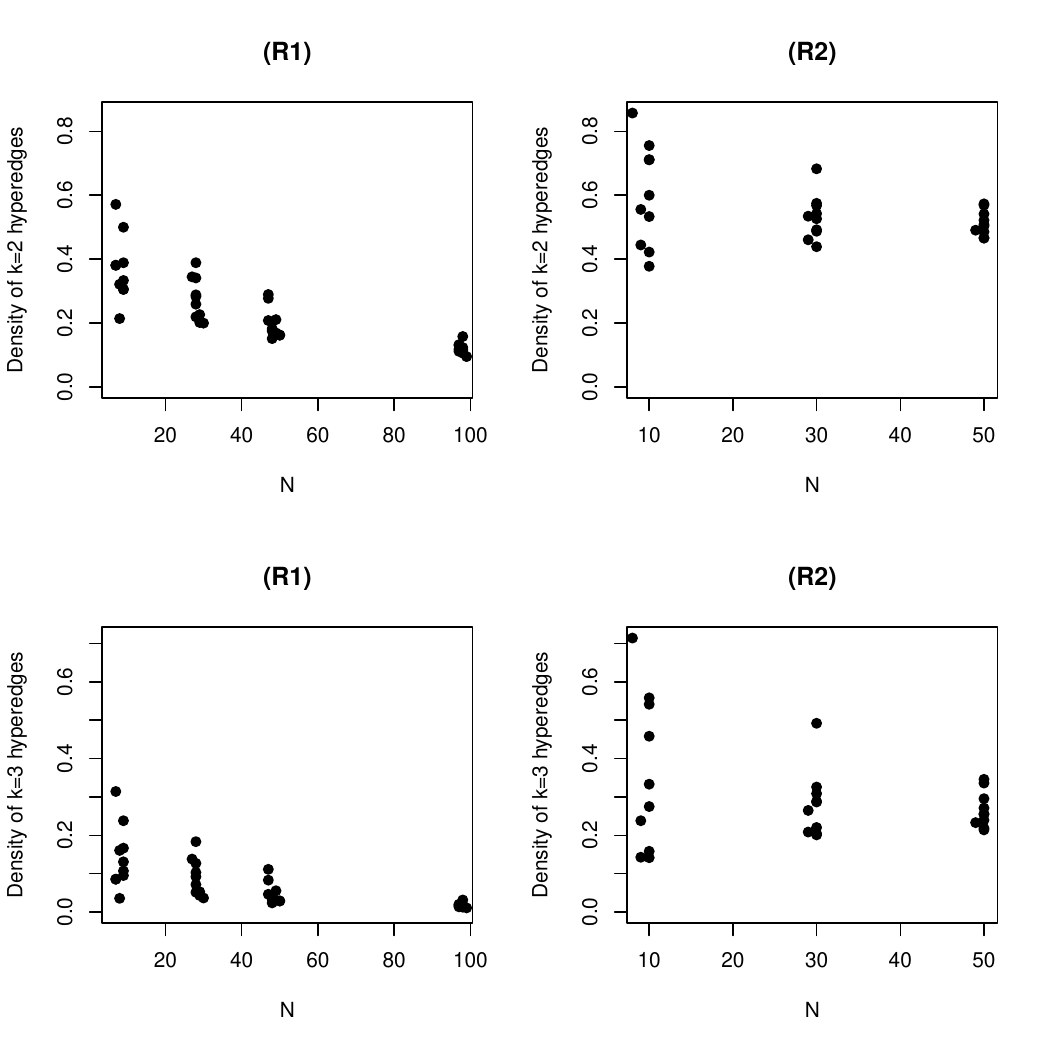}
  \caption{Density of simulated examples under regimes (R1) and (R2). Data were simulated from the generative model outlined in Algorithm \ref{alg:modhyp} with $N$ approximately equal to $(10, 30, 50, 100)$. (R1) and (R2) are controlled to be sparser and denser, respectively, and details of the parameters are given in Table \ref{table:scalability_prms}. The top row shows the density for order 2 hyperedges and the bottom row shows the density for order 3 hyperedges.} \label{fig:scalability_dens}
\end{figure}

For the study presented in Section \ref{sec:scalability}, we simulate data under regimes (R1) and (R2) from the generative model outlined in Algorithm \ref{alg:modhyp} with $N$ approximately equal to $(10, 30, 50, 100)$, $d=2$ and $K=3$. To control the density between these regimes we vary $\bm{r}$ and $\Sigma$ and details of the specific parameters are given in Table \ref{table:scalability_prms}. Note that the simulated examples do not exactly specify $N$, but instead allow hypergraphs whose largest connected component is close to $N$ to avoid situations in which the simulated data contain isolated nodes or components. For each case, we simulate 10 hypergraphs and Figure \ref{fig:scalability_dens} shows the density for each example as a function of $N$. As expected, we see that (R1) corresponds to sparser hypergraphs than (R2).

\begin{table}
\begin{center}
\begin{tabular}{ l | c | c | c | c | c }
  & approximate $N$  & $\bm{r}$ & $\mu$ & $\Sigma$ & $\bm{\upvarphi}$ \\
  \hline \hline
  \multirow{4}{1cm}{ (R1) }  & 10 & (0.75, 0.85) & (0,0) & $2 \bm{I}_2 $ & (0.001, 0.001) \\
  & 30 & (0.75, 0.85) & (0,0) & $2.5 \bm{I}_2 $ & (0.001, 0.001) \\
  & 50 & (0.75, 0.85) & (0,0) & $3 \bm{I}_2 $ & (0.001, 0.001) \\
  & 100 & (0.75, 0.85) & (0,0) & $4.5 \bm{I}_2 $ & (0.001, 0.001) \\
  \hline
    \multirow{4}{1cm}{ (R2) }  & 10 & (1.4, 1.5) & (0,0) & $2 \bm{I}_2 $ & (0.001, 0.001) \\
  & 30 & (1.4, 1.5) & (0,0) & $2.5 \bm{I}_2 $ & (0.001, 0.001) \\
  & 50 & (1.4, 1.5) & (0,0) & $3 \bm{I}_2 $ & (0.001, 0.001) \\
  & 100 & (1.4, 1.5) & (0,0) & $4.5 \bm{I}_2 $ & (0.001, 0.001) \\
\end{tabular}
\end{center}
\caption{Summary of parameter values for simulating data according to (R1) and (R2).} \label{table:scalability_prms}
\end{table}

\section{Model Misspecification}
\label{sec:misspec}
\begin{figure}[t]
  \centering
  \begin{minipage}{1.0\linewidth}
    \begin{tabular}{p{1cm}|p{9cm}|p{3.75cm}}
      Index & $\bm{U}$ misspecification & $\bm{r}$ misspecification \\
      \hline \hline
      1 & None & None \\
      \hline
      2 & $u_i \sim \sum_{c=1}^C \lambda_c \mathcal{N}( \mu_c, \Sigma_c )$, $C=2$ and distinct clusters & None \\
      \hline
      3 & $u_i \sim \sum_{c=1}^C \lambda_c \mathcal{N}( \mu_c, \Sigma_c )$, $C=2$ and fuzzy clusters & None \\
      \hline
      4 & $u_i$ sampled from a homogeneous Poisson point process & None\\
      \hline
      5 & None & Simplicial: $r_k=r_{k-1}$\\
      \hline
      6 & $u_i \sim \sum_{c=1}^C \lambda_c \mathcal{N}( \mu_c, \Sigma_c )$, $C=2$ and distinct clusters & Simplicial: $r_k=r_{k-1}$\\
      \hline
      7 & $u_i \sim \sum_{c=1}^C \lambda_c \mathcal{N}( \mu_c, \Sigma_c )$, $C=2$ and fuzzy clusters & Simplicial: $r_k=r_{k-1}$\\
      \hline
      8 & $u_i$ sampled from a homogeneous Poisson point process & Simplicial: $r_k=r_{k-1}$\\
    \end{tabular}
    \captionof{table}{Types of misspecification.} \label{table:misspec}
  \end{minipage}
\end{figure}

Here we consider the robustness of our modelling approach under different cases of misspecification, and this allows us to assess the suitability of our model for a range of hypergraph data which do not satisfy our modelling assumptions. Table \ref{table:misspec} summarises the types of misspecification we consider, where we focus on different assumptions on $\bm{U}$ and $\bm{r}$. We note that alternative cases of misspecification would be interesting to explore, such as node specific radii or non-homogeneous errors, however simulating data for reasonable $N$ with these assumptions requires significant computational cost.

For each of the 8 cases described in Table \ref{table:misspec}, we simulate a hypergraph with $N=50$ and $K=3$ and fit our model. Then, we examine the predictive distributions and compare these to the truth by inspecting 1) the degree distribution, 2) the number of occurrences of the motifs depicted in Figures \ref{fig:m3}, \ref{fig:h1}, \ref{fig:h2} and \ref{fig:h3}, 3) the density of order 3 hyperedges and 4) the number of clusters in the latent representation, as obtained by the gap statistic of a k-means clustering using the \texttt{cluster} package in R \citep{maechler2019}.

\begin{figure}[t]
  \centering
\begin{subfigure}[t]{\textwidth}
  \includegraphics[trim={1cm 1cm 0cm 0cm}, clip, width = .9\textwidth]{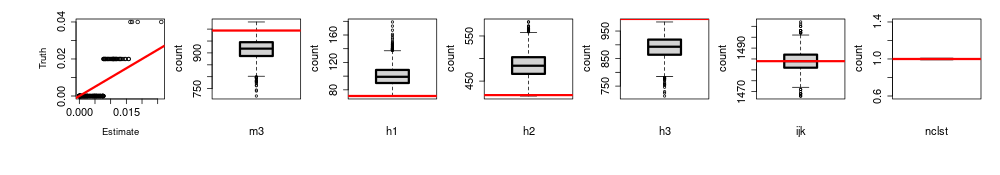}
  \end{subfigure}
\begin{subfigure}[t]{\textwidth}
  \includegraphics[trim={1cm 1cm 0cm .5cm}, clip, width = .9\textwidth]{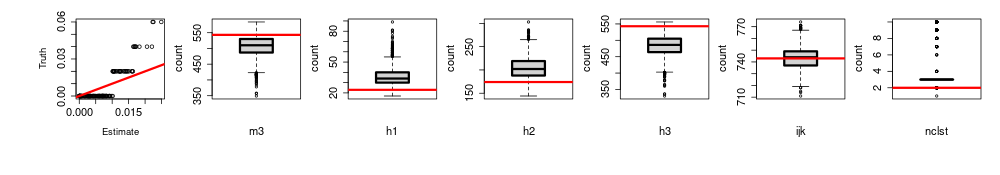}
  \end{subfigure}
\begin{subfigure}[t]{\textwidth}
  \includegraphics[trim={1cm 1cm 0cm 0.5cm}, clip, width = .9\textwidth]{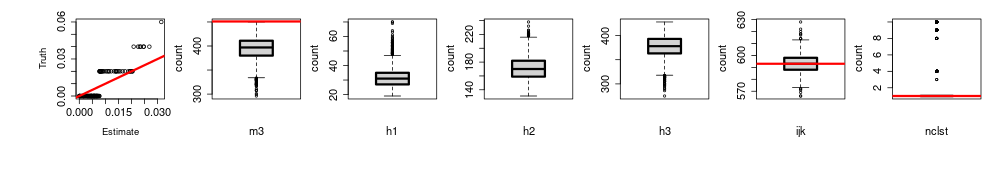}
  \end{subfigure}
\begin{subfigure}[t]{\textwidth}
  \hspace{-.7cm}
  \includegraphics[trim={0cm 1cm 0cm 0.5cm}, clip, width = .93\textwidth]{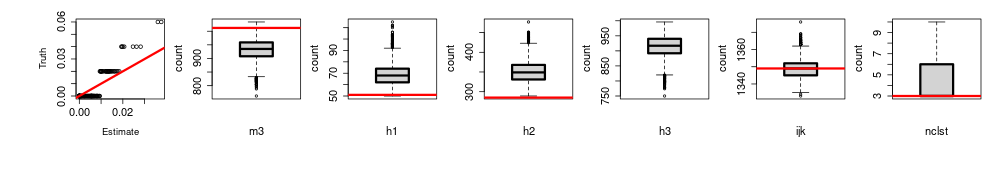}
  \end{subfigure}
\begin{subfigure}[t]{\textwidth}
  \includegraphics[trim={1cm 1cm 0cm 0.5cm}, clip, width = .9\textwidth]{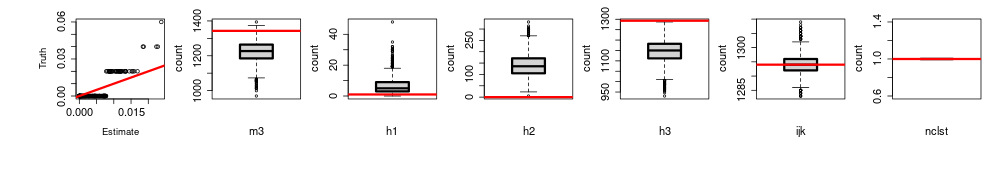}
  \end{subfigure}
\begin{subfigure}[t]{\textwidth}
  \includegraphics[trim={1cm 1cm 0cm 0.5cm}, clip, width = .9\textwidth]{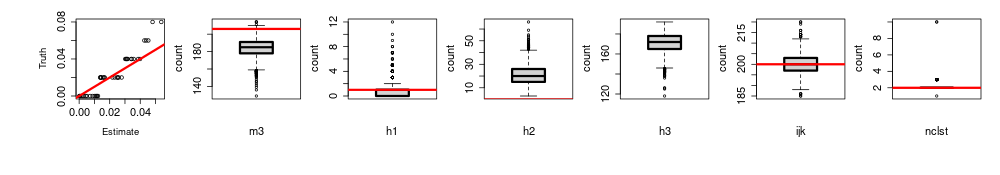}
  \end{subfigure}
\begin{subfigure}[t]{\textwidth}
  \includegraphics[trim={1cm 1cm 0cm 0.5cm}, clip, width = .9\textwidth]{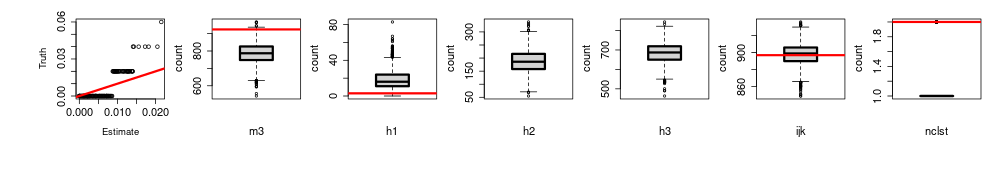}
  \end{subfigure}
\begin{subfigure}[t]{\textwidth}
  \includegraphics[trim={1cm 0cm 0cm 0.5cm}, clip, width = .9\textwidth]{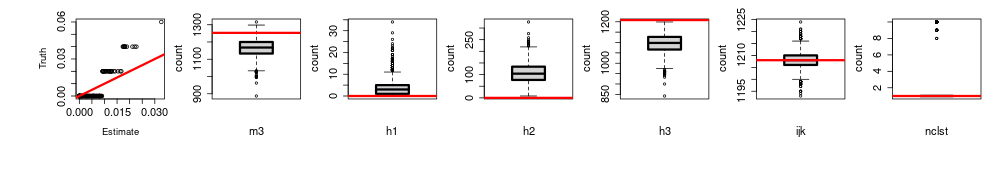}
  \end{subfigure}
  \caption{Summary of misspecification simulation study. Left to right: average degree distribution, number of triangles (Figure \ref{fig:m3}, number of Figure \ref{fig:h1}, number of Figure \ref{fig:h2}, number of Figure \ref{fig:h3}, density of order 3 hyperedges and number of clusters in latent representation. Each row corresponds to the misspecification cases summarised in Table \ref{table:misspec}. The $y$ axes and $x$ axes show the quantiles of the posterior and prior predictives, respectively, and the red lines correspond to the line $y = x$. }   \label{fig:misspec}
\end{figure}

The results of our study are presented in Figure \ref{fig:misspec}, where the $i^{th}$ row corresponds to the $i^{th}$ case detailed in Table \ref{table:misspec}. We generally observe a close correspondence between the truth and predictive distributions, however we see an overall poorer performance when the distribution of the latent coordinates differs. It is interesting to note that, although there is generally a good fit in terms of the degree distribution, we see many of the motifs are over or under predicted by the posterior predictives. This reflects the constraints implied by the latent representation. Finally, we note that there is occasional disagreement in the number of clusters of the latent coordinates. From this study it is clear that our model imposes certain properties on the predictive distributions and it is important to verify whether or not these are appropriate for a specific data example.

\section{Real Data Example Plots}
\label{app:data_examples}

\begin{figure}[t]

  \begin{subfigure}[t]{\textwidth}
    \centering
    \includegraphics[width=\textwidth]{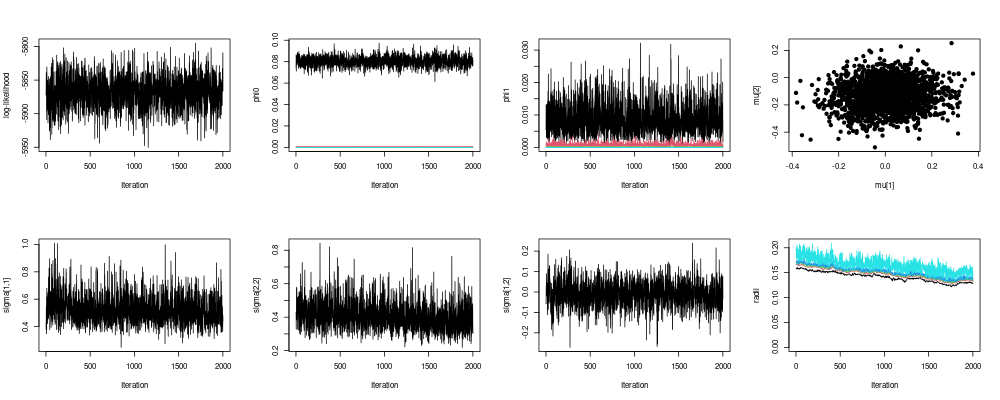}
    \caption{Grocery example from Section \ref{sec:grocery}.} \label{fig:gro_nonus_trc}
  \end{subfigure}
  \begin{subfigure}[t]{\textwidth}
    \centering
    \includegraphics[width=\textwidth]{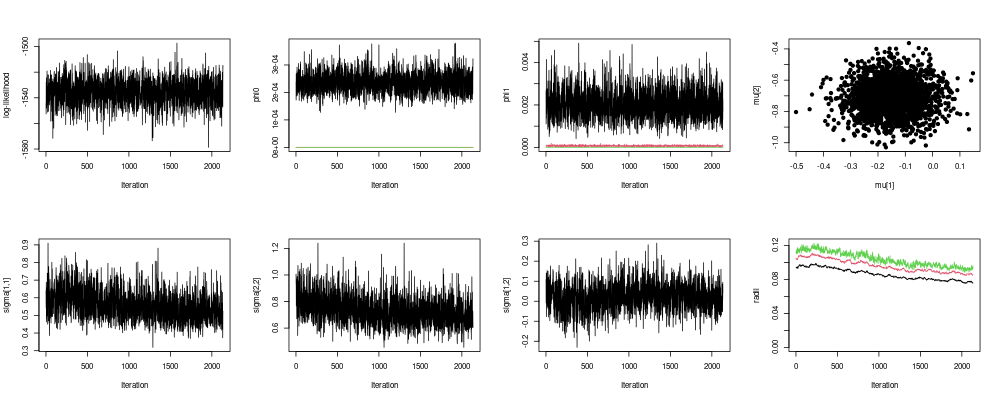}
    \caption{Coauthorship example from Section \ref{sec:coauthorship}.} \label{fig:coauth_nonus_trc}
  \end{subfigure}
  \caption{MCMC post-burn-in traceplots for all parameters except the latent positions for the Grocery and coauthorship data examples. For each figure we have: Top, left to right: log-likelihood, $\bm{\upvarphi}^{(0)}, \bm{\upvarphi}^{(1)}$ and $\mu$. Bottom, left to right: $\Sigma_{1,1}, \Sigma_{2,2}, \Sigma_{1,2}$ and $\bm{r}$, where $\Sigma_{i,j}$ indicates the $\{i,j\}^{th}$ entry of $\Sigma$.} \label{fig:coauth_grocery}

\end{figure}

\begin{figure}[t]
    \centering
  \begin{subfigure}[t]{.45\textwidth}
    \centering
    \includegraphics[width=.8\textwidth, trim={0 0 0 2cm}, clip]{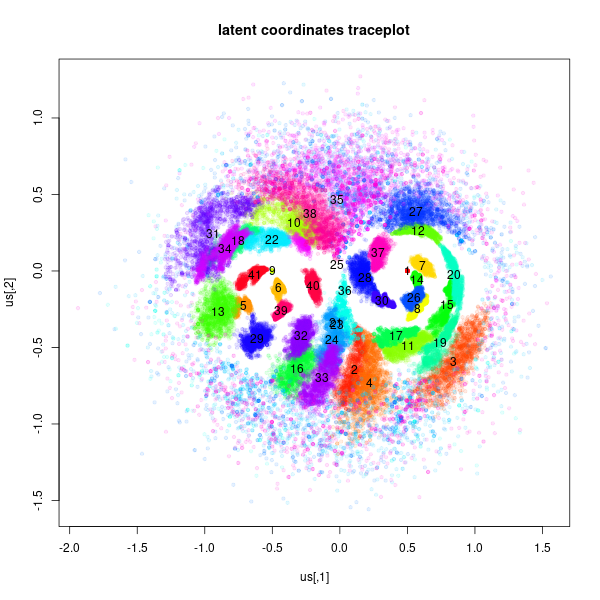}
    \caption{Grocery example from Section \ref{sec:grocery}.} \label{fig:gro_us_trc}
  \end{subfigure}
  \begin{subfigure}[t]{.45\textwidth}
    \centering
    \includegraphics[width=.8\textwidth, trim={0 0 0 2cm}, clip]{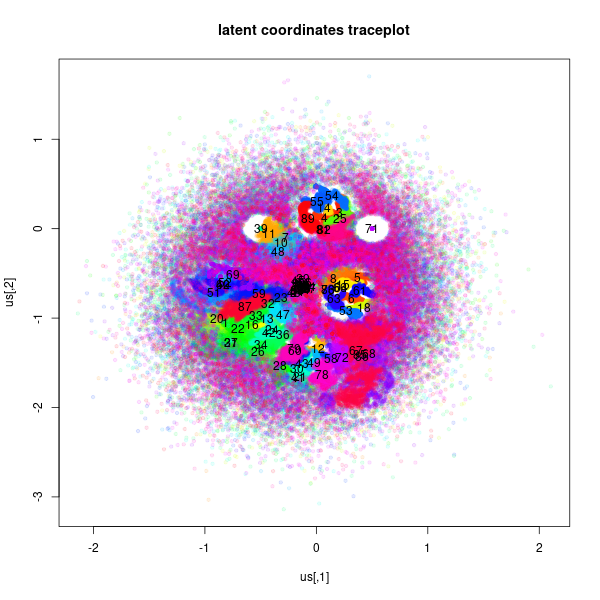}
    \caption{Coauthorship example from Section \ref{sec:coauthorship}.} \label{fig:coauth_us_trc}
  \end{subfigure}
  \caption{MCMC post-burn-in latent position $\bm{U}$ traceplots for the grocery and coauthorship dataset examples.} \label{fig:coauth_traceplot}
\end{figure}

MCMC traceplots for the grocery and coauthorship dataset are given in Figures \ref{fig:gro_nonus_trc}, \ref{fig:gro_us_trc} and \ref{fig:coauth_nonus_trc}, \ref{fig:coauth_us_trc}, respectively.

\end{document}